\documentclass[12pt,a4paper]{article}
\usepackage{amsmath}
\usepackage{amssymb}
\usepackage{amsfonts}
\usepackage{amscd}
\usepackage{delarray}
\usepackage{graphicx}
\usepackage{hyperref}
\usepackage{cancel}

\usepackage{amsmath,amsmath,amsfonts,amssymb,color,bbm}

\usepackage{graphicx}

\def\Symp#1,#2,#3,#4.{\left[\left(\begin{array}{c}#1\\#2\end{array}\right),\left(\begin{array}{c}#3\\#4\end{array}\right)\right]}
\def\Vec#1,#2.{\left(\!\begin{array}{c}#1\\#2\end{array}\!\right)}
\def\vec#1,#2.{{#1\choose{#2}}}
\newcommand{\ket} [1] {\vert #1 \rangle}
\newcommand{\bra} [1] {\langle #1 \vert}

\newcommand{\beq}{\begin{equation}}
\newcommand{\eeq}{\end{equation}}
\newcommand{\beqa}{\begin{eqnarray}}
\newcommand{\eeqa}{\end{eqnarray}}
\newcommand{\bx}{{\bf x}}
\newcommand{\by}{{\bf y}}
\newcommand{\cE}{{\cal E}}

\definecolor{redcom}{rgb}{1,0.1,0.2}
\definecolor{querycol}{rgb}{0.2,0.2,1}

\definecolor{purplerep}{rgb}{1,0.1,1}

\definecolor{geen}{rgb}{0.6,0.6,0.9}

\begin{document}

\title{Can quantum systems succumb to their own (gravitational) attraction? }
\date{}
\author{Samuel Colin\footnote{Centre for Quantum Dynamics, Griffith University, 170 Kessels Road, Brisbane, QLD 4111, Australia.
Department of Physics and Astronomy, Clemson University, 120-A Kinard Laboratory, Clemson, SC 29631-0978, USA. \texttt{email: scolin@clemson.edu}},
Thomas Durt\footnote{Ecole Centrale de Marseille, Institut Fresnel, Domaine Universitaire de Saint-J\'er\^ome, Avenue Escadrille Normandie-Ni\'emen 
13397 Marseille Cedex 20, France. \texttt{email: thomas.durt@centrale-marseille.fr}},
Ralph Willox\footnote{Graduate School of Mathematical Sciences, the University of Tokyo, 3-8-1 Komaba, Meguro-ku, 153-8914 Tokyo, Japan. \texttt{email: willox@ms.u-tokyo.ac.jp}}
}
\maketitle
\vglue -1.8truecm
 
\abstract{The gravitational interaction is generally considered to be too weak to be easily submitted to systematic experimental investigation in the quantum, microscopic, domain. In this paper we attempt to remedy this situation by considering the gravitational influence exerted by a crystalline nanosphere of mesoscopic size on itself, in the semi-classical, mean field, regime. We study in depth the self-localisation process induced by the corresponding non-linear potential of (gravitational) self-interaction. In particular, we characterize the stability of the associated self-collapsed ground state and estimate the magnitude of the corrections that are due to the internal structure of the object (this includes size-effects and corrections due to the discrete, atomic, structure of the sphere). Finally, we derive an approximated,  gaussian, dynamics which mimics several essential features of the self-gravitating dynamics and, based on numerical results derived from this model, we propose a concrete experimental setting which we believe might, in the foreseeable future, reveal the existence of gravitational self-interaction effects.}


\section{Introduction} In the framework of atomic interferometry it has been observed in several experiments that, at the atomic scale, isolated quantum systems are coupled to the terrestrial gravitational field in the standard manner. That is, they behave as test-particles imbedded in the standard, classical, gravitational field. This explains why, for instance, at the exit of a two-arm interferometer \cite{Chu},  atoms exhibit  a phase difference that is equal, up to a factor $1/\hbar$, to the integral of the Lagrangian that contains the ``usual'' gravitational potential 
$\oint (m v^2/2-m \phi(z) )~\! {d}t$ along a closed loop enclosing both arms of the interferometer (where $v=v_\mathrm{de\,Broglie}={{d}r / {d}t}={h / (m \lambda_\mathrm{de\,Broglie})}$ is the group velocity of the atoms and $\phi(z)$ the local classical gravitational potential).

This kind of experiment confirms that individual quantum systems couple to the external gravitational field in the most predictable and natural way. What it does not allow to establish however is how, in turn, gravity couples to these quantum systems. Also here, the most ``natural'' candidate, as far as one does not quantize gravity, is the mean field coupling proposed by M{\o}ller \cite{Moeller} and Rosenfeld \cite{Rosenfeld}  :

\begin{equation} R_{\mu\nu}-{1\over 2} g_{\mu\nu}R={8\pi G\over c^4} \bra{\Psi}\hat T_{\mu\nu}\ket{\Psi},\end{equation}
with $R_{\mu\nu}$ the Ricci tensor, $g_{\mu\nu}$ the space-time metric, $G$ Newton's constant, $c$, the velocity of light and $\hat T$ the stress-energy tensor. 
In the non-relativistic limit  \cite{giulini2012}, one obtains the Poisson equation
\begin{equation} \Delta V=4\pi G m |\Psi|^2,\end{equation}
where $V$ is the gravitational potential and $m |\Psi|^2$ the density of mass, when we deal with a particle of mass $m$.

As we noted before, $V$ is coupled to matter in the usual way, which allows one to derive the so-called Schr\"odinger-Newton integro-differential equation\footnote{This equation is also often referred to as the (attractive) Schr\"odinger-Poisson equation \cite{Brezzi, Illner, Arriola} or the gravitational Schr\"odinger equation \cite{Jones, Bernstein}. Throughout this paper however we shall use the name ``Schr\"odinger-Newton", which seems to be more or less standard by now in the field of quantum gravity.} from the Schr\"odinger and Poisson equations \cite{Jones} :

\begin{equation}
{i}\hbar\frac{\partial\Psi(t,{\bf x})}{\partial t}=-\hbar^2\frac{\Delta\Psi(t,{\bf x})}{2m}
-Gm^2\int {d}^3 x'\frac{\rho(t,{\bf x'})}{|{\bf x -x'}|}\Psi(t,{\bf x}).\label{NS}
\end{equation}
An immediate consequence of this equation is that even a ``free'' particle will feel its own potential, due to the gravitational source $\rho(t,{\bf x'})=|\Psi(t,{\bf x'})|^2$. In other words, the full energy now contains a contribution from
the gravitational self-energy, proportional to 
\begin{equation}
-{Gm^2\over 2}\int {d}^3 x {d}^3 x'\frac{\rho(t,{\bf x})\rho(t,{\bf x'})}{|{\bf x -x'}|}=-{Gm^2\over 2}\int {d}^3 x {d}^3 x'\frac{|\Psi(t,{\bf x})|^2 |\Psi(t,{\bf x'})|^2}{|{\bf x -x'}|},
\end{equation}
which is the average value (with weight $\rho=|\Psi|^2$) of
\begin{equation}
-{Gm^2\over 2}\int {d}^3 x'\frac{\rho(t,{\bf x'})}{|{\bf x -x'}|}~.
\end{equation}

The stationary form of equation (\ref{NS}) is known in the context of plasma physics as the Choquard equation \cite{Lieb}. It appears to be a useful tool, for instance, when describing the behaviour and properties of white dwarfs inside of which gravitational self-collapse competes with quantum diffusion. 

{ There also exist convincing theoretical arguments  \cite{giulini2012} establishing that the equation (\ref{NS}) can be derived from a general relativistic formulation of the Klein-Gordon and Dirac equations in the limit where $1/c$ and $\hbar$ simultaneously tend to 0.} It is nevertheless by no means certain that this equation correctly describes the way in which a quantum object gravitationally interacts with itself. One of the reasons is that in the microscopic and mesoscopic regimes, the gravitational self-interaction is so tiny that no experiment has been realized yet that can test the disturbance the interaction will cause to the usual Schr\"odinger equation. 

Another reason is that it is known that no such self-interaction exists when we consider the electro-magnetic interaction. For instance, if one considers the hydrogen atom, while taking into take account the Coulomb (repulsive) self-energy of the electron, one has to add a correction 
 \begin{equation}
 {Ke^2\over 2}\int {d}^3 x {d}^3 x'\frac{\rho(t,{\bf x})\rho(t,{\bf x'})}{|{\bf x -x'}|}={Ke^2\over 2}\int {d}^3 x {d}^3 x'\frac{|\Psi(t,{\bf x})|^2|\Psi(t,{\bf x'})|^2}{|{\bf x -x'}|},\label{selfC}
\end{equation}
to the energy (with $K$ the Coulomb constant, and $e$ the charge of the electron). That is, if instead of the ``usual'' Schr\"odinger equation
\begin{equation}
{i}\hbar\frac{\partial\Psi(t,{\bf x})}{\partial t}=-\hbar^2\frac{\Delta\Psi(t,{\bf x})}{2m_e}
-Ke^2\frac{1}{|{\bf x }|}\Psi(t,{\bf x}),\label{LS}
\end{equation}we consider the perturbed equation
\begin{equation}
{i}\hbar\frac{\partial\Psi(t,{\bf x})}{\partial t}=-\hbar^2\frac{\Delta\Psi(t,{\bf x})}{2m_e}
-Ke^2\frac{1}{|{\bf x }|}\Psi(t,{\bf x})+Ke^2\int {d}^3 x'\frac{|\Psi(t,{\bf x'})|^2}{|{\bf x -x'}|}\Psi(t,{\bf x})\label{LSP},
\end{equation}
it is easy to check that such a perturbation induces corrections to the spectrum of the hydrogen atom that should be measurable. For instance, an elementary computation reveals that the energy\footnote{As will become clear in section \ref{2.3} (equation (\ref{energy})), the energy of a solution of equation (\ref{LSP}) is equal to the average value of the ``non-perturbed'' Hamiltonian of equation (\ref{LS}) plus the 
perturbative self-energy (\ref{selfC}).} of ionization of a hydrogen atom would take a value, at the first perturbative order, equal to $11/16$ times the observed value of 13.6 eV (as already noticed by Schr\"odinger himself in 1927 \cite{Annalen}, see e.g. discussion in the conclusion of 
\cite{giulini2012} and references therein). Similarly, the 1S-2S transition would be characterized by an energy that differs radically from the observed value which is close to $0.75\times 13.6$ eV. However, the 1S-2S transition is the quantity for which the agreement between theory and experiment has been validated with the highest degree of accuracy ever reached throughout the whole history of physics. Indeed, it has been measured with an accuracy of the order of thirteen significant digits, and the measurements are in full agreement with the predictions made in the framework of standard quantum electrodynamics\footnote{The frequency that has been measured (thanks to techniques involving a frequency comb that made it possible to measure the Lamb transition frequency with the precision of Cesium atomic clocks) agrees with the predictions of QED within a range of experimental precision that is smaller than 100 Hz (in fact, 1.8 parts in $10^{14}$), so that it constitutes the most stringent test of QED in atoms \cite{cesium}.
}, which definitively rules out the existence of a Coulomb self-interaction of the form (\ref{selfC}). Actually, this is not so surprising as standard QED computations involve renormalisation theory which is precisely aimed at properly accounting for self-energy related effects. One of the teachings of renormalisation theory is that self-energy is, in a sense, incorporated from the beginning in the ``effective'', renormalised, electronic mass $m_e$ and charge $e$ that appear in the equation (\ref{LS}). 

Needless to say, it is still an open question whether or not gravitation is a renormalisable theory. 
It is known that general relativity is not renormalisable, and a correct quantum formulation of gravitation is still unknown. 
We believe that it is precisely for these reasons that it is worthwhile to investigate gravitational self-energy and related effects, 
because these might reveal how it is exactly that gravity couples to quantum systems\footnote{{There is a lot of confusion about what is meant by the non-relativistic limit of the M{\o}ller-Rosenfeld equation and/or semiclassical gravity. Anastopoulos and Hu for instance, showed in a very rather convincing manner \cite{Hu2,Hu1} that if one derives semiclassical gravity from quantum field theory and general relativity, one does not find the Schr\"odinger-Newton equation (\ref{NS}) but a linear equation. Now, it is clear that Rosenfeld did {\it not} accept the idea that gravity was a quantum field and considered it to be classical, with as source term the energy-momentum tensor averaged over quantum degrees of freedom of matter, which is the essence of the Schr\"odinger-Newton equation.}\label{shaky}}. Ultimately, { the main rationale} for believing in the potential validity of the Schr\"odinger-Newton equation is that the analogy with QED fails because there is {\it no} underlying Quantum Gravity of which it would be an approximation. Here, we accept the Schr\"odinger-Newton equation as an ansatz and our goal is to conceive experimental falsifications of it.  In the absence of a convincing theory of Quantum Gravity \cite{CarlipQG}, the Schr\"odinger-Newton approach remains in our eyes a valuable hypothesis, keeping in mind however that its foundational motivation is based on shaky grounds.

Another interesting feature of the Schr\"odinger-Newton equation is its essential non-linear character. Non-linearity is already present at the classical level due to the fact that the gravitational field (metric) also contributes to the stress-energy tensor. At the quantum level, non-linearity is an intrinsic feature of the mean field approximation. This has been noted by a series of physicists (Jones \cite{Jones}, Penrose \cite{penrose} and many others), who insisted on the fact that, if we want to provide a quantum formulation of the gravitational interaction, it is not consistent to neglect the intrinsic nonlinear nature of gravitational self-interaction. In parallel, it was recognised very early on\footnote{By the end of the 1980's, thanks to the works of Ghirardi, Rimini, Weber \cite{grw} and Pearle \cite{pearle}, a nonlinear and stochastic modification of the Schr\"odinger equation was proposed, 
aimed at simultaneously solving the measurement problem and at explaining the localisability of macroscopic objects. Already in the 1920's, de Broglie proposed the so-called double-solution program \cite{debroglie60}, aiming to solve the puzzle of the wave-particle duality. By the 1950's he realized that his program could only be realized thanks 
to  a nonlinear correction to the Schr\"odinger equation. His goal, phrased in a modern language, was ultimately to explain the stability of particles in terms of solitonic properties of wave packets, for which spreading would be counterbalanced by nonlinearities. de Broglie also thought that the pilot-wave theory, which he first presented at the fifth Solvay conference of 1927 \cite{bacval} 
(and which was later rediscovered by Bohm in 1952 \cite{bohm521,bohm522}) was a degenerate double-solution theory, 
in which moving soliton-like solutions have been replaced by point-particles (see ref.\cite{Fargue} for a review).} 
 that gravitational self-interaction could have something to do with spontaneous localisation \cite{Diosi84}.
The main goal of our paper is to discuss the basic properties of the Schr\"odinger-Newton equation -- such as for example the stability of its ground state -- in the light of the spontaneous localisation program, as well as  to discuss some effects that are likely to be tested experimentally.

The paper is structured as follows. 

In section \ref{2}, we revisit Derrick's no-go theorem \cite{Derrick} (subsections \ref{2.1} and \ref{2.2}) and explain why this theorem is not relevant when applied to realistic situations. We also discuss the stability problem in the case of the Schr\"odinger-Newton equation (subsection \ref{2.3}), making use of some well-established 
results for the nonlinear Schr\"odinger equation.

In section \ref{3}, we pay particular attention to the case of a self-interacting homogeneous crystalline nanosphere. We characterize the different effective regimes of self-interaction, as a function of the mass of the nanosphere.  This is done by evaluating the effective self-interaction at the level of the center of mass of the sphere, taking into account the corrections induced by the internal structure of the nanosphere. After reviewing the results  by Di{\'o}si \cite{Diosi} (section \ref{sphere}) and Schmidt \cite{Schmidt}  (section \ref{cristal}) who studied the gravitational self-interaction of, respectively, a homogeneous sphere  and a crystalline solid (taking into account the non-vanishing size of the nanosphere and its atomic nature), we extend their results and provide an exhaustive description of the different regimes of self-interaction that characterize a self-interacting solid, crystalline, nanosphere. 

The main idea underlying our analysis is that the size of the ``self-collapsed'' ground state of a homogeneous crystalline sphere is a monotonically decreasing function of its radius (or mass). In the case where the gravitational self-interaction would act as an effective localisation mechanism, it is therefore expected that for sufficiently high values of the radius (mass) of the sphere, its center of mass will behave as a classical point particle. This justifies the label ``mesoscopic'' which we borrow from Diosi's terminology \cite{Diosi} and which will be used to designate the transition at which the size of the self-collapsed ground state is equal to the radius of the sphere. Accordingly, we shall call ``classical'' and ``quantum'' those regimes for which the size of the ground state is, respectively, smaller than or larger than the radius of the sphere. It is not clear however whether or not quantum coherence ought to disappear spontaneously in the so-called classical regime \cite{Diosi,penrose} , as we shall briefly comment upon in the conclusions.

We pay particular attention to the gap between the mesoscopic regime, already considered by Di{\'o}si, and the ``quantum regime'' (or elementary particle regime) characterized by small masses, i.e. small radii of the sphere if its density is considered to be constant. Typically, we shall consider densities of the order of a ``normal'' water-like density. In the rest of the paper, ``normal'' densities will therefore always correspond to a volumetric density of mass equal to $10^3\,\mathrm{kg}\,\mathrm{m}^{-3}$. 

In section 3 we also fill the gap between the mesoscopic regime and the classical or ``macroscopic'' regime, taking into account the discrete, atomic, nature of the crystal (including Schmidt's results). Finally, in section 3.4, we describe a new, ``nuclear'', regime in which the mass of the sphere is high enough for the gravitationally collapsed ground state of the center of mass wave function to, so as to speak, explore the size of the nuclei of the crystal. These results are encoded in figure 1 which shows a log-log plot of the effective potential corresponding to the gravitational self-energy between a homogeneous crystalline sphere with a
radius of 10 nanometers and a replica of itself, translated by a distance $r$, as a function of $r$. In this figure one distinguishes four regimes, from right to left: the so-called quantum, or elementary particle regime (D) in which the internal structure of the nanosphere can be neglected  compared to the large spread of the wave function for the center of mass of the sphere (from more or less $10^{-7}$m to infinity); the mesoscopic regime (C) in which the size of the sphere is smaller than or equal to the spread of the wave function, but not too small (typically between $10^{-12}$m and $10^{-7}$m); the ``atomic'' regime (B) where the discrete, atomic, structure of the sphere plays a predominant role (between more or less $10^{-15}$m and $10^{-12}$m) and the ``nuclear'' regime (A) (for a spread $r$ between 0 and $10^{-15}$m), where it is essentially the self-energy of each nucleus that contributes to the gravitational self-interaction. 
As we shall show in section \ref{3}, these regimes are successively ``explored'', from (D) to (A), by the wave function for the center of mass of a nanosphere of normal density, when the mass (radius) of the nanosphere increases. 

\begin{figure}
\centering
\includegraphics[width=\textwidth]{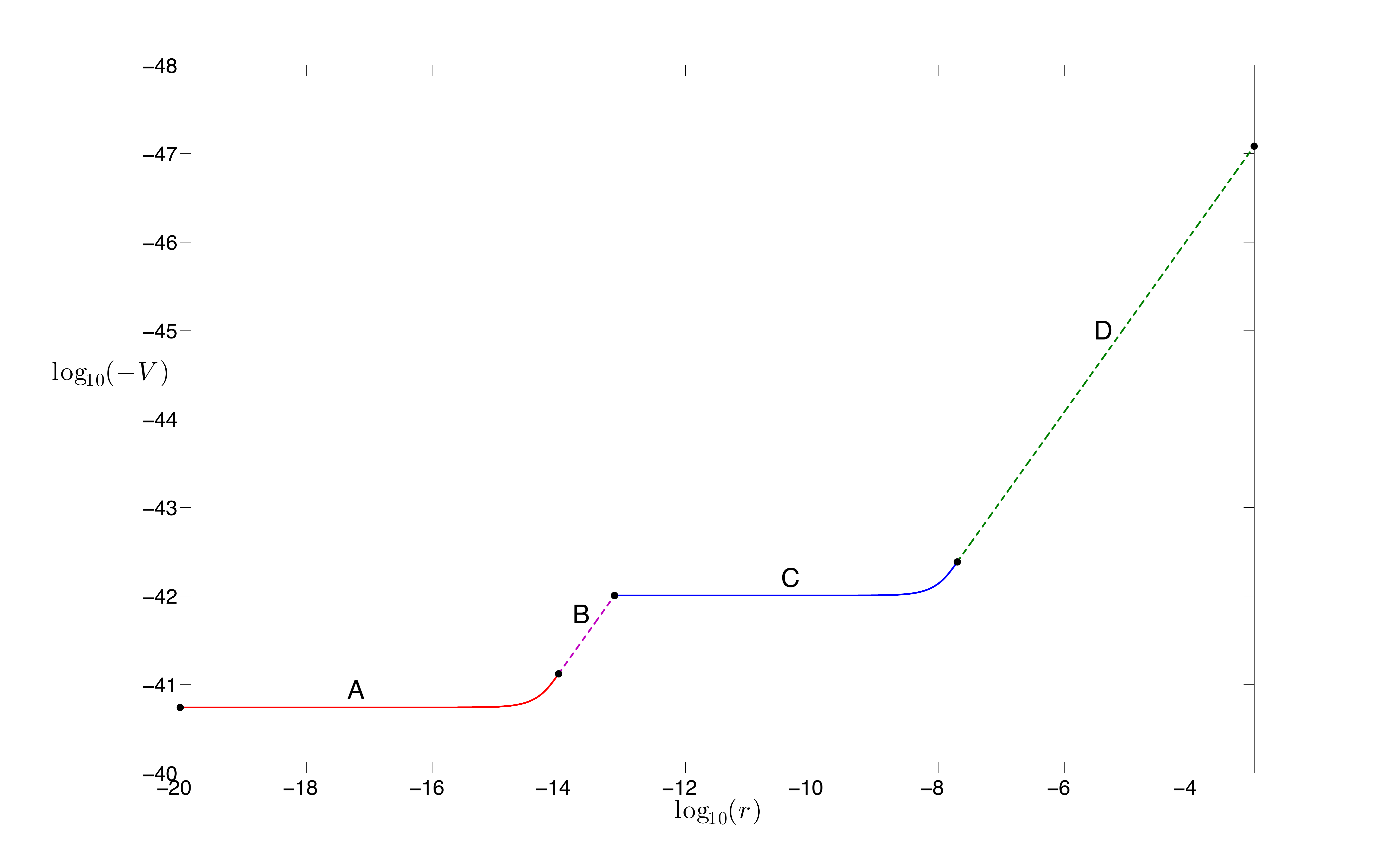}
\caption{\footnotesize\label{fig1}  A log-log plot of the effective potential (expressed in joule), corresponding to the gravitational self-energy between a homogeneous crystalline sphere with a
radius of 10 nanometers and a replica of itself, translated by a distance $r$, as a function of $r$. The density  of the sphere is that of silica (SiO$_2$), that is to say 2.65 times normal density. The labels A, B, C, D refer to the macroscopic nuclear (A), macroscopic atomic (B), mesoscopic (C) and quantum (D) regimes. These regimes are successively explored -- from (D) to (A) -- by the wave function for the center of mass when the radius of the sphere increases, while the spread $r$ of the self-collapsed bound state decreases accordingly.}
\end{figure}

In section 4 we discuss the possible relevance of gravitational self-interaction for modelling spontaneous localization in the context of quantum mechanics. Moreover, we (briefly) present an approximation scheme for tackling the localization problem for a self-gravitating nanosphere, the results of which will be used in the following section (section 5) where we discuss the possible experimental verification of self-gravitational effects. In particular, in section \ref{exp2}, we propose an experimental scheme -- inspired by recent proposals for testing the superposition principle at the level of a mesoscopic nanosphere \cite{aspelPRA,aspelPRL,maqro} -- which we believe could be used to verify the existence of a self-gravitational interaction and we  discuss it in the light of the results presented in the previous sections. Section \ref{conclusions} is devoted to the conclusions and new perspectives. 

Finally, in the Appendix we give a more detailed description of the approximation scheme for the  Schr\"odinger-Newton integro-differential equation (\ref{NS}) introduced in section 4.2, which can be used to verify the stability of the bound state at very low computational cost. The effective potential used in this approximation faithfully mimics several essential features of the Schr\"odinger-Newton equation in the quantum regime. Moreover, in this approximation, if the wave function of the center of mass of the sphere is originally gaussian, it remains gaussian at all times and its evolution reduces to an ordinary differential equation. Our approximation also makes it possible to introduce corrections that take into account the non-zero size and atomic structure of the crystalline sphere, without influencing the computational complexity of the problem.
 
  \section{Stability of a self-gravitating bound state\label{2}}
 Derrick's theorem \cite{Derrick} constituted, for decades, a real scarecrow for what concerns the de Broglie program which tries to explain the stability of well-localized matter waves in terms of an intrinsically nonlinear evolution. In \cite{Derrick} Derrick intended to show that the stability of solutions for a very large class of non-linear Klein-Gordon equations is severely compromised when space counts three or more dimensions, and this, by effectively ruling out the existence of stable, stationary, localized solutions for such equations. Since its publication it has been more or less taken for granted that this result, at the non-relativistic limit,  also prohibits the stability of soliton-like structures described by nonlinear generalizations of the Schr\"odinger equation in ``real'', three dimensional space. In this section we shall use the example of the 1+1 dimensional nonlinear Schr\"odinger equation -- the soliton system par excellence -- to show that, at the very least, Derrick's definition of stability is too narrow to be applicable to the problem he set out to settle, and that his purported no-go theorem is actually irrelevant to the entire discussion of stability of localised solutions to nonlinear evolution equations (regardless of the dimension of the ambient space).
 
\subsection{Derrick's theorem and the nonlinear Schr\"odinger  equation\label{2.1}} 
The 1+1 dimensional nonlinear Schr\"odinger equation, which for simplicity we write here in dimensionless coordinates,
\begin{equation}
{i}~\!\frac{\partial \psi}{\partial s} + \frac{\partial^2 \psi}{\partial z^2} + \big|\psi\big|^2 \psi = 0,\label{NLS}
\end{equation}
can be derived from a variational principle for the action
\begin{equation}
A_{NLS}(\psi) = \iint_{-\infty}^{+\infty}~\!\left[\frac{{i}}{2}\big(\psi^* \frac{\partial \psi}{\partial s} - \psi \frac{\partial \psi^*}{\partial s}\big) - \big|\frac{\partial \psi}{\partial z}\big|^2 + \frac{1}{2} \big|\psi\big|^4\right] {d}z {d}s.
\end{equation}
A standard analysis (see e.g. \cite{Sulem} for an excellent introduction to the subject) shows that this action is invariant
under phase transformations $\psi(z,s)\mapsto e^{{i}\mathrm\varepsilon} \psi(z,s)$, $z$ and $s$ translations, as well as Galilean transformations
\begin{gather}
(z,s,\psi) ~\rightarrow~(Z,S,\Psi)\\
Z=z- v s~\!,\quad S= s~\!,\quad\psi= {e}^{{i}(\frac{v}{2} z + \frac{v^2}{4} s)} ~\Psi,
\end{gather}
for some velocity $v$ of the moving frame $(Z,S)$.

This invariance implies (by an appropriate version of Noether's theorem) that for sufficiently localized solutions to the nonlinear Schr\"odinger equation \eqref{NLS}, the following quantities are preserved by the evolution:
\begin{align}
N(\psi) = \int_{-\infty}^{+\infty}~ \big|\psi\big|^2 {d}z\qquad &(\text{norm})\\
P(\psi) = ~\!{i}\!\!\int_{-\infty}^{+\infty}~\big(\psi^* \frac{\partial \psi}{\partial s} - \psi \frac{\partial \psi^*}{\partial s}\big) {d}z\qquad &(\text{momentum})\\
E(\psi) = \!\int_{-\infty}^{+\infty}~\left(\!\big|\frac{\partial \psi}{\partial z}\big|^2 - \frac{1}{2} \big|\psi\big|^4\right) {d}z\qquad &(\text{energy}).
\end{align}
Notice the factor $\frac{1}{2}$ in the potential energy term in the expression for the energy of the solution $\psi(z,s)$, which arises because of the nonlinearity in the equation. 

Furthermore, the Galilean invariance of the action gives rise to the following relation for the first moment of $\psi(z,s)$
\begin{equation}
R(\psi) = \frac{1}{N(\psi)} \int_{-\infty}^{+\infty} z \big|\psi\big|^2 {d}z~\!:\qquad\quad \frac{{d} R}{{d}s} = \frac{P(\psi)}{N(\psi)},\label{NLS-com}
\end{equation}
which tells us that the center of mass of any sufficiently localized solution to the nonlinear Schr\"odinger equation \eqref{NLS} travels at constant speed $ v= \dfrac{P}{N}$. Any localized structure $\big|\psi(z,s)\big|^2 = \big|\phi(z- v s)\big|^2$ can therefore be put at rest in a moving reference frame by a Galilean transformation and one can look for such structures in the form of stationary solutions
\begin{equation}
\Psi(Z,S) = {e}^{{i}\beta S} \varphi(Z).
\end{equation}
The function $\varphi(Z)$ then satisfies the stationary nonlinear Schr\"odinger equation
\begin{equation}
\frac{\partial^2\varphi}{\partial Z^2} + \big|\varphi\big|^2 \varphi = \beta \varphi,\label{statNLS}
\end{equation}
with $\beta\geq0$. The total energy of such a solution is given by 
\begin{equation}
E(\varphi) = \int_{-\infty}^{+\infty}~\left(\!\big|\frac{\partial \varphi}{\partial Z}\big|^2 - \frac{1}{2} \big|\varphi\big|^4\right) {d}Z.\label{ENLSZ}
\end{equation}

Applied to this particular case, Derrick's reasoning in \cite{Derrick} then goes as follows. For any solution $\varphi(Z)$ to \eqref{statNLS}, consider the function $\varphi_\zeta(Z) = \varphi(\zeta Z)$ for some dilation parameter $\zeta>0$ and define
\begin{align}
E_\zeta &= \int_{-\infty}^{+\infty}~\left(\!\big|\frac{\partial \varphi_\zeta}{\partial Z}\big|^2 - \frac{1}{2} \big|\varphi_\zeta\big|^4\right) {d}Z\label{Ezeta}\\
&= \zeta E_K + \frac{1}{\zeta} E_P,
\end{align}
where
\begin{equation}
E_K = \int_{-\infty}^{+\infty}~\big|\frac{\partial \varphi}{\partial Z}\big|^2 {d}Z~\!,\qquad E_P = - \frac{1}{2} \int_{-\infty}^{+\infty}~\big|\varphi\big|^4 {d}Z,
\end{equation}
are the kinetic and potential energy of the solution $\varphi_1(Z) \equiv \varphi(Z)$ to \eqref{statNLS}.
Stability of this solution with respect to arbitrary dilations would require $\varphi_1(Z)$ to be a minimum of the ``energy" functional \eqref{Ezeta} at $\zeta=1$. However, calculating the first and second derivatives
\begin{equation}
\frac{d E_\zeta}{d\zeta}\Big|_{\zeta=1} = E_K - E_P~\!,\qquad\frac{d^2 E_\zeta}{d\zeta^2}\Big|_{\zeta=1} = 2 E_P,
\end{equation}
one finds that $\varphi_1(Z)$ can only extremize $E_\zeta$ if $E_K=E_P$, which is only possible when both the kinetic energy and the potential energy are zero, an impossibility. Taken at face value, this argument would force one to conclude that no non-trivial localized, stable, stationary solution can exist for the nonlinear Schr\"odinger equation. Needless to say, this could not be further from the truth. 

As is well-known, equation \eqref{statNLS} has localized solutions of the form 
\begin{equation}
\varphi(Z) = \frac{\sqrt{2}\lambda}{\cosh(\lambda Z + \delta)}~\!,\qquad \beta=\lambda^2,\label{NLSsoliton}
\end{equation}
for real parameters $\lambda$ and $\delta$, which in the original reference frame correspond to (bright) soliton solutions for the nonlinear Schr\"odinger equation with amplitude $2\beta$ (for $|\psi|^2$), moving at (arbitrary) speed $v$. In \cite{Carr} it is shown that these solitons are in fact genuine ground states for the stationary nonlinear Schr\"odinger equation \eqref{statNLS} in the sense that they correspond to the  $j=1$ case in the spectrum of possible values for the parameter $\beta$ that correspond to localized solutions. This spectrum is described, at leading order, by the Rydberg-like formula
\begin{equation}
\beta(j) = \left(\frac{N}{4}\right)^2 \frac{1}{j^2}~\!,\qquad j = 1, 2, \hdots~\!,\label{NLSspec}
\end{equation}
where $N$ represents the (square of the) norm of the corresponding solution.

It is a well-established fact that the soliton solutions to the nonlinear Schr\"odinger equation obtained from \eqref{NLSsoliton} are not only stable with respect to the time evolution, but also that -- under suitable analytic conditions that ensure the applicability of inverse scattering techniques \cite{Zakharov} -- any initial condition will invariably decay into a radiation part and a train of solitons that are stable under mutual interaction as well as under interaction with the radiation part. 
Hence something seems to be seriously amiss with the above stability analysis \`a la Derrick. As we shall see in the following paragraph, the problem is that the criterion for stability, used above, fails to take into account some elementary truths about the nonlinear Schr\"odinger evolution, most notably that its solutions should preserve the $L^2$ norm of the initial condition (as is mentioned in passing in \cite{Kuz-Dias}).

\subsection{Stability of nonlinear Schr\"odinger-type solitons\label{2.2}}
A first, obvious, but nevertheless important fact is that the parameter $\beta$ in the stationary equation \eqref{statNLS} is completely determined by the solution $\varphi(Z)$. Indeed, multiplying both sides of the equation by $\varphi^*(Z)$ and integrating, one obtains
\begin{equation}
\beta= -\frac{E_K + 2 E_P}{N},\label{beta}
\end{equation}
for $\displaystyle N= \int_{-\infty}^{+\infty} \big|\varphi\big|^2 {d}Z \neq 0$. Furthermore, the nonlinear Schr\"odinger equation \eqref{NLS} is obviously invariant under scale transformations $(z, s, \varphi) \rightarrow (z/\zeta, s/\zeta^2, \zeta \varphi)$ for some dilation parameter $\zeta>0$. The corresponding transformation of the stationary equation \eqref{statNLS} shows that the parameter $\beta$ must scale as $\beta\rightarrow \zeta^2\beta$, i.e. as the square of the norm $N$ of a solution $\varphi$ (which scales as $N \rightarrow \zeta N$). Hence, one can express the parameter $\beta$ as $\beta = \beta_1 N^2$ for some positive constant $\beta_1$ (which is the value of $\beta$ for a solution of \eqref{statNLS} with $N=1$).

A second basic fact is that equation \eqref{statNLS} itself can be obtained from a variational problem for the action 
\begin{equation}
A = \int_{-\infty}^{+\infty} \left(\big|\frac{\partial\phi}{\partial Z}\big|^2 -\frac{1}{2} \big|\phi\big|^4\right){d}Z~+~\!\gamma \int_{-\infty}^{+\infty} \big|\phi\big|^2 {d}Z,\label{stataction}
\end{equation}
i.e., a solution $\phi(Z) = \varphi(Z)$ to \eqref{statNLS} extremizes the above action for $\gamma=\beta$. Alternatively, one can think of this variational problem as one in which one extremizes the energy functional $E(\phi)$ \eqref{ENLSZ} for a fixed value of the norm $\int_{-\infty}^{+\infty} \big|\phi\big|^2 {d}Z $ and in which $\gamma$ plays the role of a Lagrange multiplier.

A standard technique (cf. \cite{Kuznetsov,Sulem} for details) is then to define functions 
\begin{equation}
\varphi_\xi(Z) = \xi^{-1/2} \varphi(Z/\xi)\label{vardil}
\end{equation}
in terms of a solution $\varphi(Z)\equiv \varphi_1(Z)$ of equation \eqref{statNLS} that have the same $L^2$ norm $\int_{-\infty}^{+\infty} \big|\varphi_\xi\big|^2 {d}Z $ for all values of the parameter $\xi>0$. Varying the action \eqref{stataction} over the functions $\varphi_\xi(Z)$
\begin{equation}
A_\xi = \gamma N + \xi^{-2} E_K + \xi^{-1} E_P,\label{defacNLS}
\end{equation}
we find that $\displaystyle\frac{{d} A_\xi}{{d}\xi}\Big|_{\xi=1}$ should be zero (as $\varphi_1(Z)$ is a solution to \eqref{statNLS}) and hence that 
\begin{equation}
2 E_K + E_P = 0.
\end{equation}
Together with condition \eqref{beta} this yields an expression for the kinetic and potential energy for arbitrary localized solutions to \eqref{statNLS} in terms of the norm $N$ and the parameter $\beta$ : $E_K = \frac{\beta N}{3}~\!$ (which is positive), $E_P = -\frac{2\beta N}{3}~\!$ (negative). This leads to the conclusion that the total energy of such a solution varies as the third power of $N$,
\begin{equation}
E = -\frac{\beta N}{3} \equiv~\! -\frac{\beta_1~\! N^3}{3},\label{NLS-soliton-energy}
\end{equation}
and is necessarily negative. These facts can be easily verified directly on the soliton solutions themselves. For a soliton \eqref{NLSsoliton} one has $\beta = \lambda^2$, $N= 4~\! |\lambda|$ and $E = -\frac{4}{3}~\! |\lambda|^3$, which  coincides with \eqref{NLS-soliton-energy} for the value $\beta_1 = \frac{1}{16}$ (cf. $j=1$ in \eqref{NLSspec}).
This result is of course in agreement with that of the previous paragraph in that it shows that the energy of a localized solution to \eqref{statNLS} is unbounded below : the energy necessarily decreases with increasing norm. On the other hand, from the expression for the spectrum of the excited states \eqref{NLSspec}, it is clear that for given $N$ the energy $E$ does have a minimum\footnote{The first and second derivatives of the energy $E$ (and of the action $A_\xi$) relative to the parameter $\xi$ of the norm preserving variations (\ref{vardil}) are obviously equal and the second derivative is in fact positive at $\xi=1$. This indicates that the bound state of \eqref{statNLS}, discussed here, indeed minimizes the energy functional for a fixed norm.} and that it is always  the soliton \eqref{NLSsoliton} for which this minimum is attained. 

The above result is in fact even stronger as it also explains the dynamical mechanism that underlies the stability of the soliton solutions for the nonlinear Schr\"odinger equation. As mentioned above, the inverse scattering theory \cite{Zakharov} for the nonlinear Sch\"odinger equation shows that (quite) general initial conditions necessarily decay into a set of solitons (of the form \eqref{NLSsoliton}, up to Galilean transformation) and a radiation part. Therefore, if one considers a small perturbation of a soliton, it is clear that it will shed some of its $L^2$ norm in the form of radiation and that asymptotically, its norm must be smaller than that of the initial condition (which is preserved under the evolution). Hence, it follows that its energy is bounded from below by $-\frac{\beta_1}{3} N_0^3$, where $N_0$ is (the square of) the norm of the initial condition.  This ensures its (energetic) stability. It is important to note however that the exact amplitude (norm) or speed of the soliton that arises from this process cannot be obtained from mere energetic considerations, as these are the result of the full dynamics of the evolution equation. In particular, for large perturbations, instead of a single soliton, several solitons might arise in the decay process, the number of which depends heavily on the exact nature of the perturbation (as do their respective amplitudes or norms).

The above discussion might leave the impression that the stability of (stationary) localized solutions to the nonlinear Schr\"odinger equation relies heavily on its integrable character (``integrable'' meaning here, ``integrable through the inverse scattering transform"). In the next paragraph we shall see that this is in fact not the case, as the existence of infinitely many conserved quantities -- perhaps the single most important aspect of the integrability of the nonlinear Schr\"odinger equation -- is not required, per se, for the stability of its stationary solutions. Integrability of the model is only required to guarantee stability vis-\`a-vis the interactions between its solitons, and a limited number of (physical) conservation laws may indeed suffice to imply stability for simple localized solutions. This might happen even in the case of nonintegrable evolution equations.

 \subsection{The case of the Schr\"odinger-Newton equation\label{2.3}}
The Schr\"odinger-Newton equation \eqref{NS} is intimately related to a larger class of interesting physical models called Wigner-Poisson (or quantum Vlasov-Poisson) systems that describe particle density functions in phase-space \cite{Wigner, Illner}. In this context, the global existence and uniqueness of solutions to the Schr\"odinger-Newton equation was shown by Illner and Zweifel in \cite{Illner}, where the asymptotic behaviour of its solutions for the case of a repulsive (Coulomb) potential -- cf. equation \eqref{selfC} -- was also discussed. As explained in \cite{Illner}, the difference between the repulsive case and the attractive (gravitational) case is far greater than merely changing the sign of the interaction in the equation. In particular, essential differences arise when one tries to discuss the stability and asymptotics for generic solutions for these systems. The general asymptotics and stability of solutions to the Schr\"odinger Newton equation \eqref{NS} is discussed by Arriola and Soler in \cite{Arriola}, refuting a -- till then -- widely held belief that ``an attractive force might lead to a blow-up in finite time" as claimed in \cite{Brezzi} (in which the existence of solutions for the repulsive case was first proven). The stability of stationary solutions to the Schr\"odinger-Newton equation \eqref{NS} however was in fact demonstrated much earlier, in the early '80's, independently by Cazenave and Lions \cite{Cazenave} and Turitsyn \cite{Turitsyn}.

The overall stability picture for solutions to the Schr\"odinger-Newton equation \eqref{NS} is in fact remarkably similar to that of the nonlinear Schr\"odinger equation \eqref{NLS}. Equation \eqref{NS} is obtained from a variational principle for the action 
\begin{multline}
A_{SN}(\psi) = \iint\!{d}t {d}^3x~\!\Big[\frac{{i}\hbar}{2}\big(\psi^* (\bx,t)\frac{\partial \psi(\bx,t)}{\partial t} - \psi(\bx,t) \frac{\partial \psi^*(\bx,t)}{\partial t}\big) \\ - \frac{\hbar^2}{2 M}\big|\nabla\psi(\bx,t)\big|^2 + \frac{G M^2}{2} \int\!{d}^3y~\!\frac{\big|\psi(\by,t)\big|^2}{|\bx-\by|}~\!\big|\psi(\bx,t)\big|^2\Big] ~\!.
\end{multline}
The (variational) symmetries obtained from this action are, by and large, the same as those for the nonlinear Schr\"odinger equation : phase-transformations $\psi(\bx, t)\mapsto {e}^{{i}{\varepsilon}} \psi(\bx, t)$, translations in $\bx$ and $t$, and Galilean transformations\footnote{In addition to these, the action is also rotation invariant, an invariance which is of course related to the conservation of angular momentum \cite{Arriola, Giulini}.}
\begin{gather}
(\bx, t,\psi) ~\rightarrow~({\bf X}, T, \Psi)\\
{\bf X}=\bx- {\bf v} t~\!,\quad T= t~\!,\quad\psi= {e}^{\frac{{i} M}{\hbar} ( {\bf v}\cdot\bx + \frac{|{\bf v}|^2}{2} t)} ~\Psi,
\end{gather}
for some velocity ${\bf v}$ of the moving frame $({\bf X}, T)$. 

These symmetries lead to conservation laws for the (square of the) $L^2$-norm
\begin{equation}
N(\psi) = \int {d}^3x~\big|\psi(\bx,t)\big|^2~\!,\label{l2norm}
\end{equation}
and the energy
\begin{equation}
E(\psi) = \frac{\hbar^2}{2 M} \int\!{d}^3x~\big|\nabla\psi(\bx,t)\big|^2 - \frac{G M^2}{2} \iint\!{d}^3x {d}^3y~\!\frac{\big|\psi(\by,t)\big|^2}{|\bx-\by|}~\!\big|\psi(\bx,t)\big|^2~\!.
\label{energy}\end{equation}
As was the case for the nonlinear Schr\"odinger equation, invariance under Galilean transformations gives rise to a simple relation between the derivative of the first moment of a solution $\psi(\bx, t)$ and its momentum (cf. \eqref{NLS-com}), which expresses the fact that the center of mass of a sufficiently localised solution to the Schr\"odinger-Newton equation necessarily moves at constant speed.\footnote{There also exists an interesting relation between the second moment of a solution to \eqref{NS}, the kinetic energy and the energy. This relation is used to great effect in \cite{Arriola} in connection to the asympotics of certain types of asymptotically expanding solutions. It is however of no immediate relevance to the present discussion, as we are only interested in the opposite case, that of contracting solutions.} One can therefore look for a ``ground-state" solution to \eqref{NS} in the form
\begin{equation}
\psi(\bx -{\bf v} t) = {e}^{\frac{{i}\cE t}{\hbar}} \varphi(\bx)~\!,\label{redtoChoq}
\end{equation}
just as for the nonlinear Schr\"odinger equation. This leads to a stationary equation for $\varphi(\bx)$
\begin{equation}\frac{\hbar^2}{2M} \Delta\varphi(\bx) + G M^2\! \int\!d^3y ~\!\frac{\big|\varphi(\by)\big|^2}{|\bx-\by|}~\!\varphi(\bx) = \cE \varphi(\bx)~\!,\label{Choquard}
\end{equation}
which was coined the {\em Choquard} equation by E. Lieb in \cite{Lieb}. In this groundbreaking paper Lieb shows that the energy functional 
\begin{equation}
E(\phi) = \frac{\hbar^2}{2 M} \int\!{d}^3x~\big|\nabla\phi(\bx)\big|^2 - \frac{G M^2}{2} \iint\!{d}^3x {d}^3y~\!\frac{~\big|\phi(\by)\big|^2}{|\bx-\by|}~\!\big|\phi(\bx)\big|^2~\!,\label{Choquard-energy}
\end{equation}
is minimized by a unique solution $\varphi(\bx)$ of the Choquard equation \eqref{Choquard} for a given norm $N(\varphi)$. The precise statement is as follows: for fixed $\lambda\in\mathbb{R}$ it can be shown that
\begin{equation}
{E}(\lambda) = \inf\left\{ E(\phi)~\!\big|~\!  \phi(\bx)~\!:~ \|\phi\|_2\leq \lambda ,~ \|\nabla\phi\|_2~ \text{finite}~\! \right\}~\!\label{Einf}
\end{equation}
is finite ($\|~\!\|_2$ denotes the $L^2$-norm on $\mathbb{R}^3$). The minimizing $\phi$ satisfies the Choquard equation \eqref{Choquard} for some (positive) parameter $\cal E$ and is unique except for translations (in $\bx$). Furthermore, the $L^2$-norm of this minimizing solution of \eqref{Choquard} is exactly equal to $\lambda$ and the corresponding minimal value of the energy functional \eqref{Choquard-energy} is negative. In fact, increasing the norm $\lambda$ necessarily lowers the infimum ${E}(\lambda)$ defined in \eqref{Einf} and the energy functional \eqref{Choquard-energy} only possesses a minimum if one imposes an upper bound on the norm (the minimum is necessarily obtained by saturating the norm). Needless to say that this situation is strikingly similar to that of the nonlinear Schr\"odinger equation discussed in the previous paragraph. The similarity between the two equations runs even deeper. As in the case of the nonlinear Schr\"odinger equation it is clear that the parameter $\cal E$ in the Choquard equation should be related to the kinetic and potential energy:
\begin{gather}
E_K(\phi) = \frac{\hbar^2}{2 M} \int\!{d}^3x~\big|\nabla\phi(\bx)\big|^2 ~\!,\\
E_P(\phi) =- \frac{G M^2}{2} \iint\!{d}^3x {d}^3y~\!\frac{~\big|\phi(\by)\big|^2}{|\bx-\by|}~\!\big|\phi(\bx)\big|^2~\!.
\end{gather}
The relationship obtained from \eqref{Choquard}  is in fact exactly the same as that for the nonlinear Schr\"odinger equation (cf. \eqref{beta}) :
\begin{equation}
{\cal E} = - \frac{E_K(\varphi)+2 E_P(\varphi)}{N(\varphi)}~\!.
\end{equation}
The Choquard equation \eqref{Choquard}, in turn, is obtained from a variational principle for the action
\begin{equation}
A_C(\phi) = E_K(\phi) + E_P(\phi) + \gamma~\! N(\phi)~\!,
\end{equation}
which can of course also be interpreted as an extremal problem for the energy functional $E(\phi)$, for fixed $N(\phi)$, with Lagrange multiplier $\gamma$ (the extremizing function $\phi=\varphi$ satisfying \eqref{Choquard} for $\gamma={\cal E}$). If we consider deformations $\varphi_\xi(\bx) = \xi^{-3/2} \varphi(\xi^{-1} \bx)$ (for some dilation parameter $\xi>0$) of a solution $\varphi(\bx)\equiv\varphi_1(\bx)$ to the Choquard equation, we obviously have 
\begin{equation}
N(\varphi_\xi) \equiv N(\varphi)~\!,
\end{equation}
and the deformed action 
\begin{equation}
A_C(\varphi_\xi) = \gamma~\! N(\varphi) + \xi^{-2} E_K(\varphi) + \xi^{-1} E_P(\varphi)~\!,
\end{equation}
which should be extremal at $\varphi_{\xi=1}$, has exactly the same form as that obtained for the nonlinear Schr\"odinger equation (cf. \eqref{defacNLS}). Hence, the resulting virial-like constraint
\begin{equation}
\left.\frac{{d}A_C(\varphi_\xi)}{{d}\xi}\right|_{\xi=1} = - 2 E_K - E_P = 0~\!,
\end{equation}
yields the same relation between the parameter $\cal E$ and the kinetic energy
\begin{equation}
{\cal E} = \frac{3 E_K(\varphi)}{N(\varphi)}~\!,
\end{equation}
as before and the resulting relation between the total energy $E(\varphi)$ for some (sufficiently) localised solution to the Choquard equation and $\cE$, the parameter in the equation, is of course nothing but \eqref{NLS-soliton-energy} :
\begin{equation}
E(\varphi) = -\frac{1}{3} {\cal E} N(\varphi)~\!.
\end{equation}
This relation can also be found in \cite{Bernstein, Arriola, Tod} but it seems to have been obtained for the first time in \cite{Membrado} in the context of many-boson assemblies\footnote{This class of models has also been considered in depth (cf. \cite{PierreHenry}  and references therein) in the study of candidates for dark matter. Moreover, as mass conservation in the case of celestial many-boson assemblies naturally leads to $L^2$ norm conservation, virial like results like ours also appear in that context. The variational principles in that case however, are still more constrained by conservation laws than in ours, which explains for example why Derrick's theorem does not apply to the discussion of the stability of self-collapsed bosonic stars.
} that interact through (attractive) Yukawa forces. 

Scaling arguments similar to those used for the stationary nonlinear Schr\"odinger equation suggest that $E(\varphi)$ scales as $E(\varphi) = -\frac{1}{3} {\cal E}_1 N(\varphi)^3$ in terms of the square of the $L^2$-norm $N(\varphi)$, where ${\cal E}_1$ is the value of the parameter $\cal E$ that corresponds to the minimizing solution of \eqref{Choquard} for to $N=1$. This parameter can be expressed in terms of a dimensionless constant $\mathrm{~\!e~\!}$ as ${\cal E}_1 = \mathrm{e} \frac{G^2 M^5}{\hbar^2}$ and the total energy then takes the form
\begin{equation}
E(\varphi) = -\frac{\mathrm{e}}{3} N(\varphi)^3  \frac{G^2 M^5}{\hbar^2} ~\!.
\end{equation}
A simple but quite rough lower bound for the energy $E(\varphi)$ can be found in \cite{Turitsyn}:
\begin{equation}
 E(\varphi) \geq -\frac{1}{2} N(\varphi)^3  \frac{G^2 M^5}{\hbar^2} ~\!.
\end{equation}
More recently, several analytic \cite{TodMoroz, Kumar, Tod} and numerical \cite{Moroz,  Bernstein, Harrison} attempts have been made at determining the precise spectrum of parameter values $\cal E$ for localized solutions of the Choquard equation. Numerical evidence points at the existence of a spectrum of values for the parameter ${\cal E}_1= \mathrm{e}_n \frac{G^2 M^5}{\hbar^2}$, described in terms of dimensionless constants $\mathrm{e}_n$ as :
\begin{equation}
\mathrm{e}_n = \frac{a}{(n+b)^c}~\!,\quad n= 0, 1, 2, \hdots~\!,\label{Choquard-spectrum}
\end{equation}
for approximated constants \cite{Bernstein}
\begin{equation}
a = 0.096\pm0.01~,\quad b = 0.76\pm 0.01~,\quad c= 2.00\pm 0.01~\!,
\end{equation}
and which contains the so-called ``ground state" for the Choquard equation, i.e. the state with minimal energy for $N=1$, at $n=0$. Analytic results suggest that this state is spherically symmetric \cite{TodMoroz}.
The best known \cite{Bernstein, Harrison} numerical value for the parameter corresponding to this ground state is $\mathrm{e}_0 = 0.163$, first obtained in \cite{Membrado}, which fits nicely with the analytic bounds for $\mathrm{e}_0$
\begin{equation}
0.146 \leq \mathrm{e}_0 \leq \frac{32}{~\!9 \pi^2}\approx 0.36\label{e0}
\end{equation}
derived in \cite{Tod} (bounds that rule out the value of $\frac{1}{2}$ for the parameter $\mathrm{e}_0$ for the ground state proposed in \cite{Kumar}). It has been shown in \cite{Arriola} that there exist spherical symmetric solutions for the Choquard equation that correspond to excited states w.r.t. the spectrum \eqref{Choquard-spectrum} and that possess breather-like properties. Though still scarce, some numerical evidence for such states also seems to exist \cite{Ehrhardt}. In \cite{Harrison} numerical results for axially symmetric solutions to the Choquard equation show that these correspond to higher energies than that found for the spherically symmetric ground state for \eqref{Choquard}.

More interesting in the present context of gravitational collapse, is the result by Arriola and Soler \cite{Arriola} which shows that for positive values of $\cal E$ (i.e. negative values of the energy) the natural dispersion of an initial state is inhibited by the nonlinearity of equation \eqref{NS}, and that this can lead to a collapse\footnote{In fact, in \cite{Arriola} it is shown that solutions with positive energy must expand asymptotically and that inhibition of dispersion and an ensuing collapse are only possible for initial stationary states with negative total energy.\label{blobl}} to solutions that oscillate around a ground state of \eqref{Choquard}, very much as in the case of the nonlinear Schr\"odinger equation \eqref{NLS}. In a physical setting, this criterion then yields a lower bound on the mass of self-gravitating objects of given size, below which no collapse can occur. Several attempts have been made at simulating the Schr\"odinger-Newton equation \cite{Carlip, Guzman2003, Guzman2004, Giulini, Ehrhardt, vanMeter} (mostly in the spherically symmetric case but quite general schemes do exist \cite{Ehrhardt}) with the aim of demonstrating the existence of a gravitational collapse. For sufficiently massive initial conditions (so as to have negative energy), as time progresses, there indeed seems to be a clear contraction of the initial condition to a unique and stable (ground) state. In particular, in \cite{vanMeter} numerical evidence is presented for a collapse to a stable state with an energy given exactly by (50) for $\mathrm{e}=\mathrm{e}_0$. 

In \cite{Giulini} scaling arguments are combined with the numerical estimates \eqref{e0} to give a lower bound of $10^{10}$ atomic units for the mass of an initial gaussian wave packet (with a typical width of 0.5 $\mu$m) for it to undergo collapse to the ground state. Moreover, {in  \cite{vanMeter}, a precise condition is formulated, based on accurate numerical estimates, according to which a wave packet, originally prepared in a gaussian state will self-collapse (in the quantum regime) if its spatial extent $\sqrt{r^2}$ obeys the inequality 
\begin{equation}\sqrt{r^2}\leq (1.14)^3(\hbar^2/GM^3).\label{vanmeter}\end{equation}
These simulations show that the irreversible mechanism through which self-collapse occurs for the Schr\"odinger-Newton equation is seemingly of the same nature as the mechanism of radiation which is well-known to lead to the appearance of stable solitons in the case of NLS equation, as described in section \ref{2.1}: by radiating a part of their mass (norm), wave packets diminish their energy until they reach a ground state shape (solitonic in the case of NLS) for which energy conservation inhibits further radiation.} We shall come back to this important point in section 4.1.

\subsection{The case of the non-linear Klein-Gordon equation}
The discussions in the previous two sections show that an extrapolation of Derrick's analysis  \cite{Derrick} to the case of the (non-relativistic) Schr\"odinger-Newton equation is not valid. 
However, to be fair, Derrick never considered non-relativistic wave equations, only non-linear modifications of the Klein-Gordon equation, and his original results have, for some reason, been repeatedly misrepresented in the subsequent literature. 
Still, since non-relativistic wave equations are meant to emerge from relativistic ones, it seems that there should indeed be at least some intrinsic problem with Derrick's original theorem.

A first important fact that one should bear in mind is that Derrick only considered the stability of a very special class of solutions, namely purely static ones. These solutions have no time dependence (not even a factor $e^{-iEt/\hbar}$). A second important fact is that the conserved norm for the relativistic equations he considered is the Klein-Gordon norm
\begin{equation}
N_{KG}(\phi)=\frac{{i}}{2}\int {d}^3 x (\phi^*\partial_t\phi -\partial_t\phi^*\phi)~,
\end{equation}
which, as opposed to the $L^2$ norm, has the property that it is unchanged under a dilation of a static solution. Indeed, static solutions always have zero Klein-Gordon norm. One must therefore conclude that while Derrick's original theorem might be mathematically correct, it is extremely restrictive and, for example, simply multiplying a static solution 
by a time-dependent phase already provides an escape door to the theorem.

Indeed, if we consider a non-linear modification of the Klein-Gordon equation \cite{Derrick}, we must adopt a Lagrangian density of the type
\begin{equation}
\mathcal{L}(t,{\bf x})=\partial_\mu\varphi^*(t,{\bf x})\partial^\mu\varphi(t,{\bf x})-V(\varphi,\varphi^*)~.
\end{equation}
Trying an ansatz $\varphi(t,{\bf x})=e^{-iEt}\phi({\bf x})$, we obtain
\begin{equation}
\mathcal{L}(x)=E^2|\phi({\bf x})|^2-|{\bf \nabla}\phi({\bf x})|^2-V(e^{-iEt}\phi,e^{iEt}\phi^*)~.
\end{equation}
Moreover, if $V(e^{-iEt}\phi,e^{iEt}\phi^*)=V(\phi,\phi^*)$ the action is given by
\begin{equation}\label{lag}
S=\int dt\int d^3x\left(E^2|\phi({\bf x})|^2-|{\bf \nabla}\phi({\bf x})|^2-V(\phi({\bf x}),\phi^*({\bf x}))\right)~.
\end{equation}
Now, the Hamiltonian density is defined as 
\begin{eqnarray}
\mathcal{H}(x)=\partial_t\varphi^*\partial_t\varphi+|{\bf \nabla}\varphi |^2+V(\varphi,\varphi^*)~,
\end{eqnarray}
and for the above ansatz, we obtain
\begin{equation}
H=\int d^3x\mathcal{H}(x)=\int d^3 x (E^2|\phi({\bf x})|^2+|{\bf \nabla }\phi({\bf x})|^2+V(\phi,\phi^*))~.
\end{equation}
It is only in the static case ($E=0$) that the Lagrangian and the Hamiltonian are simultaneously extremized. Otherwise, Derrick's analysis does not apply.
We are of course always free to depart from the functional (\ref{lag}) and to repeat Derrick's analysis after introducing the functional \begin{equation}
L_\alpha=\int d^3x[E^2|\phi(\alpha{\bf x})|^2-|{\bf \nabla}\phi(\alpha{\bf x})|^2-V(\phi(\alpha{\bf x}),\phi^*(\alpha{\bf x}))]
\end{equation}
but this analysis does not say anything about the stability of the wave function $e^{-iEt}\phi({\bf x})$ because $S$ as well as its second derivative
$\partial^2_{\alpha} L_\alpha$ are devoid of physical meaning and in particular must not be interpreted as energies.

 \section{Self-interaction of a structured quantum system\label{3}}
The overwhelming majority of quantum systems of interest possess an internal structure. Atoms for instance consist of a nucleus around which electrons orbit. 
The nucleus itself is constituted of nucleons and the nucleons themselves (neutrons and protons) are constituted of quarks. It is not a simple problem to describe the influence of self-gravitation on such structured systems because we must then also take into account internal interactions between the subsystems (for instance electro-magnetic interaction between nucleus and electrons, or between electrons in an atom). The equation (\ref{NS}) must therefore be generalized to take all these contributions into account:
\begin{align}
&{i}\hbar\frac{\partial\Psi(t,{\bf x}_1,{\bf x}_2,\hdots,{\bf x}_i,\hdots{\bf x}_N)}{\partial t}=
-\hbar^2\frac{\sum_{i=1,2...N}\Delta\Psi(t,{\bf x}_1,{\bf x}_2,\hdots,{\bf x}_i,\hdots{\bf x}_N)}{2m_i}\nonumber\\
&\qquad- G\!\sum_{i,j=1,2...N}m_im_j\int d^{3N} x'\frac{|\Psi(t,{\bf x}'_1,{\bf x}'_2,\hdots,{\bf x}'_j,\hdots{\bf x}'_N)|^2}{|{\bf x}_i -{\bf x}'_j|}\Psi(t,{\bf x}_1,{\bf x}_2,\hdots,{\bf x}_i,\hdots{\bf x}_N)\nonumber\\
&\hskip9cm+V^{\mathrm{ext.}}+V^{\mathrm{int.}},\label{NSgen}
\end{align}
where $N$ denotes the number of particles in the system and $V^{\mathrm{ext.}}$ represents the interactions that originate outside the system, while $V^{\mathrm{int.}}$ represents all internal interactions that are not of a gravitational nature (weak, strong, electro-magnetic).

We believe that in the case of crystalline nanospheres, which contain a large number of nuclei, a systematic study of the gravitational self-energy is fully justified and that self-gravitational effects are likely to be observable for such nanospheres. 
In the next section we shall study the mesoscopic regime for which the extent of the self-collapsed ground state is of the order of the size of the nanosphere. However, we shall first briefly return to the regime in which equation (\ref{NS}) is still valid.

\subsection{The quantum or elementary particle regime\label{far}}
The Schr\"odinger-Newton equation (\ref{NS}) describes the evolution of a free quantum system under its gravitational self-interaction. 
It is valid when the system is not a composite system -- i.e. when it does not possess any kind of internal structure -- or when it is completely oblivious to its internal structure. 
For example, as far as we know, this is the case for elementary particles such as electrons and neutrinos, but we can, in general, consistently neglect the structure of a quantum system when the typical size of its internal structure is significantly smaller than the extent of its ground state. In what follows we shall refer to this regime as the {\em quantum} regime. In fact, in \cite{Giulini} several estimates are given for the standard deviation $A=\sqrt{<r^2>}$ of the Choquard self-collapsed state\footnote{\label{ABCD}One estimate for the width of the Choquard bound state given in \cite{Giulini}  is $2\hbar^2 / G m^3$, which corresponds to an energy of $E\approx -0.125~\! G^2m^5/3\hbar^2$ and which fits roughly with the estimate of the energy $E(\varphi)\approx -0.163~\! G^2m^5/3\hbar^2$ mentioned in Sec.2.3. Accordingly, we shall from here on assume that the quantity 
$A_o = 2 \hbar^2/G m^3$ is a good estimate of the size of the self-collapsed state.}. They all indicate that $A$ is of the order of  $\hbar^2 / G m^3$. For an object of mass $m$, the quantum  regime is thus likely to occur whenever $\hbar^2 / G m^3$  is considerably larger than the size of the object. 
Extrapolating to the case of protons and neutrons, we obtain an extent of the order of $3\times 10^{23}$ meters; in the case of the electron, we obtain an extent of the (cosmological) order of $3\times 10^{33}$ meters, which shows why the non-linearity may be neglected in normal conditions. 
It is also consistent in this context to neglect the perturbation that could possibly appear if elementary particles were to have an internal structure. 
The electron, for instance, has been probed with energies $E$ up to the TeV and these experiments have failed to reveal any internal structure. Hence, if it has a radius, it 
has to be smaller than ${h c}/{E}\approx 10^{-18}\text{m}$, a quantity which is obviously negligible compared to $3\times10^{33}$ meter. 
To conclude, self-gravitation is so weak that it is not likely to be measurable at the level of elementary particles. 

\subsection{The self-interacting homogeneous sphere\label{sphere}}
In what follows it is assumed that the wave function of equation \eqref{NSgen} can be written as the product of two wave functions:
one for the center-of-mass degrees of freedom and one for the relative and internal degrees of freedom. This is justified because internal interactions are invariant under translations, and hence the center of mass decouples from the relative coordinates in the following sense. Imposing a factorizable solution of equation  \eqref{NSgen} of the form $\Psi(t,{\bf x}_1,{\bf x}_2,\hdots,{\bf x}_i,\hdots{\bf x}_N)$ = $\Psi_{CM}(t,{\bf x}_{CM})$$\Psi_{rel}(t,{\bf x}_{rel,1},\hdots,{\bf x}_{rel,i},\hdots{\bf x}_{rel,N})$ as an ansatz, with ${\bf x}_{rel,i}={\bf x}_{i}-{\bf x}_{CM}$, we obtain (see appendix for more details) the coupled system of equations:
\begin{align}
&{i}\hbar\frac{\partial\Psi_{CM}(t,{\bf x}_{CM})}{\partial t}=
-\hbar^2\frac{\Delta_{CM}\Psi_{CM}(t,{\bf x}_{CM}))}{2M}- ~\!\Psi_{CM}(t,{\bf x}_{CM})~\times\!\!\!\!\!\! \sum_{\tiny i,j=1,2...N}\!\!\!Gm_im_j\nonumber\\
&\int d^{3}x'_{CM} |\Psi_{CM}(t,{\bf x'}_{CM})|^2 \int d^{3N-3} x_{rel}\int d^{3N-3} x_{rel}' ~\frac{|\Psi_{rel}({\bf x'}_{rel,1},\hdots,{\bf x'}_{rel,i},\hdots{\bf x'}_{rel,N})|^2}{|{\bf x}_i -{\bf x}'_j|}\nonumber\\
&\hskip9cm\label{NSgengen}\nonumber\\
&{i}\hbar\frac{\partial\Psi_{rel.}(t,{\bf x}_{rel,1},\hdots,{\bf x}_{rel,i},\hdots{\bf x}_{rel,N})}{\partial t}=
-\hbar^2\frac{\sum_{i=2...N}\Delta\Psi_{rel}(t,{\bf x}_{rel,1},\hdots,{\bf x}_{rel,i},\hdots{\bf x}_{rel,N})}{2\mu_i}\nonumber\\
&+V^{\mathrm{int.}}\Psi_{rel}(t,{\bf x}_{rel,1},\hdots,{\bf x}_{rel,i},\hdots{\bf x}_{rel,N}).
\end{align}
In the above we made use of the fact that self-gravity is negligible in comparison to other internal potentials like e.g. the Coulomb potential. On the other hand, its effect is cumulative at the level of the coordinates of the center of mass, on which, as noted by Penrose \cite{penrose}, ``usual'' potentials like the Coulomb potential or the potential in usual Newtonian gravity do not depend. It is worth noting that, contrary to the result of the analysis by Adler \cite{Adler2} who considered a wave function separable in individual coordinates ${\bf x}_{i}$, it is not the individual contribution to the self-interaction of the type $Gm_i^2/|{\bf x}_i -{\bf x}'_i|$ (considered to be `suspect'  by Adler \cite{Adler2}) which differentiates self-gravity from usual, linear, gravity. As is corroborated at several places in our paper (e.g. section \ref{numericSam} and the Appendix), in the regimes that we are interested in, individual self-interaction is small and is rapidly overwhelmed by interactions with neighbours when the number of atoms increases. Our analysis therefore suggests that it is essentially the non-linearity of the semi-classical self-interaction that marks the difference with ``usual'', linear quantum mechanics. Hence, if the mean width of the centre-of-mass wave-function is considerably larger than the distance between neighbouring nucleons (atoms) in the crystalline nanosphere, a coarse graining approach in which the crystal is treated as an homogeneous medium is fully justified\footnote{This is confirmed by the numerical estimates which we shall present in section \ref{numericSam}.}.

\subsubsection{The Di{\'o}si potential}
Di{\'o}si already considered the problem of a self-interacting sphere in \cite{Diosi84}, where he showed the following: 
\begin{itemize}
\item in order to describe how the wave function of the center of mass of a homogeneous sphere evolves under the influence of the gravitational self-interaction, one must replace the kernel ${-GM^2}~\! {|{\bf x}_{CM} -{\bf x}'_{CM}|^{-1}}$ in the potential in equation (\ref{NS}) by the net contribution of the (classical) self-interaction of the sphere, i.e. by
\begin{equation}
-G({M\over {4\pi R^3\over 3}})^2\int_{|\tilde{\bf x}| \leq R, |\tilde{\bf x}'| \leq R} {d}^3 \tilde x {d}^3 \tilde x'\frac{1}{|{\bf x}_{CM} +\tilde{\bf x}-({\bf x}_{CM}'+\tilde{\bf x}')|}~,\label{self-sphere}
\end{equation}
where $R$ is the radius of the sphere and $M$ its mass.
\item when the mean width of the center-of-mass wave function is small enough in comparison to the size of the sphere, the self-interaction \eqref{self-sphere} reduces, in a first approximation, to a non-linear harmonic potential:  
\begin{align}&-G({M\over {4\pi R^3\over 3}})^2\int_{|\tilde x| \leq R, |\tilde x'| \leq R} {d}^3 \tilde x {d}^3 
\tilde x'\frac{1}{|{\bf x}_{CM} +\tilde{\bf x}-({\bf x}_{CM}'+\tilde {\bf x}')|}\nonumber\\[-3mm]\nonumber\\
&\hskip1.5cm\approx {GM^2 \over R}\left[-\frac{6}{5}+\frac{1}{2}\left(\frac{|{\bf x}_{CM}-{\bf x}'_{CM}|}{R}\right)^2
+{\cal O}\left(\left(\frac{|{\bf x}_{CM}-{\bf x}'_{CM}|}{R}\right)^3\right)\right].
\end{align}

As the symmetry arguments which were used to obtain the Choquard equation \eqref{Choquard} from the Schr\"odinger-Newton equation are still valid, even for the above effective potential, one obtains the following equation for the ground state wave function of the center of mass, where $\bx$=$\bx_{CM}$ :
\begin{equation}
\frac{\hbar^2}{2M} \Delta\varphi(\bx) - {G M^2\over 2R^3}\! \int\!d^3y ~\!\big|\varphi(\by)\big|^2
~|\bx-\by|^2~\!\varphi(\bx) = -\cE^D\varphi(\bx)~.\label{eqDiosieff}
\end{equation}
Here we introduced the effective parameter  $\cE^D$ as
\begin{equation}
\cE^D= {6GM^2\over 5R} - \cE~\!,\label{EpsD}
\end{equation}
with respect to the parameter $\cE$ used in the reduction \eqref{redtoChoq} (\`a la Choquard) of the time-dependent problem resulting from \eqref{NSgen}, to the time-independent equation \eqref{eqDiosieff}. All of the above is only valid, of course, to the extent that Di{\'o}si's approximation to a harmonic potential is valid, i.e. for widths of the ground state that are smaller than the radius $R$ of the sphere.
\item Di{\'o}si noted that, in this regime, the characteristic dependence of the ``energy'' $\cE^D$ of a normalized ground state on the size $A =\sqrt{<r^2>}$ of the nano-object, should be of the order of
\beq
\cE^D \approx {\hbar^2\over 2 MA^2}
+{GM^2\over 2 R^3}A^2~.
\eeq 

The radius of the ground state (denoted as $A_o^{\mathrm{meso}}$) for which this expression attains its minimum is of the order
\begin{equation}
A_o^{\mathrm{meso}}\approx ({\hbar^2\over GM^3})^{{1\over 4}}R^{{3\over 4}}=({A_o\over 2})^{{1\over 4}}R^{{3\over 4}}~\!,\label{A0meso}
\end{equation}
where, as mentioned in footnote \ref{ABCD}, $A_o$ as  is a rough estimate of the characteristic width of the ground-state wave function for the Choquard equation in the so-called quantum regime, i.e. when the object is not a composite system but behaves like an elementary particle (cf. section $\ref{2.3}$). 
\item Considering objects of normal density, Di{\'o}si derived 
a critical radius for the sphere, of the order of $10^{-7}$ m,  for which its size is half that of the Choquard ground-state wave packet (i.e., $R = {1\over 2}A_o\approx ({\hbar^2\over GM^3}))$. 
The corresponding number of particles, $N$, at normal density, and for a sphere composed of $\mathrm{H_2O}$,
  is $N_c^{\mathrm{meso}}\approx 4\cdot 10^8$ atoms. For larger (heavier) spheres, one should assume that the object will behave as a classical object, in the sense that it will localize spontaneously on a spatial distance that becomes small compared to its size, due to the collapse induced by self-gravitational interactions. 
In this sense, this critical radius characterizes the micro-macro transition.

In the regime studied by Di{\'o}si and which corresponds to the region (C) in figure \ref{fig1}, the ground state wave function of the center of mass of the sphere obeys equation (\ref{eqDiosieff}). We expect that this regime will be reached  when the mass of the nanosphere is such that, say, $A_o^{\mathrm{meso}}\approx ({\hbar^2\over GM^3})^{{1\over 4}}R^{{3\over 4}}\leq 10^{-1}R\approx 5\cdot 10^{-2}A_0$. For normal densities and a sphere made of $\mathrm{H_2O}$ molecules this would mean that $N\geq (10^{+4})^{1\over 3}N_c^{\mathrm{meso}}\approx 9\cdot 10^9$ atoms. 
\end{itemize}
\subsubsection{Stability of the ground state for the Di{\'o}si equation}
It is straightforward to repeat the analysis already performed in the case of the non-linear Schr\"odinger (NLS) 
and Schr\"odinger-Newton equations in the case of Di{\'o}si's equation (\ref{eqDiosieff}).

The associated action is now 
 
\begin{equation}
A_D(\phi) = E_K(\phi) + E^D_P(\phi) + \gamma~\! N(\phi)~\!,
\end{equation}
where, as before, the kinetic energy $E_K(\phi)$ is equal to $ \frac{\hbar^2}{2 M} \int\!{d}^3x~\big|\nabla\phi(\bx)\big|^2 ~\!,$ and $N(\phi)$ denotes the (square of the) $L^2$-norm defined by (\ref{l2norm}), while the potential energy is now given by
\beq E^D_P(\phi) = \frac{G M^2}{4R^3} \iint\!{d}^3x {d}^3y~\big|\phi(\bx)\big|^2~\!\big|\phi(\by)\big|^2~|\bx-\by|^2~\!.\label{potdiosi}\eeq
This can once again be interpreted as an extremal problem for the energy functional $E(\phi)$ for fixed $N(\phi)$ with Lagrange multiplier $\gamma$ (the extremizing function $\phi=\varphi$ satisfying \eqref{eqDiosieff} for $\gamma=-{\cal E^D}$) and ${\cal E^D} =  \frac{E_K(\varphi)+2 E^D_P(\varphi)}{N(\varphi)}~\!.$

If we again consider norm preserving deformations $\varphi_\xi(\bx) = \xi^{-3/2} \varphi(\xi^{-1} \bx)$ (for some dilation parameter $\xi>0$) of a solution $\varphi(\bx)\equiv\varphi_1(\bx)$ to the Di{\'o}si equation, we find the deformed action 
\begin{equation}
A_D(\varphi_\xi) = \gamma~\! N(\varphi) + \xi^{-2} E_K(\varphi) + \xi^{2} E^D_P(\varphi)~\!,
\end{equation}
which should be extremal at $\varphi_{\xi=1}$. This results in the virial-like constraint
\begin{equation}
\left.\frac{dA_D(\varphi_\xi)}{d\xi}\right|_{\xi=1} = - 2 E_K +2E^D_P = 0~\!,
\end{equation}
from which we derive a relation between the parameter $\cE^D$ and the kinetic energy
\begin{equation}
{\cE^D} =\frac{ E_K(\varphi)+2E^D_P(\varphi)}{N(\varphi)}= \frac{3 E_K(\varphi)}{N(\varphi)} > 0~\!.\label{EpsDpos}
\end{equation}
The resulting relation between the total energy $E^D(\varphi)$ for some (sufficiently) localised solution to the Choquard equation and $\cE^D$, the parameter in equation \eqref{eqDiosieff}, is now
\begin{equation}
E^D(\varphi) = 2 E_K(\varphi)= {2\over 3}~\! {\cal E^D} N(\varphi)~\!.
\label{posit}\end{equation}
It should be noted that  the total energy for such a solution is necessarily positive. Furthermore, as happened for the non-linear Schr\"odinger and Choquard equations, the first and second derivatives of the energy $E^D$ and of the action $A_\xi$, relative to the parameter $\xi$ of the norm preserving variations we consider here, are equal, and the second derivative is positive, which suggests that (the norm being constant) the bound state that minimizes the action $A_D(\varphi)$ also minimizes the energy. Scaling arguments similar to those used in the previous sections also suggest that $E^D(\varphi)$ as well as ${\cE^D}$ are both decreasing functions of the norm. One should bear in mind however that the ``genuine'' energy of the wave function related to \eqref{NSgen}, is obtained from $E^D$ by subtracting ${6GM^2\over 5R}$.

\subsubsection{{Analytic expression for the} ground state in the mesoscopic regime.\label{bla1}}
We shall now show that there exists a solution to Di{\'o}si's equation (\ref{eqDiosieff}) endowed with the following properties:
\begin{itemize}
\item It has a gaussian shape (cf. also \cite{lefinlandais}).
\item It minimizes the value of $\cE^D$ that can be reached on the set of wave functions whose average values of $x$, $y$ and $z$ are equal to zero.
This set contains all wave functions with amplitudes that only depend on $x, y$ and $z$ through their absolute values: $\Psi (x,y,z)=\Psi(|x|,|y|,|z|)$; 
purely radial wave functions, obviously, also belong to this set.
\item The extent $A$ of the gaussian solution and its ``effective'' energy $\cE^D$ agree with Di{\'o}si's predictions, i.e. it is of the order:
\beq
\cE^D= -\cE+{6GM^2\over 5R}\approx~\! {\hbar^2\over MA^2}+{GM^2\over R^3}A^2~,\nonumber
\eeq
for $A=A_o^{\mathrm{meso}}$ as given by \eqref{A0meso}.
\end{itemize}
In order to see this, let us rewrite equation (\ref{eqDiosieff}) as :
\begin{align}
\cE^D \Psi({\bf x})&=-\frac{\hbar^2}{2M}\Delta\Psi({\bf x})
+{GM^2 \over 2R^3}\int {d}^3 x'|\Psi({\bf x'})|^2 \left(|{\bf x}|^2 -2{\bf x}\cdot{\bf x'}+|{\bf x'}|^2\right)\Psi({\bf x})\nonumber\\
&=-\frac{\hbar^2}{2M}\Delta\Psi({\bf x})
+{GM^2 \over 2R^3}\bigg[\big(\int {d}^3 x'|\Psi({\bf x'})|^2\big)~\!|{\bf x}|^2~\!\Psi({\bf x})\nonumber\\ 
&\hskip1.5cm -2~\! \big(\!\int {d}^3 x'|\Psi({\bf x'})|^2{\bf x'}\big)\cdot{\bf x}~\!\Psi({\bf x})
 +\big(\int {d}^3 x'|\Psi({\bf x'})|^2|{\bf x'}|^2\big)\Psi({\bf x})\bigg]~.
 \end{align}
If the average values of $x$, $y$ and $z$ are equal to zero, i.e. $\int {d}^3 x'~\!{\bf x'}|\Psi({\bf x'})|^2=0$, we obtain the equation for a quantum mechanical oscillator:
\begin{eqnarray}
-\frac{\hbar^2}{2M}\Delta\Psi({\bf x})+{k|{\bf x}|^2\over 2} \Psi({\bf x}) = E^{osc.}\Psi({\bf x})\label{Qoscill}~,
\end{eqnarray}
with 
\begin{equation}
E^{osc.}=\cE^D-{GM^2 \over 2R^3}\int {d}^3 x' |{\bf x'}|^2 |\Psi({\bf x'})|^2\qquad\text{and}\quad k={GM^2 \over R^3}\int{d}^3 x'|\Psi({\bf x'})|^2~\!.
\end{equation} 
We are interested in wave functions normalised to unity, in which case we obtain $k={GM^2 \over R^3}$. 
It is of course well-known that in this case the minimal value for $E^{osc.}$ for a bound state is equal to ${3\over2}\hbar\sqrt{{k\over M}}$ and that this value is obtained for a gaussian wave function
\begin{equation*}
\Psi({\bf x}) = \frac{1}{(\sqrt{\pi} A)^{3/2}}~\! \exp\big({-\dfrac{|{\bf x}|^2}{2 A^2}}\big)~\!,
\end{equation*}
with width $A=\sqrt{{\hbar\over \sqrt{kM}}}$. One immediately finds that the minimal value of $E^{osc.}$ is
$$E^{osc.} = \frac{3}{2}\hbar\sqrt{\frac{GM}{R^3}}$$
and that the width of the corresponding gaussian, $A= ({\hbar^2\over GM^3})^{{1\over 4}}R^{{3\over 4}}$, is of the order of $A_o^{\mathrm{meso}}$.
It is easily verified that
$$\cE^D = \frac{9}{4} \hbar\sqrt{\frac{GM}{R^3}} = \frac{3}{2}E^{osc.}~,\quad E_K = \frac{1}{2} E^{osc.} = E^D_P\quad\text{and hence}\quad E^D = E^{osc.}~,$$
in accordance with the well-known equipartition of the energy of an oscillator between kinetic and potential energy. We therefore find that, indeed, $\cE^D = {9\over 8}\left({\hbar^2\over MA^2}+{GM^2\over R^3}A^2\right)$ and that the above gaussian wave function minimizes not only $E^{osc.}$, but  also $\cE^D$ and $E^D$, as far as Diosi's approximation is valid. 

\subsubsection{Between the mesoscopic and quantum regimes}
\begin{figure}
\centering
\includegraphics[width=\textwidth]{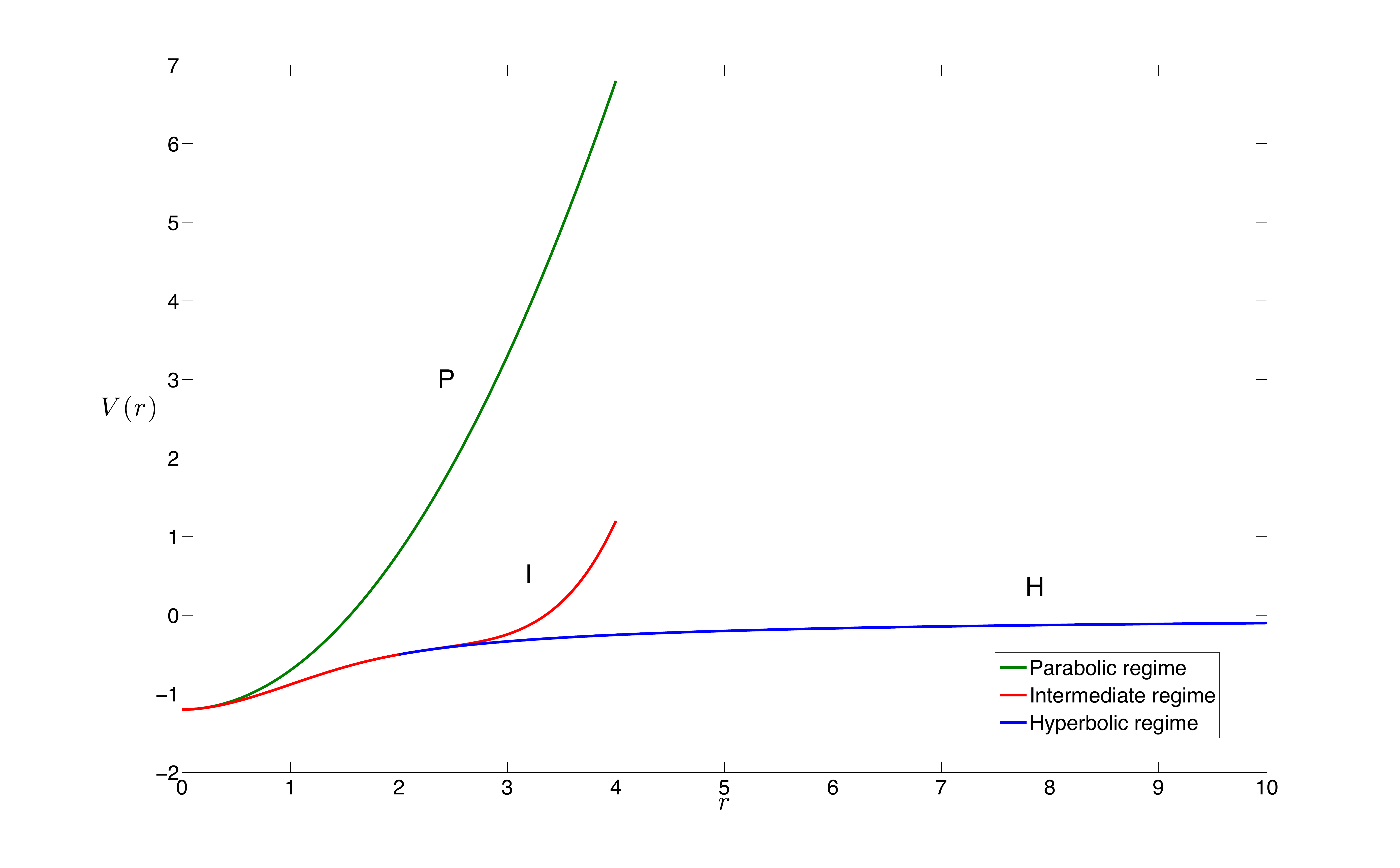}
\caption{\label{unknown}Plot of the gravitational self-energy (divided by $GM^2/R$) between a homogeneous crystalline sphere and a replica of itself, translated by a distance $r$, as a function of $r/R$, in the parabolic (P), intermediate (I) and hyperbolic (H) regime.}
\end{figure}

In the intermediate domain, where $R$ is comparable to the distance $|{\bf x_{CM} -x_{CM}'}|$, no numerical studies have been realized so far.
As it happens, the effective potential that should be used in an extension of \eqref{eqDiosieff} to this region, can be calculated exactly\footnote{There are quite a number of papers in which the authors purport to calculate this potential. Unfortunately, most of these results are incorrect. To the best of our knowledge, the only paper to give the correct expression is \cite{Iwe82}, where it is given without proof. Expression \eqref{fullpot} was obtained independently, by the present authors, by direct calculation of the classical integral \eqref{self-sphere}. More details can be found in appendix 3.} from \eqref{self-sphere}, in terms of $d=|{\bf x}_{CM}-{\bf x}'_{CM}|$, {we find for 0$\leq d\leq 2R$:}
\begin{equation}
V^\mathrm{eff}(d) = \frac{GM^2}{R}~\!\left(-6/5+\frac{1}{2}\left(\frac{d}{R}\right)^2-\frac{3}{16}  \left(\frac{d}{R}\right)^3+\frac{1}{160}\left(\frac{d}{R}\right)^5\right).\label{fullpot}
\end{equation}
{Otherwise, when $d$ is larger than twice the size of the object, the integration is straightforward. Making use of Gauss's theorem we recover the usual Coulomb-like shape:}
\begin{equation}
V^\mathrm{eff}(d) =- \frac{GM^2}{d}\quad (d\geq 2R).\label{fullpot2}
\end{equation} The fifth degree polynomial (in $d\over R$) \eqref{fullpot}  agrees, up to its 4th derivative, with the Newtonian potential in $1/d$ at the transition point ($d=2R$)  \eqref{fullpot2} (see also figure \ref{unknown}).

However, although we have an exact expression for the effective potential that will replace the harmonic potential in \eqref{eqDiosieff} in this domain,  the resulting equation is quite complicated and certainly not easier to solve than, say, the Choquard equation (\ref{Choquard}). It is beyond of the scope of this paper to tackle this problem in detail, but {interpolating the already established results in the macroscopic and quantum regimes, we may infer that, at the mesoscopic transition, the ground state} will possess ``intermediate'' properties when compared to the ground states of equations  (\ref{Choquard}) and (\ref{eqDiosieff}). 
Amongst others, we infer that
\begin{itemize}
\item[-]its energy belongs to the range $[-{1\over3 }~\! 0.163 {G^2M^5\over \hbar^2},~\!0.3 {G^2M^5\over \hbar^2}]$, i.e.: it is comprised, roughly, between the energies derived from equations (\ref{Choquard}) and (\ref{eqDiosieff}) when $R={\hbar^2\over G M^3}$ (which implies that $M=M_c^\mathrm{meso}$ is of the order of $4\cdot10^{-18}$ kg at normal densities and that the energy for \eqref{eqDiosieff} is $E^{osc.}-{6\over 5} {G^2 M^5\over\hbar^2} = {3\over10}{G^2 M^5\over\hbar^2}$). 
\item[-]{it has a pseudo-gaussian shape with an extent that is, roughly, of the order of ${\hbar^2\over G M^3}$. }
\end{itemize}
Actually, Di{\'o}si's equation (\ref{eqDiosieff}) is fully justified when the extent of the wave function $A$ obeys, say, $A<10^{-1} R$, so that $A\approx A_o^{\mathrm{meso}}= ({\hbar^2\over GM^3})^{{1\over 4}}R^{{3\over 4}}<10^{-1} R$, that is when ${\hbar^2\over G M^3}<10^{-4} R$.  On the other hand, the Choquard equation (\ref{Choquard}) is justified when $A$ obeys, say, $A>10R$, that is, when $A\approx A_o={{2}\cdot \hbar^2\over GM^3}>10 R$. 

 Therefore the mesoscopic transition occurs ``in between'' the {mesoscopic} and the quantum regimes, in the region where ${\hbar^2\over G M^3}\approx 10^{-2} R$, {rather than when ${\hbar^2\over G M^3}\approx R$, as was predicted by Di\'osi in \cite{Diosi}.} Admittedly, this difference will not have much influence on the critical radius of the nanosphere for which the transition occurs, since at normal density, the ratio ${\hbar^2\over G M^3R}$ scales as $R^{-10}$.
We shall not go into further details here concerning this regime but, in the Appendix we shall present a method aimed at solving, in an approximative fashion, the time-dependent evolution equation of a self-interacting nanosphere. In particular, we shall find that the mesoscopic transition is characterized by $A\approx 10^{-7} m$, as can be seen in figure \ref{figbound}. 
 
\subsection{Self-interacting crystal in the sub-atomic regime\label{cristal}}\subsubsection{Regime dominated by the atom-atom self-energy} In \cite{Schmidt} Schmidt considers a crystalline structure in which the atoms are separated by distances of the order of 1 {\AA}. He first postulates, as is usually done, that the wave function of the full object 
can be factorized into the product of a contribution from the center of mass of the object and a wave function that contains the contributions of the relative distances of the $N$ atoms (with $N$ of the order of the Avogadro constant) as well as all the other degrees of freedom (electronic levels, nuclear and electronic spins and so on). Moreover, Schmidt assumes that the width of the wave function of the center of mass is smaller than the distance between two nuclei. In that case, according to Schmidt, the main contribution to the gravitational self-interaction is that due to the self-interaction of individual nuclei with themselves\footnote{Electrons do not contribute for two reasons. They are considerably lighter than nucleons (by a factor of the order of $10^3$) and thus much lighter than nuclei. Moreover, their wave functions are spread over distances considerably larger than those of the nuclei (by a factor of the order of $10^5$).} because one can neglect the possible overlap of neighbouring nuclei. In this regime, the self-collapsed wave function of the center of mass of the crystal (\ref{NSgen}) typically obeys the reduced equation
\begin{equation}
-\frac{\hbar^2}{2M}\Delta_{CM}\Psi({\bf x}_{CM})-GNm^2\int{d}^3 x' \frac{|\Psi({\bf x}'_{CM})|^2}{|{\bf x}_{CM}-{\bf x}'_{CM}|} \Psi({\bf x}_{CM}) = \cE^S\Psi({\bf x}_{CM})~\!,
\label{schmidt}
\end{equation}where $m$ represents the mass of one nucleus (or the sum of the masses of the nucleons that constitute one elementary ``molecule'' of the crystal, in the case that the crystal is not monoatomic). Here $N$ represents the number of elementary atoms (molecules) in the crystal and $M$ the total mass of the crystal (to a good approximation one has $M=Nm$). 

Making use of the Galilean invariance and scaling properties of the Schr\"odinger-Newton equation \eqref{NS}, Schmidt then finds that the typical radius  $A_0^\mathrm{atomic}$ of the self-collapsed ground state of the center of mass wave function is of the order of ${\hbar^2\over Gm^3}{1\over N^2}$.  

One conclusion to be drawn from these results is that the second basic assumption made by Schmidt -- according to which the spread of the wave function is smaller than the distance between the nuclei -- is consistent, provided the crystal is massive enough. In fact, the simple model presented here makes it possible to derive, as a function of the atomic number (the number of nucleons present in an elementary component of the crystal), a critical number of atoms, $N_c^{atomic}$, beyond which a self-collapse will localise the center of mass onto distances of the order of the crystal period (for $N\approx N_c^\mathrm{atomic}$, $A_0^\mathrm{atomic} \approx \delta$ where $\delta$ is the mean distance between neighbouring nuclei). For instance, for the elements C, Fe and Au, $N_c^\mathrm{atomic}$ is respectively of the order of $10^{15}$, $10^{14}$ and $10^{13}$ (cf. \cite{Schmidt}). These values characterize the transition to what we call the atomic regime, that is, to the regime in which the extent of the self-collapsed ground state of the wave function for the center of mass becomes smaller than the inter-atomic distance. {For instance, at normal density for which $m={{18\times10^{-3}\,\mathrm{kg}}\over N_A}$ with $N_A\approx 6\cdot 10^{23}$, assuming that the self-energy is dominated by the self-energies of the nuclei and that the width  of Choquard's bound state ($\hbar^2/(G m^3 N^2)$) is approximately 1 \AA, we find that $N_c^\mathrm{atomic}$ is of the order of} {{$10^{14}$.}

\subsubsection{Between the atomic and mesoscopic regimes\label{numericSam}} 
Schmidt's regime is valid when the extent of the ground state is smaller than the typical distance $\delta$ between two neighbouring nuclei 
of the crystal, but it is not clear how much smaller it {must be for the self-energy of the individual nuclei to dominate the self-energy contributed ``by the rest of the sphere''}. 
In order to tackle this question we numerically estimated the value of the gravitational potential energy that arises between two slightly separated crystalline spheres 
with cubic symmetry.  The nuclei belonging to the first crystalline sphere are located at positions $\delta(k,l,m)$ where $\delta$ is the length of an edge of the cubic unit cell, 
$k,l,m$ are integers such that $\delta\sqrt{k^2+l^2+m^2}\leq R$ and $R$ is the radius of the sphere. The total number of nuclei $N$ is the total number of triplets $(k,l,m)$ for which $\delta\sqrt{k^2+l^2+m^2}\leq R$ ($N$ is approximately  $\frac{4 R^3}{\delta^3}$). The second sphere is a replica of the first one but translated 
by $(x\delta,0,0)$, with $x\in]0,1[$.
Figures \ref{fig2} and \ref{bisbis} encapsulate the main conclusions. In figure \ref{fig2} we present the variation of the energy for two different numbers of nuclei: 
$N_1=515$ (N1 curve, corresponding to a radius $R_1=5\delta$),  and $N_2=4169$ (N2 curve, corresponding to a radius $R_2=10\delta$).
The energy  is normalised in such a way that the value 1 corresponds to Di{\'o}si's self-interaction at zero distance (that is to say it is divided by ${V_D=-}6GM^2/5R$ =$-6N^2m^2/5R$). 
 \begin{figure}
\centering
\includegraphics[width=\textwidth]{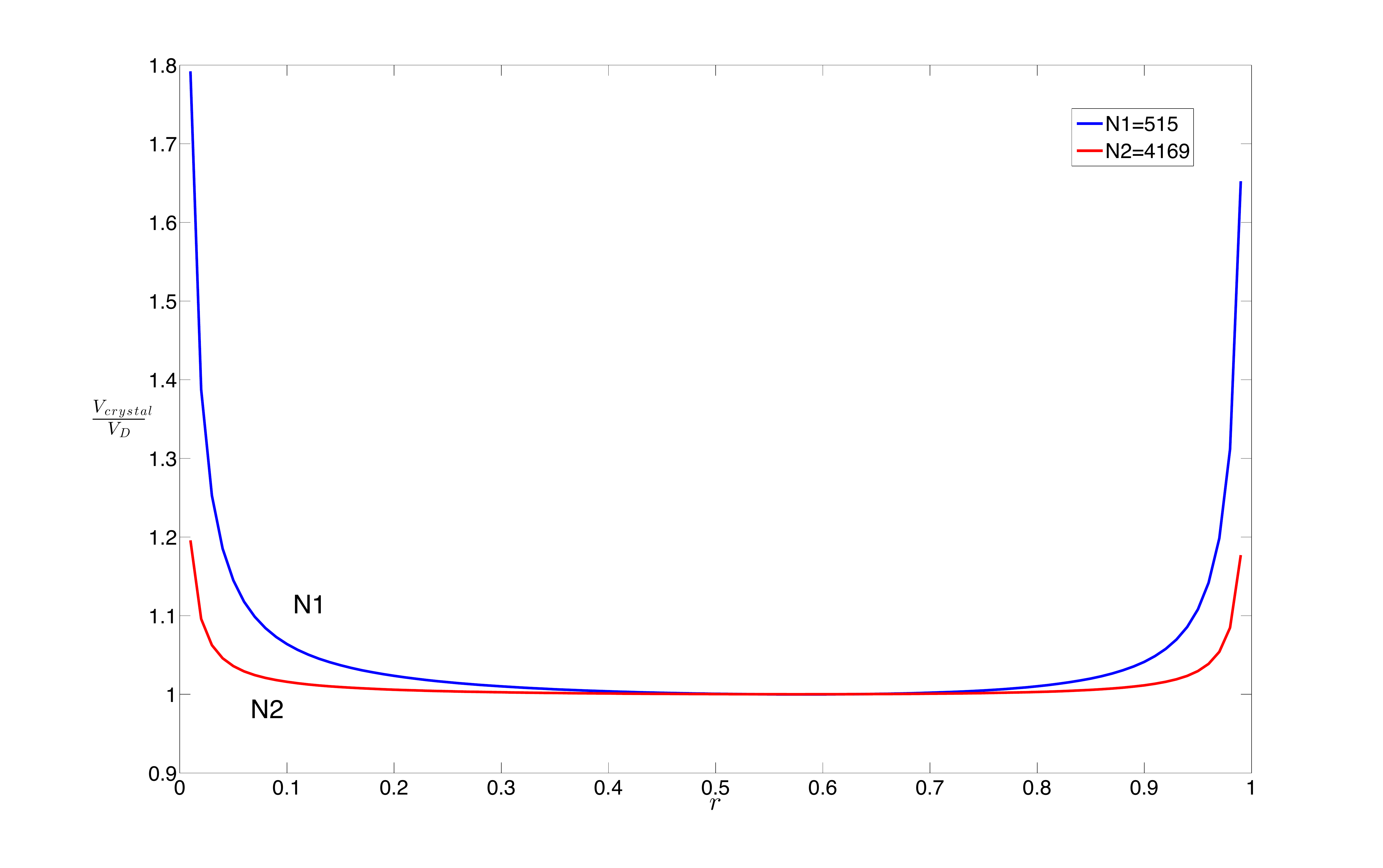}
\caption{\label{fig2}Potential energy between a crystalline sphere and its replica translated by a distance $x\delta$ along an edge of the cubic unit-cell, divided by  $V_D=-6GM^2/5R$.
The N1 (resp. N2) curve corresponds to a crystalline sphere containing $515$ (resp. $4169$) nodes. }
\end{figure}
 \begin{figure}
\centering
\includegraphics[width=\textwidth]{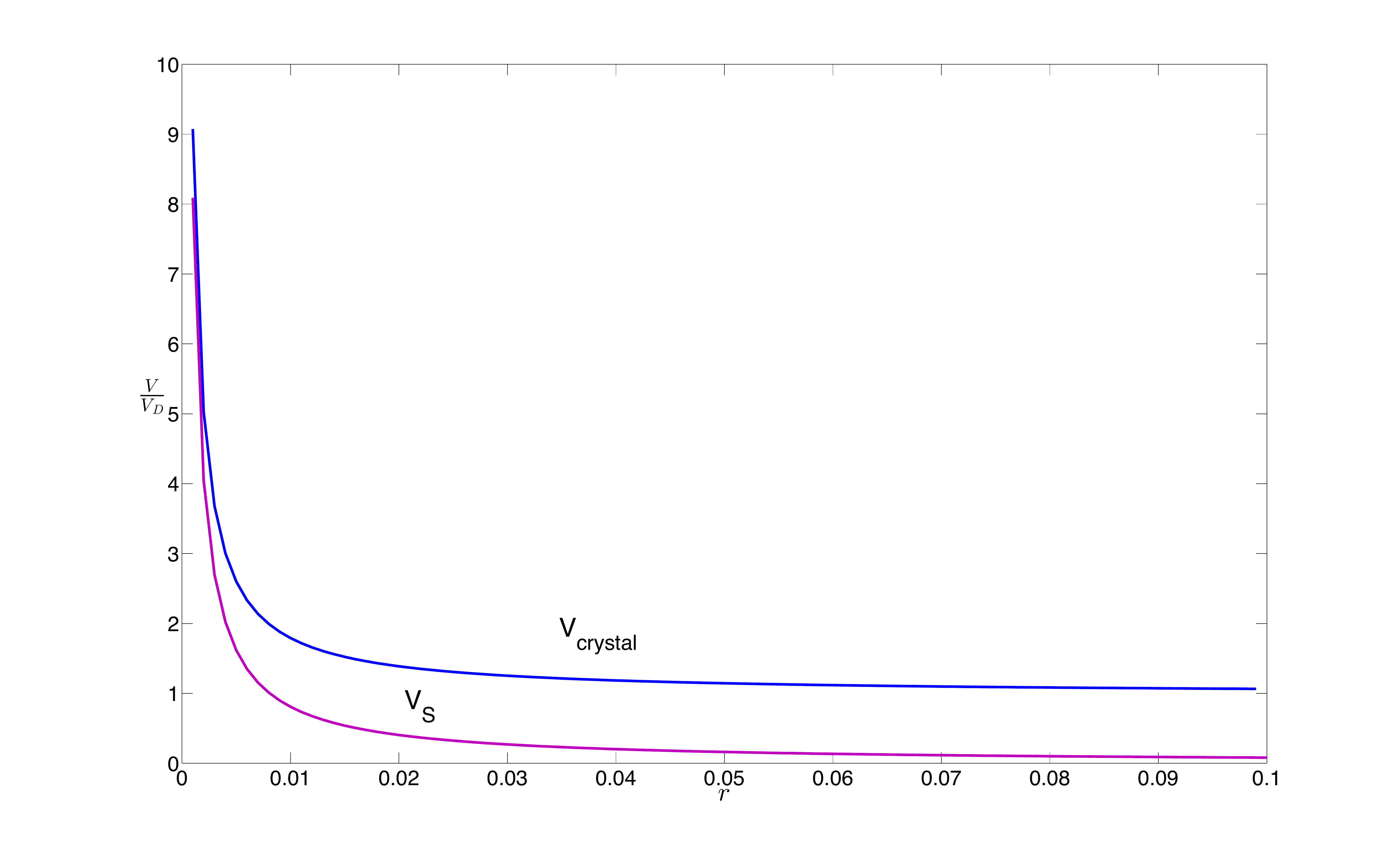}
\caption{\label{bisbis}Comparison of $V_{crystal}$, the potential energy between a crystalline sphere and its replica translated by a distance $x\delta$ along an edge (evaluated numerically, with $N_1=515$), and the ``Schmidt'' potential $V_S={-GNm^2\over{x\delta}}$, both normalized by $V_D=-6GM^2/5R$. The curve labelled by ``$V_{crystal}$'' interpolates between the Schmidt and Diosi potential $V_D$.}
\end{figure}

In figure \ref{fig2} we clearly recognize Schmidt's regime in the zones close to x=0 and x=1, which correspond to the case where the lattice sites of ``both''\footnote{It is important to bear in mind that we are of course always dealing with the same sphere and that we are simply applying a mathematical trick to calculate the effective potential, but that the limitations of classical language require us to talk about two spheres located in different positions, as if they were different.} spheres nearly overlap.  In both regions, it can be seen that the numerical estimate of the self-interaction varies essentially like the inverse of the distance to the closest edge.  Now, if one averages this behaviour over a region comparable to the inter-nucleus distance, we expect that there will be no singularity in the region $x\approx 0$, as it is an effect in $1/r$ which is integrable around the origin in the case of an isotropic (purely radial) distribution.  The apparent singularity at the right of figure \ref{fig2} ($x\approx 1$) and, in fact, in subsequent regions $x\approx 2, 3, \cdots$ not shown in the figure, is also an artifact of the simulation and is not expected to lead to any kind of measurable, physical effect. Contributions to the self-energy in these regions should in fact be scaled by a factor $(d/\delta)^3$, expressed in terms of the ratio of a certain characteristic distance $d$ and the mean distance $\delta$ between neighbouring nuclei. The characteristic distance $d$ can be estimated as that for which the contribution of the self-energies of the individual nuclei ($-N G m^2/d$) is equal to the self-energy of a homogenous sphere at small distance ($-6 G m^2 N^2 / (5R)$): $d=5R/(6N)$. Working at normal densities, with an atomic number equal to 18, one obtains that  $d$ is {of the order of $10^{-29}/R^2$} which means that, for instance, when $R \approx 100 nm$, $d$ is of the order of $10^{-15}$ m. This distance should be compared with a $\delta$ of the order of 1\AA, which yields a scale factor $d^3/\delta^3$ of the order of $(10^{-15}/10^{-10})^3=10^{-15}$, which is very small.

We observe that when the center of the second crystal is located at the middle point of an edge of the first unit cell of the first crystal (this corresponds to the value $x=0.5$ on the picture)  the two curves overlap and  agree with Di{\'o}si's prediction (${-6GM^2/5R}$) for the self-energy. One also immediately notices that when the number of atoms increases, the extent of the zone adjacent to the edges of the figure ($x=0$ and $x=1$), that is, the zone where discretisation effects \`a la Schmidt prevail and where Di{\'o}si's approximation is no longer valid,  is strongly reduced. This is to be expected because the departure from the parabolic regime (Di{\'o}si's regime) is proportional to the number of atoms, while the interaction between neighbours is proportional to $N^2$. This observation is corroborated by the above estimate of the width $d$ of the zone where these discretisation effects prevail, which scales like $1/R^2$.

In conclusion, the accuracy of Di{\'o}si's coarse graining increases when the size of the crystal increases. Once the crystal is big enough, it becomes wrong to assume that the self-interaction of each atom with itself dominates the interactions with its neighbours. The lesson to draw from our analysis is that, typically (for nanospheres close to the mesoscopic regime) the transition between the Schmidt and Di{\'o}si regimes is not as sudden as was supposed by Schmidt. 
This is confirmed by the results presented in figure \ref{bisbis}, where it is clear that over a vast region, the interaction between the entire crystal largely dominates that predicted by Schmidt for the nuclei. Further confirmation of this fact can be found in the numerical estimates sketched in the Appendix where we explicitly included corrections \`a la Schmidt in our numerical code and checked that they could most often be neglected. This can be seen for instance in figures \ref{simulation2} and \ref{simulation1} where the trajectories depicted are very close, whether we consider self-gravity with the Schmidt potential or without it.

On the other hand, it is important to note that for sufficiently massive objects, a new transition will occur to another regime in which Schmidt's model loses its validity. This new regime (which we call the nuclear regime) will be reached when the spread of the center-of-mass wave function becomes smaller than the size of the nuclei, in which case it is necessary to take into account the typical quantum spread of the nuclei. In this new regime, which we shall consider next, the gravitational self-interaction of the center-of-mass wave function crucially depends on the structure of the nuclei (and in particular on their effective size). The transition to this regime corresponds to the region A-B in figure \ref{fig1}, {where we considered a nanosphere counting more or less 10$^5$ SiO2 molecules. The atomic corrections are important whenever $GNm^2/r \geq (5/6)G(Nm)^2/R$, i.e. when $r \leq (6/5) (R/N)\approx 10^{-13}$ meter, which corresponds to the region $B$ in figure \ref{fig1} (here $N$ is the number of elementary atoms/molecules in the crystal, here 10$^5$, $m$ is their mass and $R$ is the radius of the sphere, here 10$^{-8}$ meter).}

\subsection{Beyond the atomic regime: nuclear localisation\label{nuclear2}}
The atomic (Schmidt) regime is valid provided the crystal that we consider is not too massive. 
The reason is that if the mass of the crystal is sufficiently large, Schmidt's radius of self-localisation will become smaller than the extent of the nucleus of the atoms that compose the crystal. In this case it is no longer correct to treat the self-interaction of each nucleus as a function in $1/r$ because we must take into account the structure of the nuclei as well. 
It is natural to treat each nucleus as a homogeneous spherical distribution of mass and hence we can repeat Di{\'o}si's analysis as presented in section \ref{sphere}. I.e.: instead of being hyperbolic, the self-interaction becomes parabolic. 
Combining the results outlined in the sections \ref{sphere} and \ref{cristal}, it is easy to check that in this regime (which we shall refer to as the {\it nuclear} regime), the ground state of the center of mass of the crystal has to obey
\begin{align}
&-\hbar^2\frac{\Delta_{CM}\Psi({\bf x}_{CM})}{2M}\nonumber
+GNm^2\int{d}^3 x' |\Psi({\bf x}'_{CM})|^2\frac{1}{r_\mathrm{nucleus}}\left(-\frac{6}{5}+\frac{1}{2}(\frac{|{\bf x}-{\bf x}'|}{r_\mathrm{nucleus}})^2\right)\Psi({\bf x}_{CM})\\ &\hskip9cm = e_\mathrm{nuclear}\Psi({\bf x}_{CM}),
\label{nuclear}
\end{align}
{which corresponds to the region $A$ in figure \ref{fig1}, where we conservatively chose for ${r_\mathrm{nucleus}}$ a value of $10^{-15}$ meter. However, as recognized by Chen {\it et al.} \cite{Chen}, a more realistic value would be of the order of $10^{-12}$ meter, due to the vibrational smearing of the atomic positions inside the crystal, as we shall discuss next.}
{\subsubsection{Conservative estimate of the critical transition.}
First, let us (rather conservatively) assume that the typical nucleus size is of the order of $10^{-15}$ meters, in accordance with the models considered in the previous sections in which nuclei are assumed to be clamped.}
If we consider spherical crystals of normal density, the transition between the atomic and nuclear regimes occurs when the number of atoms reaches a critical value, $N^\mathrm{nuclear}_c$, such that for $N\approx N_c^\mathrm{nuclear}$ one has that: $A_0^\mathrm{nuclear} \approx 10^{-5}A_0^\mathrm{atomic} =10^{-5}\delta$. Now, we showed previously that, in normal conditions, i.e. for normal densities, $N_c^\mathrm{atomic}\approx 10^{14}$. Hence, since $N\hbar^2/(GM^3)$ scales likes $1/N^2$, one finds that  $N_c^\mathrm{nuclear}=\sqrt{10^5}N_c^\mathrm{atomic}$. Thus, $N_c^\mathrm{nuclear}\approx 10^{17}$.
Typically, the nuclear regime will therefore be effective when the mass of the nanosphere is such that, say, {$ ({\hbar^2\over Gm^3N^2})^{{1\over 4}}(R)^{{3\over 4}}\leq 10^{-1}(10^{-5}\delta)$, the lefthand side in which scales like $1/N^{1/4}$. In normal conditions this implies that $N\geq (10^{+4})N_c^{nuclear}\approx$ }{ $10^{21}$ atoms. } {Now, one could wonder whether, in the same spirit, it would also be necessary to consider the substructure of the nuclei (e.g. protons, neutrons, quarks and so on) which would lead to subsequent transitions (nucleonic regime, quarkonic regime and so on). Fortunately, the situation is not so complex. Actually, we are erring on the conservative side by assuming that nuclei are clamped and the same remark is valid for nucleons and quarks. As is well-known, in a crystal, nuclei vibrate around their equilibrium position so that their wave function is de facto smeared out. This fact inspired us to develop a less conservative estimate of the structural corrections due to the inhomogeneous mass distribution inside the crystal, which we set out in detail in the next paragraph.}

{\subsubsection{Realistic estimate of the critical transition.\label{compa}}
In \cite{vanMeter} , van Meter writes  the following about the Schr\"odinger-Newton equation: {\it ...this theory predicts significant deviation
from conventional (linear) quantum mechanics. However, owing to the difficulty of
controlling quantum coherence on the one hand, and the weakness of gravity on the
other, definitive experimental falsification poses a technologically formidable challenge....} To our knowledge, only one realistic experimental proposal has been formulated, so far, to address this challenge. We have in mind here the proposal by Chen {\it et al.} \cite{Chen} in which it is suggested that one realize a tomographic measure of the state of a mesoscopic harmonic oscillator. Chen et al. consider a harmonic oscillator, cooled to a state the size of which is small compared to the quantity $\Delta x_{zp}~\!$: the square root of the so-called zero-point uncertainty which, at low temperatures, constitutes an estimate of the size of the wave packet of the nuclei inside the crystal. For an Si crystal,  $\Delta x_{zp}$ is of the order of $5\cdot 10^{-12}$ meter.}

{The basic idea in their proposal is that in addition to the external harmonic potential imposed on the system, because of its own self-gravitational interaction, it would also be subjected to a quadratic potential with a spring constant equal to $\frac{GM}{(\Delta x_{zp})^3}$, which is effectively the case in the Di\'osi regime, as we discussed in a previous section.  Indeed, if we consider the equation (\ref{fullpot}) which expresses the effective self-interaction potential of a homogeneous microsphere, it is straightforward to check that, at short distance ($d=|{\bf x}_{CM}-{\bf x}'_{CM}|\ll R)$ , the effective self-interaction reduces to a harmonic potential. Now, the spring constant $k^{hom}$ for a homogeneous microsphere is equal to $\frac{GM}{R^3}$, where $R$ is the radius of the nanosphere and $M$ its mass. We find thus a spring constant of the order of $\frac{Gm}{(a_0)^3}$, where $a_0$ is the Bohr radius (10$^{-10}$ meter) and $m$ the atomic mass of the crystal. The authors of \cite{Chen} suggest, roughly paraphrased, to replace  the Bohr radius in the expression $\frac{GM}{(a_0)^3}$ by $\Delta x_{zp}$, arguing that the density is considerably higher at the level of the nucleus, which would therefore yield the main contribution to the self-gravitational interaction. This idea is very similar to Schmidt's idea, except for the fact that the authors consider the macroscopic regime ($d=|{\bf x}_{CM}-{\bf x}'_{CM}|\ll \Delta x_{zp})$ and not the atomic regime as in Schmidt's case.}

{In order to model this situation, let us (as explained in appendix 3)  express $k$ , the effective spring constant which characterizes the self-interaction of a crystalline microsphere, as a sum of the contribution between different nuclei ($k^{hom}$) and the contribution from individual nuclei ($k^{nucleic}$):}
\begin{equation}k=k^{hom}+k^{nucleic},\label{chen}\end{equation}
{with  $k^{hom} = \frac{GM}{a_0^3}$ and where $k^{nucleic}={1\over M}\cdot N\cdot \frac{G(M/N)^2}{ \Delta x_{zp}^3}=\frac{GM}{N\cdot \Delta x_{zp}^3}$ represents the structure corrections resulting from the atomic structure of the mass distribution inside the crystal ($N$ is the number of nuclei that individually interact with themselves, as explained in detail in appendix 3).  Equation (84) illustrates a very general feature of structure corrections: the latter scale like $N$ while the contribution of the homogeneous sphere scales like $N^2$  so that structure corrections must be taken in consideration only for low masses.
Clearly, the structure corrections will be dominant ($k^{hom} \leq k^{nucleic}$) only if $N\leq ({a_0\over \Delta x_{zp}})^3\approx 8000$. The atomic mass of silicium being equal to 28 a.m.u. (44 a.m.u. in the case of SiO2), structural corrections will therefore prevail, only for macromolecules counting less than, say, 224000 nuclei (352000 for a SiO2 crystal). Now clearly, at this scale, self-gravity is so weak that it is impossible to realize any experiment that would allow us to reveal the existence of self-gravitational interaction as, for example, even the cosmic background radiation would overwhelm the self-gravitational corrections (as can be shown using the expressions (\ref{uh1}) and (\ref{uh2}) given in conclusion). Structure corrections can therefore be fully neglected if we consider a mass range of 1 g$<$M$<$10 kg, as is the case in ref.\cite{Chen},  excluding any possibility of observing manifestations of self-gravity in the hz-khz regime, and contradicting the suggestion made by Chen {\it et al.} Structure corrections can also be neglected in the mesoscopic range of masses that is of interest to us (10$^{-20}$ kg$<$M$<$10$^{-11}$ kg) and for which we see that it is consistent to treat the sphere as an object of homogeneous mass distribution. In particular, if instead of the conservative estimate of $10^{-15}$ meter one chooses a realistic value of $5\cdot 10^{-12}$ meter for the nucleic radius, the height of the step $B$ in figure 1 (obtained for a mass of $10^{-20}$ kilogram) will be almost unnoticeable and the region $A$ will lie in the continuation of region $C$. Our analysis therefore shows that, in the regime of mass beyond the mesoscopic transition, one can distinguish essentially two limiting cases for large and small ratios (width of CM- state)/(diameter of sphere) and two types of centre-of-mass dynamics, corresponding to  Schr\"odinger-Newton (\ref{NS}) in the first,  and to a harmonic potential (\ref{eqDiosieff}) in the second case (see also Ref.\cite{giulio} for a very recent and extensive discussion of these two limiting cases). However, as we shall explain in section \ref{timing}, this does not imply that all attempts at measuring manifestations of self-gravity, experimentally, should therefore be automatically doomed.

\section{Gravitationally induced self-collapse\label{timing}} 
\subsection{Spontaneous localisation\label{lah}}
In section 2 we studied the subtle questions Derrick's result raises concerning the stability of nonlinear evolution equations such as the Schr\"odinger-Newton equation, and we arrived at the conclusion that, provided unitarity is guaranteed (in other words, provided the unitarity of the Schr\"odinger-Newton-equation is a fundamental feature of the gravitational self-interaction\footnote{It is worth noting here that each kind of noise that is expressed by a real potential will also respect unitarity.}) then the ground states that we consider throughout this paper are indeed minimal energy states. More precisely, we explained that they minimize the energy functional for the (stationary) Schr\"odinger-Newton equation within a class of functions with bounded $L^2$ norm and, moreover, that conservation of this norm during the evolution yields a lower bound on the energy of the ground state.

From the literature on the Schr\"odinger-Newton equation  (\ref{NS}) however, it is not clear how it is exactly that a quantum system will (spontaneously) collapse to a ground state, under the action of self-gravitation. An important result in this respect is that of Arriola and Soler \cite{Arriola}, who show that if the energy of an initial state for the Schr\"odinger-Newton evolution is positive, then there will necessarily be an asymptotic expansion of the initial wave packet. However, when the initial energy is negative, a self-collapse can occur as (among other reasons) there is an energetic gain in doing so. Indeed, a collapse of an initial condition to the ground state, accompanied by an oscillatory `radiative' part in the solution (to preserve the norm and energy), will naturally lead to a decrease in energy for the collapsed state, as the radiative part of the solution will account for a large part of the total kinetic energy\footnote{It should also be noted that, contrary to the common intuition that prevails in the case of the linear Schr\"odinger equation, a multiple of the ground state is no longer a ground state. In particular, splitting a ground state into two or more parts will increase the total energy of the system.}. In our mind, it is this type of intrinsically non-linear mechanism, similar in spirit to the soliton-forming mechanism for the NLS equation, that explains how it is possible that the wave function will spontaneously evolve to a self-collapsed ground state. In \cite{Schmidt} Schmidt proposed gravitational dipolar emission as an extrinsic origin for this process, but we do not agree with this explanation. (The reason being that in the case where two packets are separated in space, there is no dipolar emission because the interference beating has zero amplitude in the region where the packets do not overlap. Hence there cannot be dipolar gravitational emission when the wave function consists of two non-overlapping packets.) Most importantly, we believe that the `relaxation' process of an initial state towards the ground state does not require any external source of dissipation: each time it is the non-linearity itself that will provide the mechanism for self-collapse. This idea is in fact confirmed by numerical studies of the Schr\"odinger-Newton equation  (\ref{NS}) and in particular by the work of van Meter \cite{vanMeter} who established, for a class of gaussian initial conditions, that the self-collapse process is indeed accompanied by radiation of excess mass (norm). In section 4.2 we shall give a rough approximation of the stability criteria of Arriola and Soler \cite{Arriola} and van Meter \cite{vanMeter} for a collapse to occur (see e.g. equation (\ref{vanmeterapprox})). The details of this approximation will be discussed in the Appendix.

Summarizing, our main hypotheses concerning spontaneous localisation of the wave function for the Schr\"odinger-Newton are the following:}
\begin{itemize}
\item[(A)] gravitational self-interaction is present in nature.
\item[(B)] this self-interaction is non-linear and non-local (in the sense that it obeys an integro-differential non-linear PDE, very close to equation (\ref{NS})). 
\item[(C)] self-interaction is the sole mechanism responsible for self-localisation\footnote{Hypothesis (C) excludes Stochastic Spontaneous Localisation mechanisms {\it\`a la} GRW \cite{Bell,grw,Diosi,pearle,CSL,Bassi}, a problem we shall briefly address in the conclusion. The precise interplay of such mechanisms with self-gravity is studied in \cite{GRWSELF}.} (as we believe  that self-localisation is likely to be enhanced in the presence of other, external, dissipation mechanisms).
\end{itemize}

Although hypothesis (C) could in principle be derived from hypotheses (A) and (B), this is beyond the scope of the present paper. One reason is that obtaining accurate numerical results on temporal evolutions that are described by non-linear PDEs is a highly nontrivial matter. Even in the case of well-known, local, non-linear evolution equations such as the NLS of KdV equations it is very difficult to give an accurate numerical estimate of the asymptotic state to which a wave packet collapses (e.g. of its $L^2$ norm), let alone of the finer detail of the transitory regime. In order to circumvent the complexity of such calculations, we propose an approximation scheme that allows us to replace the non-linear PDE by a system of coupled non-linear ODEs, as explained in detail in the Appendix. In the next section we shall show that this approximated dynamics constitutes a non-dissipative approximation of the non-linear PDEs (\ref{NS}) and (\ref{NSgen}), in the case of gaussian initial states.

\subsection{Non-dissipative approximation of the self-collapsing dynamics\label{appro}}
Let us consider a nanosphere that evolves in free space and that is subjected to no other interaction than its own self-gravitational attraction. In the quantum regime -- i.e., in the case where we can neglect the size of the nanosphere in comparison to the spread of the wave function of its center of mass -- this nanosphere will undergo a force which we can roughly approximate by a radial harmonic potential, the same as that which would be generated, at short distance, in case the wave function of the center of mass of the nanosphere is a Heaviside function that is constant between the origin and a distance roughly equal to the extent of the wave function:
$r^\mathrm{{max}}=\alpha\sqrt{<r^2>}$ (where $\alpha$ is an adjustable parameter taken to be of the order of unity when we remain confined to the vicinity of the mesoscopic transition and/or in the quantum regime; see the Appendix for a more detailed treatment). The resulting harmonic potential can be written as $k^\mathrm{{self}}  r^2/2$, where
\begin{equation}
k^\mathrm{{self}}= GM^2/(\alpha L)^3,
\end{equation} 
with $L= \sqrt{<r^2>}$ (here the index `$\mathrm{self}$' refers to the fact that we are dealing with self-interaction). A second essential element in  our approximation is that we assume that this potential remains valid, even for larger values of $r$ ($r>r^\mathrm{{max}}$).

As we shall explain in the Appendix, in the absence of external forces,  at this level of approximation, we find that the radius $r_\mathrm{self}$ of the static bound state obeys  the following constraint:
{\beq
L^\mathrm{self}=\sqrt{<r^2_\mathrm{self}>}=\sqrt{{3\hbar\over 2\sqrt{(k^\mathrm{self})M}}}=\sqrt{{3\hbar\sqrt{(\alpha L^\mathrm{self})^3}\over 2\sqrt{( GM^2)M}}}~,
\eeq 
so that $L^\mathrm{self}=9\alpha^3\hbar^2/(4GM^3)$. The parameter $\alpha$ can be adjusted in various ways. For instance, in the quantum regime, one could choose $\alpha^3$ such that the width of the ground state fits the rough estimate of $2\hbar^2/GM^3$ derived in \cite{Giulini}. This would correspond to the choice $\alpha^3=8/9$ (or $\alpha\approx 0.96$). However, one is also free to choose the value of $\alpha^3$ so as to reproduce the minimal energy of the bound state. As shown in the Appendix, we can associate a conserved energy $E_{eff}$ to the approximated Schr\"odinger-Newton equation
\begin{equation}
{i}\hbar\frac{\partial\Psi(t,{\bf x})}{\partial t}=-\hbar^2\frac{\Delta\Psi(t,{\bf x})}{2M}
+{ GM^2\over 2(\alpha\sqrt{<r^2>})^3}r^2~\! \Psi(t,{\bf x}),\label{TDfirst}
\end{equation}
obtained from (\ref{NS}) by replacing the self-gravitational potential by
\begin{equation}
V_{harm}= { GM^2\over 2(\alpha\sqrt{<r^2>})^3}r^2.
\end{equation}
In the case of a gaussian state  $(1/(\pi a^2)^{3/4})\cdot \exp(-r^2/2a^2)$, this energy $E_{eff}$ is equal to
\begin{equation}
3\hbar^2/4Ma^2-GM^2/(\alpha^3\sqrt{3/2}a),
\end{equation}
and $E_{eff}$ is therefore minimized when $a=a_0=(3/2)^{3/2}\alpha^3\hbar^2/GM^3$, for which it attains the value $E^{min}_{eff}$= $-(2/9)G^2M^5/\alpha^6\hbar^2$. If we impose that this value of the energy coincides with the aforementioned minimum of $-(0.163)G^2M^5/\hbar^2$, we obtain the constraint}
\begin{equation}\alpha^6\approx 0.222/0.163~\!,\end{equation}
which corresponds to a value $\alpha\approx 1.05~\!$.
 Besides, as is also shown in the Appendix, a gaussian initial state is unstable in the sense that it will spread to infinity\footnote{To be precise: we showed that a gaussian initial state is stable when its energy is negative, which is slightly different, but it is worth noting however that a linear stability analysis of the system of coupled non-linear differential equations (\ref{redsys1}--\ref{redsys3}) given in the Appendix supports the same conclusion as that obtained by Arriola and Soler \cite{Arriola}.} when its energy $E_{eff}$ (w.r.t. equation \eqref{TDfirst}) is positive\footnote{Intuitively, this corresponds to the threshold above which kinetic energy overcomes potential binding energy.}, which means that $3\hbar^2/4Ma^2\geq GM^2/(\alpha^3\sqrt{3/2}a)$ or equivalently $a\leq a_0/2$. If we reformulate this stability criterion in terms of the spread $\sqrt{r^2}$, which is equal to $\sqrt{3/2}\,a$, we obtain
  \begin{equation}\sqrt{r^2}\leq (9/8)(\alpha)^3(\hbar^2/GM^3)\approx (9/8)\sqrt{222/163}(\hbar^2/GM^3)\approx 1.31 (\hbar^2/GM^3)\label{vanmeterapprox}\end{equation}
It is instructive to compare this condition with the criterion (\ref{vanmeter}) found by van Meter \cite{vanMeter}, which reads $\sqrt{r^2}\leq 1.48 (\hbar^2/GM^3)$. It is of course not suprising that the exact numerical values in both expressions differ slightly, as van Meter considered the case in which the exact dynamics (\ref{NS}) for a gaussian initial state becomes unstable and we, on the other hand, give a gaussian approximation for the resulting bound state (which is of course not gaussian). Another difference is that our approximated dynamics predicts that gaussian states that obey the stability criterion $\sqrt{r^2}> (9/8)(\alpha)^3(\hbar^2/GM^3)$ will oscillate around the bound state during an infinitely long time, while in van Meter's approach they collapse sooner or later onto the bound state by radiating some excess norm (mass). In view of the results by Arriola and Soler \cite{Arriola} however, such oscillations do not seem to be completely implausible and further numerical experiments will be necessary to settle this question. Altogether, we believe that our approximated dynamics constitute a rather faithful non-dissipative approximation of the ``real'' self-collapsing process, as it seems to describe the same phenomena as the real Sch\"odinger-Newton equation, for critical values of the spread and energy of the wave packets that are surprisingly close to those obtained through other approaches. Furthermore, as we also show in the Appendix, our approximation scheme has the immense advantage that it can be enlarged in order to incorporate finite structure effects, which are necessary when tackling the self-collapsing process in the mesoscopic regime and beyond, without increasing the computational complexity.

\section{Experimental manifestations of self-gravity\label{timing}} 

\subsection{How to isolate self-gravitation from environmental frictions}
To some extent, the question of localization of quantum macro-objects through gravitational collapse is purely academic. This is mainly due to two reasons. Firstly, there are numerous sources of decoherence \cite{Haroche,schlosshauer} that can explain why coherent superpositions of macro-objects only survive during a very short time. Decoherence due to scattering by a mesoscopic object of environmental thermal photons, or of residual gas molecules, provides for instance a very plausible mechanism for localization.
The second reason is that localized macro-objects are liquid or solid and that, in general, they interact strongly with their environment through all kinds of frictious forces. We do not need for instance to resort to gravitational self-collapse in order to explain why the furniture in a room (chairs, tables and so on) stays localized at the same place in the absence of external intervention. This is simply because movement requires energy and without a certain energy supply, movement is impossible. This elementary truth was already known to Aristotle and it remains, despite the Newtonian revolution, a valid explanation of many everyday life experiences. Actually, Newton's ideas impose themselves only rarely in every day life, for instance we need to resort to them in order to explain images that were captured during a mission of the space shuttle, where people and objects are floating inertially in space. Similarly, in order to observe self-gravitational effects, which are very weak, one has to get rid of friction forces and/or sources of decoherence that are typically much stronger than self-gravitation. 

This is realizable nowadays thanks to optical tweezer technology which makes it possible to let nano-objects levitate \cite{aspelreview}. Recently, Aspelmeyer {\it et al.} proposed to prepare levitating (silicate) nanospheres with radii of the order of 40 nanometers, in a coherent quantum state of size comparable to the size of the sphere \cite{aspelPRA,aspelPRL}. Typically, such nanospheres comprise in the order of $10^8$ atoms, which situates them close to the mesoscopic transition\footnote{At normal densities, the mesoscopic transition is likely to occur for nanospheres of radius equal to more or less 130 nanometers. In the proposal made in  \cite{aspelPRA,aspelPRL}, the density is more or less 2,6 times the normal density and the critical mesoscopic radius $R_c^{mesoscopic}$ scales like the density to the power $-0.3$ so that $R_c^{mesoscopic}\approx 100$ nanometer, which means that the proposal of Aspelmeyer's group aims at testing the ``close'' quantum regime. } (in fact, on the quantum side of the transition, in our terminology). A slightly different proposal \cite{maqro,kalten} aims at embarking nano-objects on a spaceship, in order to imbed them in a gravitation-free, inertial, environment. Such proposals essentially aim at testing the superposition principle in the mesoscopic domain, a task that seemed to be unrealizable some years ago \cite{frontiers} by more conventional techniques (atomic and molecular interferometry for instance). The essential reason for this seeming impossibility is that the de Broglie wavelength of nano-objects is so small that uncontrollable de-phasing will occur along the arms of the interferometer, which makes it impossible to preserve coherence and to observe interferences.  This limitation is less stringent in the case of light objects, which explains why the superposition principle has been tested with various elementary particles (electrons, neutrons) and molecules (bucky ball porphyrine and so on), up to about 1000 units of atomic mass  \cite{Arndt,Arndt2,Arndt3}. Actually, this range of masses corresponds to the far quantum regime (far away from the mesoscopic transition) which we addressed in section \ref{far}. In this regime, the self-interaction is so small (and, correspondingly, the radius of the ground state so huge) that the self-gravitational interaction can consistently be neglected, and it is to be expected that the evolution is essentially the usual linear one. 

In the mesoscopic and close quantum regimes however, it is possible in principle, as we shall  soon show, to test the validity of hypotheses (A) and (B) formulated in section \ref{lah}, which is already very challenging in itself. Indeed, as we discussed in the introduction, electronic electrostatic self-energy is absent from Schr\"odinger's equation, because in QED electro-magnetic self-energy is incorporated from the beginning in the renormalised effective parameters that characterize the electron (mass, charge and so on). 
It could well be that for similar reasons the Schr\"odinger-Newton equation is not valid, and that gravitation is negligible compared to other types of interactions, at the quantum and mesoscopic scales \cite{Adler2}, in which case the evolution of the wave function of the center of mass of a crystalline nanosphere is linear, in agreement with the usual Schr\"odinger equation
\begin{equation}
i\hbar{\partial \Psi({\bf x_{CM}},t)\over \partial t}=-\frac{\hbar^2}{2M}~\Delta_{CM}\Psi({\bf x_{CM}},t)
\label{lineartemp}\,.
\end{equation}
Now, as we have argued before ({see e.g. footnote 4 in the introduction}), if self-gravitation {\em is} present, the wave function $\Psi({\bf x_{CM}})$ will on the contrary obey a non-linear equation characterized by two limiting cases (cf. the comments at the end of section 3.4.2), the quantum and ``harmonic oscillator'' regimes \cite{giulio}. Ultimately, this kind of question can only be settled through experiments. Hence, it would be very interesting to be able to test the mere existence of self-gravitation itself, which brings us to the experimental proposal in the next paragraph.

\subsection{Experimental proposal\label{exp2}}

In this section, we shall consider an experiment realized in a very weak gravitational field, typically in a satellite. Our aim is to reveal the existence of a self-gravitational potential of the type considered in the previous sections, which would influence the distribution of positions of the center of mass of a nanosphere in the vicinity of the mesoscopic regime. Our proposal is inspired by recent experimental protocols (e.g. by Aspelmeyer {\it et al.} as described in \cite{aspelPRA,aspelPRL}) during which a nanosphere -- of radius comprised, say, between 40 and 100 nanometers -- is trapped in an optical trap and gets cooled to its ground state. 
Similar to what is done with atoms trapped in optical lattices, we can in principle measure the position of the nanosphere by turning off the levitating potential and by letting the nanosphere fall in a microgravity environment \cite{maqro,kalten} onto a precisely calibrated surface \cite{aspelPRA,Arndt}. Our goal is to realize this experiment repeatedly, in order to estimate the dispersion of the lateral position. If we release the trap\footnote{Magneto-static, mechanical and other ways to initially trap and localize the nanosphere could be imagined at this level. For instance, the initial confinement could be obtained thanks to a device similar to those used in the building of quantum corrals \cite{corral}, 
in which the nanosphere is localised through point effects on a distance smaller than 1\AA. }, and if there is no self-gravitation, the state of the center of mass of the nanosphere will diffuse according to the free Schr\"odinger evolution (\ref{lineartemp}) for which it is well-known that in the absence of any forces, the dispersion of an initially gaussian packet at time $t$, $\delta x_t$, obeys:
 \begin{equation}
 \delta x_t=\delta x_0\sqrt{1+({\hbar t\over \delta x_0^2 M})^2}.\label{libre}
 \end{equation} 
 If we wait long enough, the ratio $\hbar t/\delta x_0^2 M$ will eventually become significantly larger than 1 and we have that $\delta x_t\approx \hbar t/\delta x_0 M$, in accordance with the Heisenberg uncertainties  ($M\delta v\approx \hbar/\delta x_0 $) and with the fact that if we let a wave packet evolve freely  during a sufficiently long time, the measurement of its position amounts\footnote{Using the properties of the propagator of the free Schr\"odinger equation, Feynman and Hibbs established \cite{Feynman}  that $\lim_{t\rightarrow \infty} |\Psi(x,t)|^2dx|_{x=pt/m}=|\tilde \Psi(p/\hbar,t=0)|^2dp$ where $\tilde \Psi(k,t=0)$ is the Fourier transform of the initial wave function: $\tilde \Psi(k,t=0)$=$(1/\sqrt{2\pi})\int_{-\infty}^{+\infty}exp(-ikx)\Psi(x,t=0)$. This also means that, whatever the shape of the initial wave packet is, equation (\ref{libre}) provides (when $t$ is large enough) a lower bound for $\delta x_t$, because real gaussian wave packets saturate the Heisenberg uncertainty relations.} to measuring a velocity ($\delta x\approx \delta v\cdot t$). If on the other hand self-gravitation {\em is} present, the dispersion will slow down and could even be inverted in case self-gravitation prevails, as shown in the results of our numerical simulations plotted in figure \ref{fig5}. As we shall now show, in the mesoscopic regime, such an experiment makes it possible to discriminate between an evolution with or without gravitational self-interaction (i.e., between the free linear Schr\"odinger evolution and the Schr\"odinger-Newton evolution)\footnote{{In the present paper our aim is only to sketch an ideal experiment that would be realizable in ideal conditions (perfect vacuum, zero temperature, no external decoherence, etc.). The discussion of the robustness of free fall experiments of this type with respect to experimental imperfections can be found elsewhere in the literature. For instance, the realizability of extreme pressure and temperature conditions as well as the stability of the satellite are dealt with in Refs.\cite{kalten,maqro}, while the interplay between decoherence and self-gravitation is studied in Ref.\cite{GRWSELF}.}}. 

\subsubsection{Silicate nanospheres}
Let us first consider the case where the density of the nanosphere is equal to more or less twice the normal density (which characterizes the silicate nanospheres previously considered by Aspelmeyer et al. in their experimental proposals \cite{aspelPRA,aspelPRL}). 
 
To tackle the physical problem set out in the previous section, we numerically simulated  the time evolution of initial gaussian states under equation (\ref{TDfirst}) following the approximation scheme outlined in the Appendix and section \ref{appro}.
In the quantum (hyperbolic) regime of section 3.1,  the influence of self-gravity is maximal around the bound state, that is to say when $\delta x_0\approx {\hbar^2\over G M^3}$ so that ${\hbar \over \delta x_0 M}\approx {GM^2 \over \hbar }$. In order to maximize the difference between the free and self-gravitating trajectories, one should therefore maximize the factor ${GM^2 \over \hbar }$. If the density is kept fixed, it is obvious that one should try to maximize  the radius of the nanosphere $R_S$. Now, the maximal mass (radius) in which the quantum regime is still valid corresponds to the mesoscopic transition $\hbar^2/GM^3=R$ introduced in sec. 3.2.4. Beyond this transition, i.e. for larger masses and radii, one has that the spread of the bound state satisfies $\delta x_0\approx ({\hbar^2\over GM^3})^{{1\over 4}}R^{{3\over 4}}$, so that 
${\hbar \over \delta x_0 M}\approx {\hbar^{1/2}G^{1/4} \over ({4\pi\over 3}\rho)^{1/4}R^{3/2} }$, in which case one should try to minimize the mass (radius) of the nanosphere in order to optimally discriminate between the free and self-gravitational cases.

The experiment should thus, at twice the normal density, be realized with a nanosphere of radius $R$ more or less equal to $110$ nanometers (for which the values of $\hbar^2/(G M^3)$ and $({\hbar^2\over GM^3})^{{1\over 4}}R^{{3\over 4}}$ are both equal { to $110$ nanometers, which corresponds to the mesoscopic transition).}
Then, if the initial extent of the ground state wave function of the center of mass is of the order of $110$ nanometers, it will remain so throughout time, in case self-gravitational effects are indeed present (see figures \ref{fig5} and \ref{fig4.5}). Otherwise (see figure \ref{fig4.5}, plain lines), in the free case, the extent of the wave function will be of the order of $10 \delta x_0\approx 10^{-6}\,\mathrm{m}= 1$ micron after $10^4\,\mathrm{s}$. 

\begin{figure}
\centering
\includegraphics[width=\textwidth]{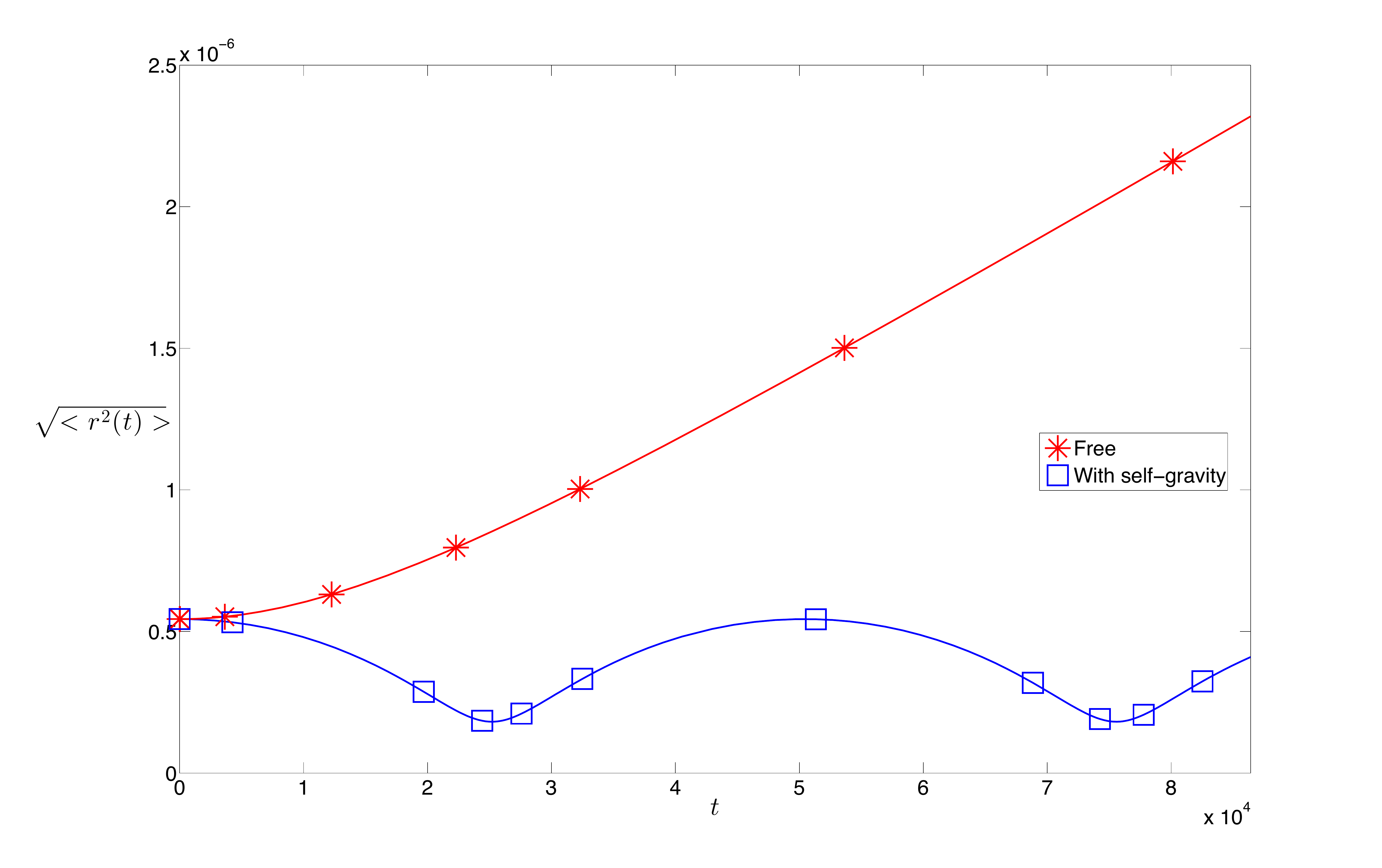}
\caption{\label{fig5} $\sqrt{<r^2>}_t$ expressed in $\mathrm{m}$ as a function of $\mathrm{t}$ (in s) for a homogeneous nanosphere of radius $10^{-7}~\mathrm{m}$ and mass density of $2650~\mathrm{kg}\,\mathrm{m}^{-3}$.
The upper curve corresponds to the free case (\ref{libre}), whereas the lower curve corresponds to the self-gravitating case (\ref{NS}). }
\end{figure}

One can in principle measure the dispersion of the positions by repeated accurate measurements of a collection of impacts of freely falling nanospheres on a surface. In order to observe the effect we are after, rather long times will be required, which explains why it is necessary to embark the experiment on a satellite, i.e. in an almost gravitation-free environment.
Of course, such an experiment is not so easy to realize: to prepare the nanosphere in an initial state for which $\delta x_0 \approx 10^{-7} m$ and then to measure its position with a precision of the order of one micron, after a free fall of $10^4$ s, constitutes a serious experimental challenge. In practice, even in a satellite it is impossible to attain zero gravity as the attraction exerted by the walls of the satellite itself cannot be neglected. Microgravity in such an environment \cite{maqro} is therefore typically, in the best case, of the order of $10^{-9}g$, where $g$ is the gravitational acceleration at the surface of the earth.  A free fall of $10^4$s, in such conditions, represents a distance of 0.5 meter, which is at the limit of distances available in a satellite.

Another possibility would be to prepare an even larger value of $\delta x_0$, as shown in figure \ref{fig5}, in which case the gaussian state shrinks due to self-gravity, but then (as discussed above) the free expansion is slowed down too, so that no benefit results from this approach. 

Yet another strategy would be to initially localize the nanosphere with strong contact interactions, for which we expect that $\delta x_0$ can be made very small. It is interesting then to work in the parabolic regime ($r<R$), for which the effective potential is (in first approximation) a radial harmonic potential ${kr^2\over 2}$, characterized by a constant $k=GM^2/ R^3$ that does not depend on the size of the wave packet, until it reaches the mesoscopic transition $\delta x_t \approx R$. At fixed density, the constant $k$ in this effective potential obviously increases as $R^3$, the mass being proportional to $R^3$. This implies that when the radius of the nanosphere increases, packets of decreasing size $\delta x_0$ will remain trapped by the potential. {In particular, if initially the wave packet is prepared with a width $\delta x_0$, it will remain confined (see footnote \ref{blobl}) inside the region $r<2R$ whenever  its initial kinetic energy  is smaller than the height of the potential evaluated at $r=2R$, i.e.: whenever $3\hbar^2/ 4M \delta x_0^2<2GM^2/ R$, which imposes that $8/3~\!GM^3/R=4G(4\pi\rho/3)^3R^8>\hbar^2/  \delta x_0^2$. Corrections for internal structure (see figure \ref{unknown}) impose the more realistic bound $1.4 GM^3/R=1.4 G(4\pi\rho/3)^3R^8>\hbar^2/  \delta x_0^2$.
The minimal value of $\delta x_0$, that for which the bound above is saturated, is
\begin{equation}
\delta x_0=\sqrt {\hbar^2R\over 1.4 GM^3}~\!.
\end{equation}
If we again consider the mesoscopic transition {($R\approx 110$ nanometers), we obtain that $\delta x_0\approx R$, as discussed above. Now, one can see that the factors ${\hbar \over \delta x_0^2 M}$ and ${\hbar \over \delta x_0 M}$ increase when the radius $R$ of the sphere increases. This suggests, for instance, to consider the case $R=10^{-6}$ m which imposes, at twice the normal density, that $M\approx 10^{-14}$ kg and $\delta x_0\approx10^{-11}$ m, such that $\delta x_t\approx {\hbar t\over \delta x_0 M}\approx 10^{-9}t$ provided $t>10^{-2}$ s. One therefore has to wait for at least $10^2$ s before the free and self-interacting situations can be discriminated by standard techniques\footnote{Laser interferometry makes it possible for instance to measure positions with a precision of the order of the wavelength of the light emitted by the laser, and $10^{-7}$ m is a standard wavelength for such applications. In \cite{Arndt} a sub-Rayleigh precision of 40 nanometer is reported in similar experiments. A precision of 1 nanometer is reported in \cite{fionax}. These very precise localisation techniques require to fit the recorded data with the point-spread-function (PSF) of the optical device used to measure them, which makes it possible to get rid of the Abbe-Rayleigh limit ($\delta x\approx \lambda$). Nowadays, they are routinely implemented in biophotonics \cite{fionax}. From now on we shall take for granted that a measure of position with an accuracy of 10 nanometer is reachable experimentally.}}, which is still difficult to realize experimentally but not impossible. 

On the other hand, a radius $R=10^{-5}$m of 10 micron  imposes, at twice the normal density, that $M\approx 10^{-11}$kg and $\delta x_0\approx10^{-15}$m, so that $\delta x_t\approx {\hbar t\over \delta x_0 M}\approx 10^{-8}t$ provided $t>10^{-7}$s, in which case $\delta x_t\approx  10^{-7}$m after 10s.} 
It is obviously difficult to initially localize any further the nanosphere. In figures \ref{simulation2} and \ref{simulation1} we show the results of our simulations in the corresponding conditions ($R$= one micron and ten micron). In order to obtain the results presented in these figures, we scanned a regime of parameters situated in the mesoscopic transition {($R\approx 100$ nanometers) and we minimized the free-fall time necessary for discriminating between the free and self-interacting regimes, imposing that the free and self-collapsed trajectory differ by more or less 100 nanometer.

\begin{figure}
\centering
\includegraphics[width=1.0\textwidth]{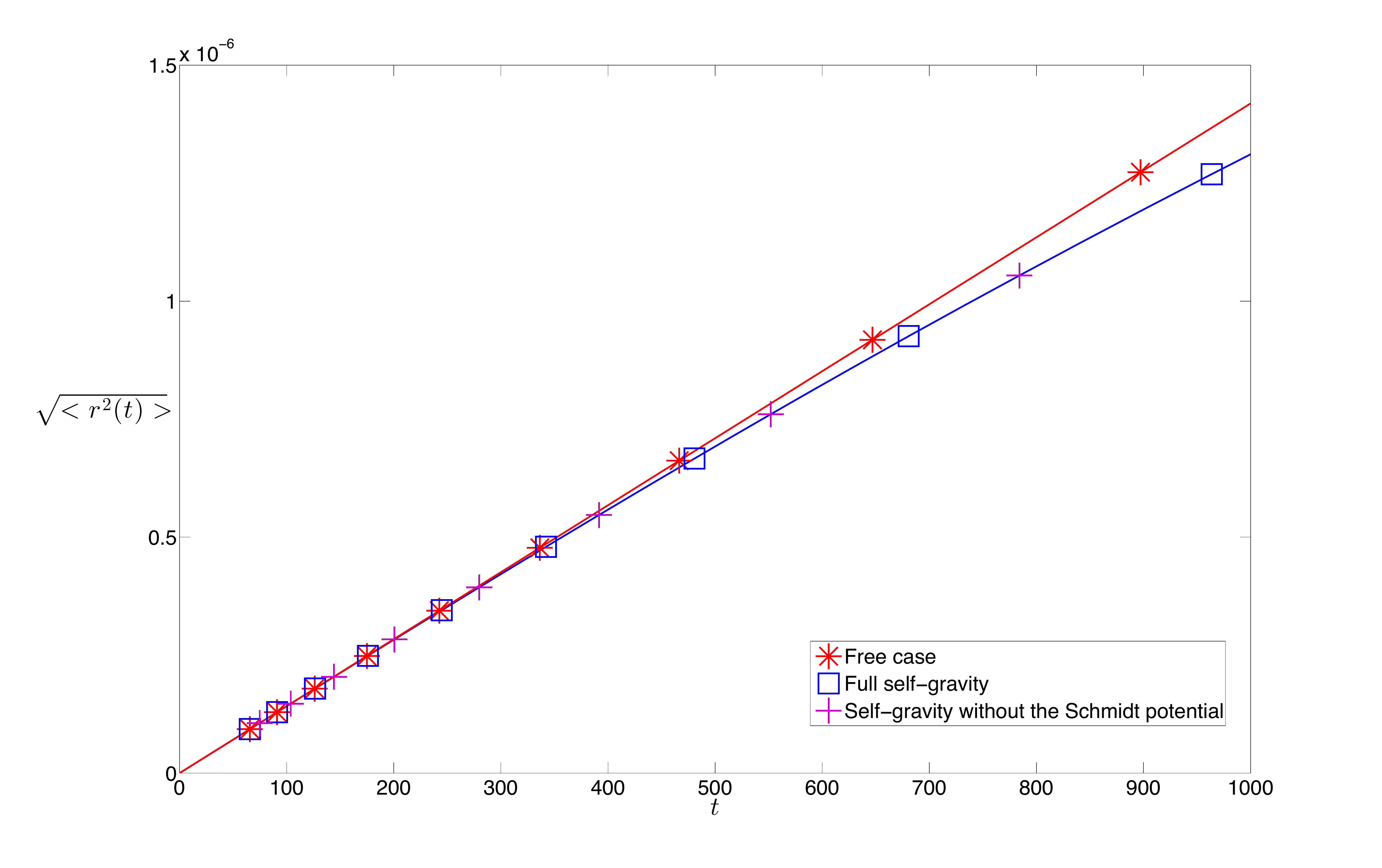}
\caption{\label{simulation2} Separation of the free and self-interacting packets as a function of time (in $\mathrm{s}$). $R=10^{-6}\,\mathrm{m}$ and $\delta x_0=10^{-11}\,\mathrm{m}$.
The mass density of the nanosphere is $2650\,\mathrm{kg}\,\mathrm{m}^{-3}$. 
A separation of $10^{-7}~\mathrm{m}$ between the two mean spreads occurs approximately after $t=970\mathrm{s}$.}
\end{figure}
\begin{figure}
\centering
\includegraphics[width=1.0\textwidth]{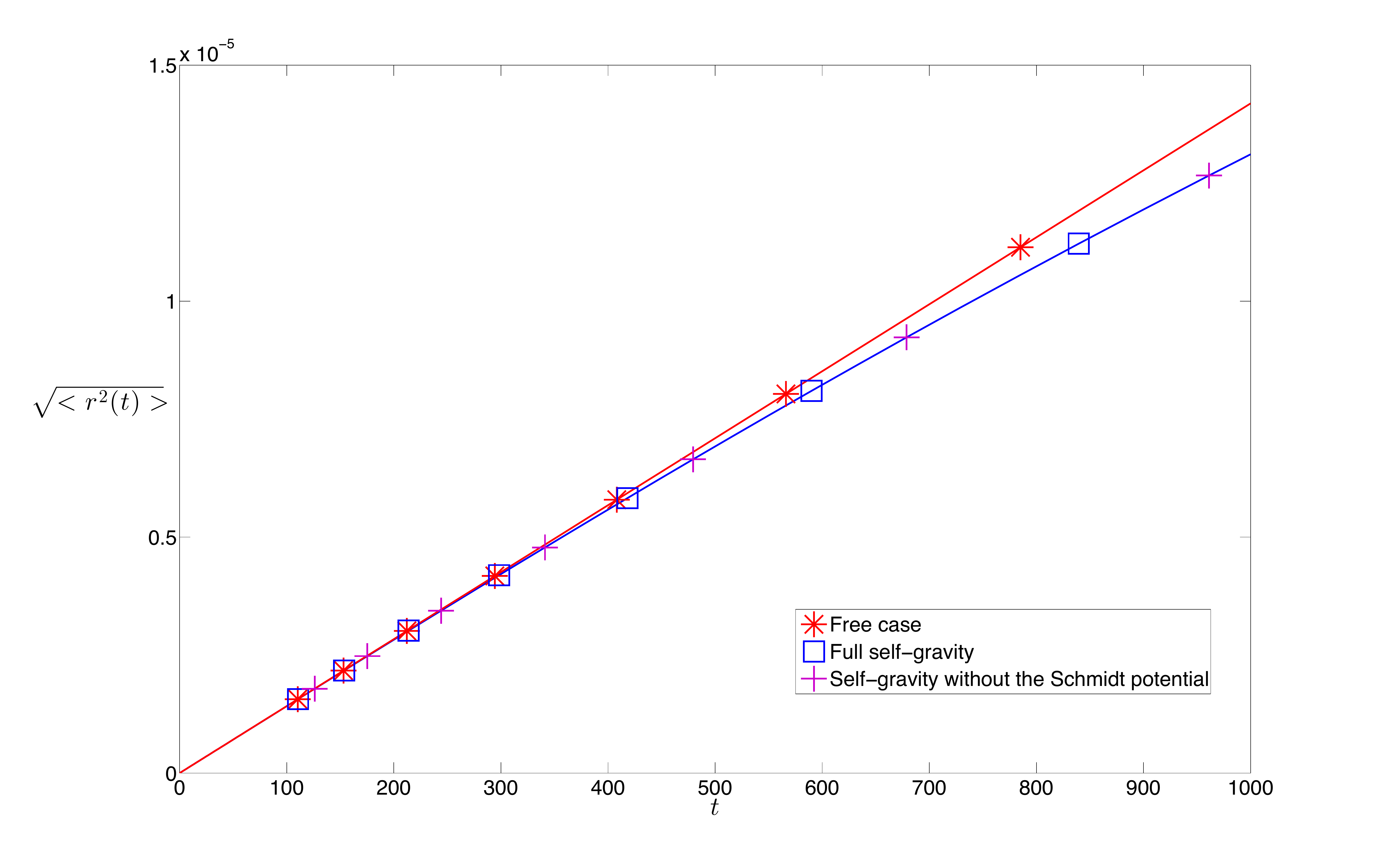}
\caption{\label{simulation1} Separation of the free and self-interacting packets as a function of time (in $\mathrm{s}$). $R=10^{-5}\,\mathrm{m}$ and $\delta x_0=10^{-15}\,\mathrm{m}$.
The mass density of the nanosphere is $20000\,\mathrm{kg}\,\mathrm{m}^{-3}$. 
A separation of $10^{-7}~\mathrm{m}$ between the two mean spreads occurs approximately after $t=410\mathrm{s}$.}
\end{figure}

To conclude, in the case of silicate nanospheres, our second strategy, valid in the classical, near mesoscopic, regime requires us to be able to reach free-fall times of the order of at least 400 s and to measure the dispersion of positions of freely falling nanospheres with an accuracy in the range of $10^{-8}$m $\sim$ $10^{-7}$m.

\subsubsection{Gold nanospheres}

In the case of gold nanospheres (more or less seven times more dense than silicate, the density of gold being equal to 19.32 g/$\rm{cm}^{3}$), self-gravity is enhanced so that, even when the free diffusion is nearly frozen due to the heavy weight of the object, the attraction of the bound state is such that it results into clearly distinguishable experimental effects. This can be seen for instance at the level of the lower curves of figure \ref{fig5-09-11} in which the difference between the free and self-gravitating spreads is larger than 50 nanometers after 500 seconds, which is not out of reach\footnote{{The displacement we predicted is so small that at first sight the required measurement precision and the required shielding from any sort of external influence, even in outer space, might seem completely unrealistic. However, it fits nicely with well-documented proposals of optomechanical experiments embarked on a satellite or the ISS, aimed for example at testing the existence of spontaneous localisation mechanisms \cite{kalten, maqro,Romero,aspelPRL}. It is beyond the scope of the present paper to address these very technical questions (and in particular the influence of decoherence) here, but we do address these questions in a separate publication \cite{GRWSELF}. At this point, it suffices to mention that an accuracy of 50 nanometer in the measurement of positions, after a free fall of 500 seconds is entirely compatible with the requirement of a 5 nanometer accuracy after a time of flight of 200 seconds such as required for the MAQRO proposal \cite{maqro}. In a last resort, the time of flight is bounded from above by the average time separating two collisions of the sphere with the residual gas atoms/molecules, which is of the order of 200-300 seconds in extreme vacuum (and temperature) conditions \cite{maqro,Romero}.}} of presently available sub-Rayleigh localisation measures \cite{Arndt,fionax}. If on the contrary we try to enhance the diffusion by diminishing the initial spread, then the kinetic energy dominates the self-gravitational binding so that the trajectories with and without self-gravity are no longer distinguishable, as can be seen in figure \ref{fig4-11-100k8} 
(lower curves). Contributions due to decoherence will discussed in the conclusions. The results plotted here were derived from the work developed in \cite{GRWSELF}.

\begin{figure}
\centering
\includegraphics[width=\textwidth]{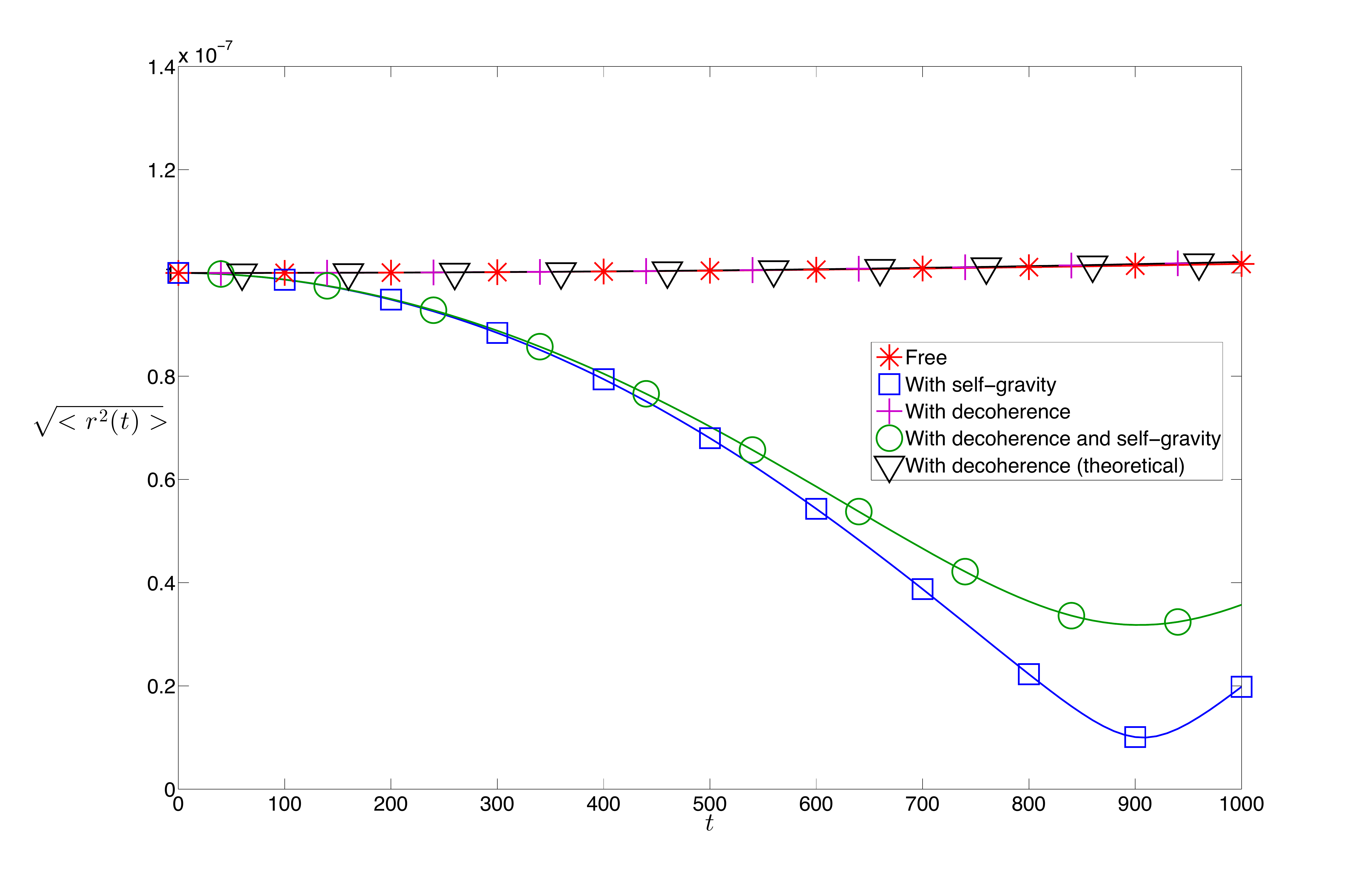}
\caption{\label{fig5-09-11}Spread (in $\textrm{m}$) of the center-of-mass density matrix for $t\in[0,1000]$, for a nanosphere of radius 100 nm and mass density 
$\rho=20000\,\mathrm{kg}\,\mathrm{m}^{-3}$ (density of gold), with initial spread $\delta r_0$=$10^{-7}$ m. The curves marked by squares and stars correspond to the absence of decoherence, respectively with and 
without self-gravity. The decoherence parameters (\ref{Lambda}) are $\alpha=10^{11}$ m$^{-2}$ and $\gamma=1$ s $^{-1}$. The curves marked by plus symbols and circles respectively correspond to the absence and presence of self-gravity, while the curve marked by triangles is the asymptotic estimate of the curve marked by plus symbols \cite{GRWSELF}.}
\end{figure}

 \begin{figure}
\centering
\includegraphics[width=\textwidth]{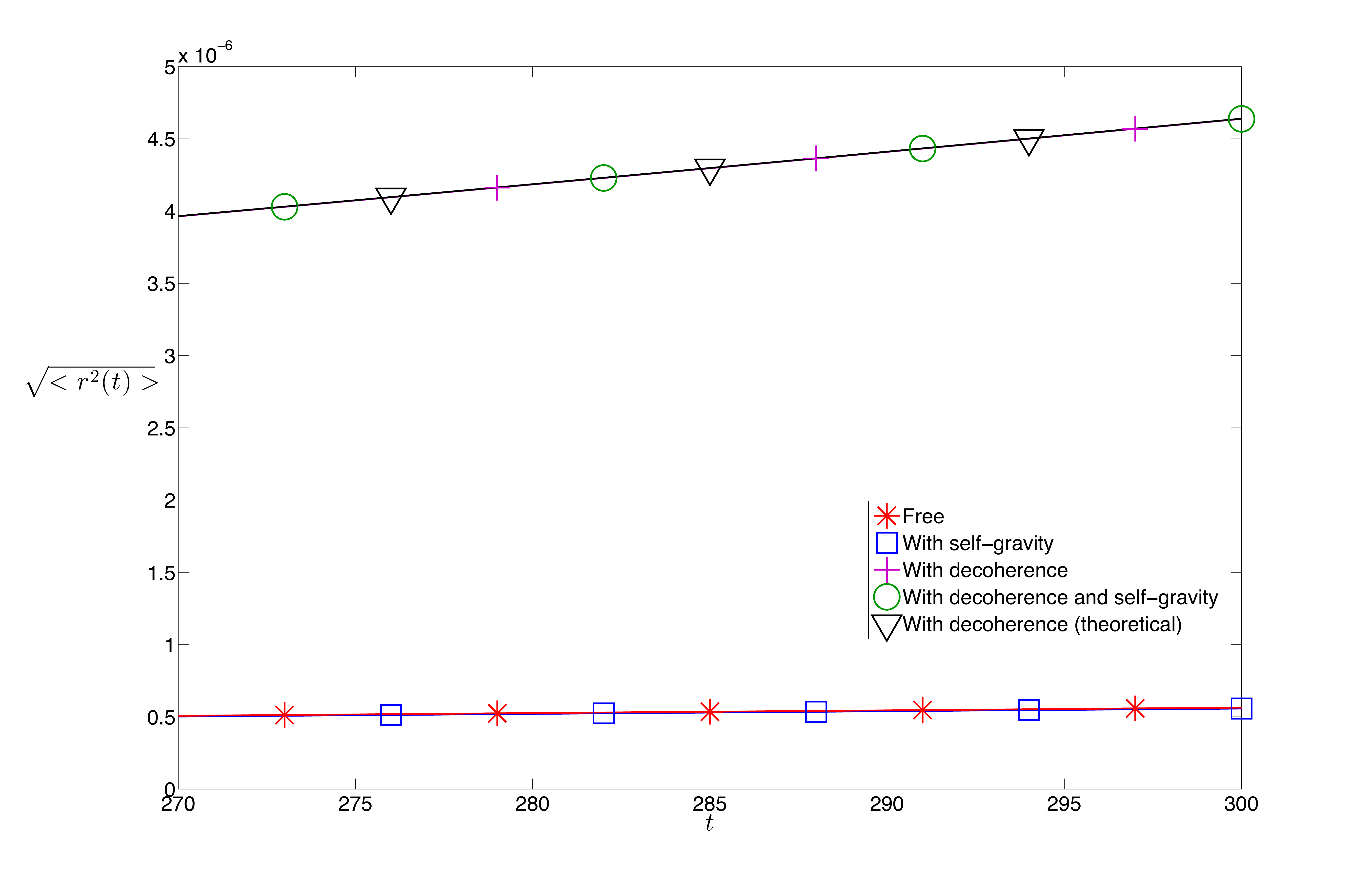}
\caption{\label{fig4-11-100k8}Spread (in $\textrm{m}$) of the center-of-mass density matrix for $t\in[270,300]$, for a nanosphere of radius 100 nm and mass density 
$\rho=20000\,\mathrm{kg}\,\mathrm{m}^{-3}$ (density of gold), with initial spread $\delta r_0$=$10^{-9}$ m. The curves marked by squares and stars correspond to the absence of decoherence, respectively with and 
without self-gravity. The decoherence parameters (\ref{Lambda}) are $\alpha=10^{18}$ m$^{-2}$ and $\gamma=1$ s $^{-1}$. The curves marked by plus symbols and circles respectively correspond to the absence and presence of self-gravity, while the curve marked by triangles is the asymptotic estimate of the curve marked by plus symbols \cite{GRWSELF}.}
\end{figure}

\section{Conclusions}\label{conclusions}
In this paper we have characterized the different regimes that can occur when one wishes to study the self-collapsed ground state of a crystalline nanosphere. These range from the extreme classical (nuclear) regime to the far quantum (elementary particle) regime, passing through the atomic, mesoscopic and quantum regimes. An important corollary of our analysis is that beyond the mesoscopic transition, i.e. for sufficiently massive spheres -- as the effective self-gravitational potential scales as $N^2$ in this regime, as opposed to the nuclear regime in which it scales as $N$ (see e.g. equation (\ref{chen})) --  it is consistent to treat the nanosphere as a homogeneous object and to neglect structural corrections related to its atomic (discrete and composite) structure. There remain then essentially three regimes: the quantum regime (\ref{NS}) and the macroscopic (``harmonic potential'') regime (\ref{eqDiosieff}), separated by a rather complicated interpolation polynomial that we derive in appendix 3.
We also discussed in depth the stability of the ground state in these different regimes.

In the final part of the paper we gave an outline of what we believe to be a realistic experimental proposal, aimed at revealing the existence of a gravitational self-interaction in the vicinity of the mesoscopic scale.
We have not addressed here other questions that were already tackled in previous experimental proposals \cite{maqro,aspelPRL,Romero},  such as the validity of the superposition principle and in particular the preparation of Schr\"odinger cat states. Actually, in the experiment that we proposed, the situation is much simpler than in the case where a system has to be prepared in a coherent superposition state of two spatially separated wave packets (cat state). This has spared us the extra complications of having to analyse the localization process induced by the self-gravitational interaction, which is an extremely complex problem that remains beyond the scope of this paper. In our experimental proposal, we considered an initial state that is sufficiently (but not too strongly) localised to begin with, in which case the self-gravitational interaction is similar to a negative pressure field that slows down its expansion.  An approximated picture of the dynamics enabled us to grasp the essential features of the problem (see the Appendix for more details). 

All this explains why, in our proposal,  it is not required to be able to measure the coherence length of the wave function of the center of mass of the nanosphere. In our approach it suffices to be able to measure its extent ($<r^2>$), which is a much simpler task than in other proposals, for instance than in those proposals aimed at revealing mechanisms of spontaneous localization through their influence on decoherence \cite{maqro,aspelPRL,aspelPRA}. 

Although such problems remain out of the scope of the present paper, it is worth noting that the experiment proposed by us is also sensitive to certain ``exotic'' models of spontaneous localisation which invoke an external mechanism (such as the Ghirardi-Rimini-Weber (GRW) \cite{grw}, Pearle \cite{pearle} and Continuous Spontaneous Localisation (CSL) \cite{CSL} models) aimed at explaining why in the classical world objects are localized \cite{Bassi}. These models  (which are far more ambitious than ours, as they also aim at solving the quantum measurement problem \cite{Bell,Bassi}) differ from the model studied in our paper in the sense that they are intrinsically indeterministic\footnote{Whether this distinction is fundamental is still an interesting open question. It is known for instance that in certain circumstances non-linearity can act as a noise amplifier. This is the case for instance with optical rogue waves \cite{opticalrogue,Kharif,Akhmediev}, but is not yet clear whether self-gravitation possesses similar features. It could be that the long range behaviour of a self-gravitating system is, to some extent, unpredictable for the same reason that certain chaotic dynamical systems are unpredictable.} (they are sometimes referred to as Stochastic Spontaneous Localisation (SSL) models). It is beyond the scope of the present paper to try to tackle these questions and all what needs to be known, at this level of analysis, is that in these SSL models it is assumed that a stochastic process is present in nature which causes, from time to time, spontaneous quantum jumps to occur, during which the wave function of quantum objects localizes in a region of space of finite size.  

According to Di\'osi and Penrose \cite{diosi87,Diosi,penrose}, who proposed that there is a connection between stochastic spontaneous localisation and self-gravitation, one could postulate the existence of a dissipative mechanism through which self-gravitation will induce macroscopic systems to spontaneously collapse after a typical time \cite{shan2012,Adler2}, of the order of ${\hbar\over \delta E}$ (where $ \delta E$ is the energetic gain associated to the localization process).  Making use of the estimate (\ref{e0}), one would expect the collapse time between a zero energy state and the collapsed ground state, to be of the order of $\hbar/(0.163 G^2M^5/\hbar^2)$.  At the mesoscopic transition ($R\approx 10^{-7}$ m) and at twice the normal density, this provides a collapse time of the order of 10$^{4}$ seconds (i.e., of the order of an hour). This is still a bit too long to be measurable in a realistic experiment, but this collapse time decreases as $R^{-15}$ so that at a radius $R$ of the order of 1$\mu m$ the Di\'osi-Penrose collapse time ought to be nearly instantaneous\footnote{Similarly, if we resort to GRW's original model \cite{grw}, according to which the rate of spontaneous collapses (jumps) is of the order of 10$^{-16}$ s$^{-1}$, multiplied by the number of nucleons (which, in the present case, is of the order of 10$^{10}$), we arrive at a collapse time close to 10$^6$ s. This makes it very unlikely that related effects can be observed in the mesoscopic domain. However, there are several reasons to think that the jump rate should in fact be proportional to the square of the number of nucleons (see e.g. \cite{Tumulte} and references therein), in which case the average time between two jumps would be of the order of 10$^{-4}$ s, which would then lead to measurable effects.}. Therefore, it is not excluded that our experimental proposal would make it possible to discriminate between ``naked'' self-gravitational models on the one hand, similar to those that we considered in the present paper, and spontaneous models on the other hand. It is also worth stressing that previous proposals for discriminating exotic SSL models focus on the decrease of interference due to decoherence (see e.g. \cite{Romero} and references therein) rather than on the direct influence of non-standard mechanisms on trajectories, which is the strategy we chose in our approach. {Despite this difference, the requirements that the experimental setup would demand (the amount of precision needed in the various parts of the experiment, the extent of isolation from the environment that is needed, the availability of experimental control at these scales, the stabilisation of the satellite and so on) are essentially the same in our approach  as in interferometric approaches \cite{maqro}. }

In particular, we studied in depth the interplay between decoherence and self-gravitation in another paper \cite{GRWSELF} . The main by-product of this study is that we were able to show that, to some extent, self-gravitation is robust (insensitive) to decoherence. More precisely, as shown in \cite{GRWSELF}, when the decoherence time is significantly smaller than the duration of the free-fall experiment, the relative influence of decoherence, as compared to self-gravitation, is measured by a dimensionless parameter of the form $\Lambda_{deco}/\Lambda_{crit}$ where
\begin{equation}
\Lambda_{crit.}={G^4M^{11}\over \hbar^7}\label{uh1}
\end{equation}
 in the quantum regime $\sqrt{<r^2>}>2R$, and
 \begin{equation}
 \Lambda_{crit.}={GM^{2}R^{-3}\over \hbar}\label{uh2}
 \end{equation}
 in the mesoscopic regime $\sqrt{<r^2>}<2R$, while $\Lambda_{deco}$ measures the strength of decoherence:
 \begin{equation}
 \Lambda_{deco}=\gamma \cdot \alpha \label{Lambda},
 \end{equation} 
 where $\gamma$ represents the localisation rate (i.e., the inverse of the average time between two spontaneous jumps) and where $\alpha$ is the inverse of the square of the localisation length. What is important is that with our criterion for robustness it can be seen that there exists a window of parameters around the mesoscopic transition which remains open to experimental investigation, even in the presence of environmental decoherence (see figure \ref{fig5-09-11}), provided it is not too strong (see figure \ref{fig4-11-100k8}). As a result, the experiment proposed by us makes it possible to probe not only the possible existence of self-gravitational interactions but also the possible existence of exotic decoherence sources, for instance of the GRW, CSL or Diosi-Penrose type, which would be the manifestation of a universal mechanism of spontaneous localisation {\it \`a la} GRW.

Last but not least, it is worth noting that we systematically assumed, as is usually done in similar studies \cite{penrose}, that the wave function of the center of mass decouples from the other degrees of freedom (see e.g. equation (\ref{NSgengen}) and appendix 2). It would be highly interesting to consider the problem of entanglement in the presence of self-gravitational mechanisms. This would be relevant for instance for tackling the measurement problem, but we would then face serious problems in connection to the interplay between non-linearity and non-locality \cite{gisin,czachor}, a question which has not yet been fully elucidated although serious progress has been made in this direction over the last decade in the framework of the Spontaneous Localisation theories \cite{Bassi}.

It is our hope that experiments will soon make it possible to probe gravitational self-interaction in the mesoscopic regime and we believe that this constitutes a very promising research field, as it allows us to investigate a sector in which quantum gravity effects are likely to be present, thereby effectively bringing gravitation into the quantum realm.

\section*{Acknowledgements}
This work was made possible through the support of a grant from the John Templeton Foundation. 
TD and RW acknowledge support in the past from a mobility grant FWOKN184 
``Solitonen en solitonachtige oplossingen van gedeeltelijk integreerbare niet-lineaire partieel differentiaalvergelijkingen met corpusculair gedrag''.
SC was supported by a PIAF fellowship (Perimeter Institute - Australia Foundations) for the period during which most of this research has been conducted. 
The authors wish to express their gratitude to professor Wiseman (Griffith University) for logistic and administrative support, 
and to professor Lambert (Vrije Universiteit Brussel) for his encouragement during the initial phases of this project.
One of us (TD) acknowledges fruitful discussions with A. Rahmani (University of Technology of Sydney). He also thanks M. Aspelmeyer and the members of his team (Quantum foundations and information on the nano-and microscale)Vienna) for bringing his attention to the recent development of optomechanics and the possibility to trap and cool nanospheres by (quantum) optical methods.

\section*{Appendix 1: non-linear differential approximation of the (non-linear integro-differential partial derivative) Schr\"odinger-Newton equation} \subsection*{An approximated evolution for a self-gravitating nanosphere in the quantum regime}

No exact solution of equation (\ref{NS}) is known, even in the static regime. Moreover, even in the case of the non-linear Sch\"odinger equation (NLS) which has been studied in considerable depth by mathematicians and physicists alike, it is a non-trivial problem to correctly approximate the temporal behaviour of the solutions by means of numerical methods. A fortiori, it might seem illusory to try to accurately approximate the solutions of equation (\ref{NS}) by numerical methods, and the same limitation persists when we consider equation (\ref{NSgen}) and even more so when the self-interaction takes the form of (\ref{fullpot}).  Furthermore, these  equations being at the same time non-linear and non-local, any numerical treatment will be highly time consuming \cite{Giulini,vanMeter}. It is therefore worthwhile to try to develop perturbative schemes that are sufficient to grasp the qualitative behaviour of solutions to \eqref{NS}, if possible at a minimal computational cost. 

{In order to achieve this goal, we approximated (first of all in the quantum regime) the self-interaction of the nanosphere by a harmonic potential, the same that would be generated at short distance in the case where the wave function of the center of mass of the nanosphere is a Heaviside function that is constant between the origin and 
$r^\mathrm{{max}}=\alpha\sqrt{<r^2>}$, with $\alpha$ an adjustable parameter (as outlined in section \ref{appro}). The parameter $\alpha$ is of the order of unity when we confine ourselves to the quantum regime. The resulting potential is harmonic and equal to 
$k^\mathrm{{self}}r^2/2$ where $k^\mathrm{{self}}= GM^2/(\alpha L)^3$ with $L= \sqrt{<r^2>}$. 

We also assumed that, to a good approximation, it remains so for larger values of $r$. We realized that, whenever gaussian wave packets are considered, this perturbative scheme, which essentially consists of replacing the self-gravitational potential of interaction by a harmonic (parabolic) potential of which the spring constant is adjusted in function of the spread of the wave function, allows us to drastically simplify the treatment of the problem.}
Indeed, the corresponding evolution equation reads, in the quantum regime,
\begin{equation}
{i}\hbar\frac{\partial\Psi(t,{\bf x})}{\partial t}=-\hbar^2\frac{\Delta\Psi(t,{\bf x})}{2M}
+{ GM^2\over 2(\alpha\sqrt{<r^2>})^3}r^2 \Psi(t,{\bf x}),\label{TD}
\end{equation}where  $<r^2>=\int\!{d}^3x \big|\psi(\bx,t)\big|^2r^2~$.
This equation is appealing from more than one point of view.
Firstly, its static version possesses a minimal energy ground state of gaussian shape, with width {$\sqrt{<r^2>}=\sqrt{{3\hbar \over 2\sqrt{Mk^\mathrm{{self}}}}}$, where $k^\mathrm{{self}}={ GM^2\over (\alpha\sqrt{<r^2>})^3}$. This yields the value $\sqrt{<r^2>}=9\alpha^3\hbar^2/4GM^3$}, which fits qualitatively with more accurate numerical estimates (see footnote \ref{ABCD}, the discussion at the end of section \ref{2.3} and references therein and the last paragraph of section \ref{appro}). 
Secondly, the equation is separable in Cartesian coordinates when the initial wave function is the product of three functions with identical dependence on $x$, $y$ and $z$ (as is the case for radial gaussian functions). 
Furthermore, as we shall subsequently show, it possesses a remarkable property:
if the initial state is a (complex) gaussian function, its time evolution remains a gaussian function at all times and equation (\ref{TD}) in fact reduces to a system of coupled ordinary differential equations (cf. eqns. (\ref{redsys1}--\ref{redsys3})). 
Thirdly, the equation (\ref{TD}) still grasps, in our view, the most essential features of the original equation  (\ref{NS}). For instance, non-linearity manifests itself through the spring constant $k^\mathrm{{self}}={ GM^2\over (\alpha\sqrt{<r^2>})^3}$, which also intrinsically incorporates non-locality through its dependence on $\sqrt{<r^2>}$ which is a quantity that depends on the global features of the wave function. Finally, there is the added benefit that the numerical integration of the approximate system of ordinary differential equations (\ref{redsys1}--\ref{redsys3}) is obviously much less time consuming than that of the (non-linear integro-differential partial derivative) equation (\ref{NS}). 

Of course, exact solutions of equation (\ref{NS}) do not remain gaussian throughout time, but we nonetheless believe that our approximation is faithful enough to differentiate between the behaviour of wave packets in the case of gravitational self-interaction and in the ``free'' case (i.e., without self-interaction). In fact, equation (\ref{TD}) possesses an invariant $E^{\mathrm{eff}}$,
\begin{equation}
E^\mathrm{eff}=\frac{\hbar^2}{2 M} \int\!{d}^3x~\big|\nabla\psi(\bx,t)\big|^2 - \frac{G M^2}{\alpha^3\sqrt{<r^2>}}~\!,\label{Eeff}
\end{equation}
which mimics the exact energy (\ref{energy}) associated to equation (\ref{NS}). 
In particular, the kinetic energy term in \eqref{Eeff} being always positive, it is easily shown that solutions to equation (\ref{TD})  with negative energy $E^\mathrm{eff}$, necessarily remain trapped by their own (approximated) gravitational potential. Indeed, at any time $t$, 
$$0\leq \frac{\hbar^2}{2 M} \int\!{d}^3x~\big|\nabla\psi(\bx,t)\big|^2 = \frac{G M^2}{\alpha^3\sqrt{<r^2>}} + (\frac{\hbar^2}{2 M} \int\!{d}^3x~\big|\nabla\psi(\bx,0)\big|^2 - \frac{G M^2}{\alpha^3\sqrt{<r^2>}|_{t=0}} )$$
or,
$$ -\frac{G M^2}{\alpha^3\sqrt{<r^2>}}\leq (\frac{\hbar^2}{2 M} \int\!{d}^3x~\big|\nabla\psi(\bx,0)\big|^2 - \frac{G M^2}{\alpha^3\sqrt{<r^2>}|_{t=0}} )< 0~\!,$$
which shows that $<r^2>_t$ is bounded from above for all times $t$. This is confirmed by numerical simulations, the results of which are given in figure \ref{fig4.5}.

\begin{figure}
\centering
\includegraphics[width=\textwidth]{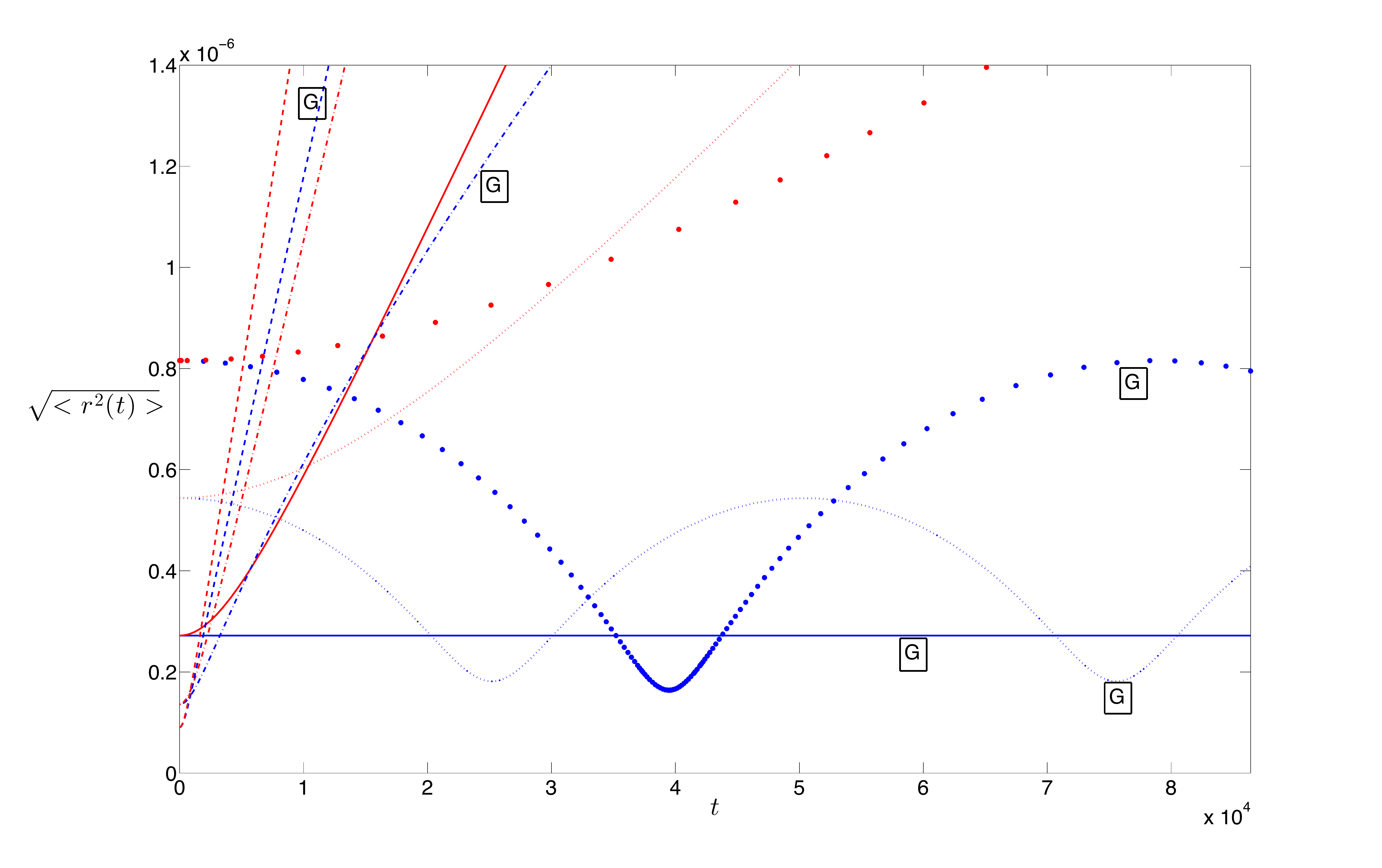}
\caption{\label{fig4.5} $\sqrt{<r^2>}_t$ as a function of $t$ for a homogeneous nanosphere of radius $10^{-7}~\mathrm{m}$ and mass density 
$2650~\mathrm{kg}\,\mathrm{m}^{-3}$. Identical markers are shared by the curves with the same initial conditions, for both the self-interacting case (\ref{NS}), marked by a boxed $G$, as well as the free case (\ref{libre}) (unmarked)
for various initial values for the mean width of the state. 
The final time corresponds to $24$ hours. }
\end{figure}

{Moreover, for a (real) gaussian wave function $(1/(\pi a^2)^{3/4})\cdot \exp(-r^2/2a^2)$, the kinetic energy is equal to $3\hbar^2/4Ma^2$ and its ``effective'' potential energy yields $-GM^2/(\alpha^3\sqrt{3/2}a)$. The energy $E^\mathrm{eff}$ is minimal for $a_0 = (2/3)^{1/2}~\! \sqrt{<r_0^2>}=(3/2)^{3/2}\alpha^3\hbar^2/GM^3$ which corresponds\footnote{This is to be compared with the aforementioned numerical estimate \cite{Bernstein, Harrison} of $-0.163G^2M^5/\hbar^2$ and a rough estimate of $-(1/8)G^2M^5/\hbar^2$ derived  in ref.\cite{Giulini}, in a very similar fashion (for $a_0\approx 2\hbar^2/GM^3$), on the grounds of dimensional arguments.} to an effective energy $E^\mathrm{eff}=-(2/9)G^2M^5/\hbar^2\approx -0.222~\!G^2M^5/\hbar^2$ (for $\alpha=1$).} It is easily confirmed that the gaussian wave packet that minimizes the energy is also the equilibrium point for the equations (\ref{redsys1}--\ref{redsys3}). Finally, independent of the value of $\alpha$, we find that the effective energy of a gaussian wave packet becomes positive for $a\leq a_0/2$ and that this renders its localisation unstable (in accordance with the results in \cite{Arriola} which we remarked upon in footnote \ref{blobl}). This is confirmed by the plots in figure \ref{fig4.5}.

\subsection*{Approximated evolution for a self-gravitating nanosphere valid in all regimes}

In principle, the method we chose to treat the evolution of gaussian wave functions in the quantum or ``hyperbolic'' regime (corresponding to the bound $\sqrt{<r^2>}\gg R$ and depicted by the ``H'' curve in figure 2 as well as by the region D in figure 1) is also valid in the mesoscopic or ``parabolic'' regime (corresponding to the bound $\sqrt{<r^2>}\ll R$ depicted by the ``P'' curve in figure 2, as well as by the region C in figure 1), because in that case it is fully justified to approximate the self-gravitational potential by a harmonic potential. Indeed, as we have shown in section \ref{bla1}, the bound state in this (mesoscopic) region is the same as that we would obtain if the potential were harmonic and of the form $V_{\alpha=1}( r)=(GM^2/R)(-6/5+d^2/2)$ with $d= r/R$.  The associated conserved quantity then reads
\begin{equation}E_{\alpha=1}^\mathrm{eff}=\frac{\hbar^2}{2 M} \int\!{d}^3x~\big|\nabla\psi(\bx,t)\big|^2 -6GM^2/5R+ \frac{G M^2(\sqrt{<r^2>})^2}{R^3}.
\end{equation}

In the region, say, $0.1R\leq \sqrt{<r^2>} \leq R$, the effective potential  depicted by the I curve in figure 2 interpolates between the parabolic ($\sqrt{<r^2>}\ll R$) and hyperbolic regimes ($\sqrt{<r^2>}>2R$), and is given by the equation
\begin{equation}V^\mathrm{eff}_{\alpha=1}( \sqrt{<r^2>})=(GM^2/R)(-6/5+\tilde d^2/2-3\tilde d^3/16+\tilde d^5/160),\label{ouf}\end{equation} 
with $\tilde d= \sqrt{<r^2>}/R$. Moreover, one can show that the equation \begin{equation} {i}\hbar\frac{\partial\Psi(t,{\bf x})}{\partial t}=-\hbar^2\frac{\Delta\Psi(t,{\bf x})}{2M}
+ k^\mathrm{{self}}(\sqrt{<r^2>})/2)r^2 \Psi(t,{\bf x}),\label{TDgen}\end{equation} where  $<r^2>=\int\!{d}^3x \big|\psi(\bx,t)\big|^2r^2~$ possesses a conserved quantity 
\begin{equation}
E^\mathrm{eff}=\frac{\hbar^2}{2 M} \int\!{d}^3x~\big|\nabla\psi(\bx,t)\big|^2 -V^{eff}(\sqrt{<r^2>}),
\end{equation}
where 
\begin{equation}
V^{eff}(z)=\int z\cdot k^\mathrm{{self}}(z) dz~.
\end{equation}
All this suggests we generalize our method by introducing (in the case $\alpha=1$) a harmonic potential defined as:
 \begin{equation}
 V_{\alpha=1}(\sqrt{<r^2>},r)=(GM^2d^2/R)(1/2-(9/32)\tilde d+(1/64)\tilde d^3),
 \end{equation}
 with  $d=r/R$ and $\tilde d=\sqrt{<r^2>}/R$  when $0\leq\sqrt{<r^2>}\leq 2R$, and 
 \begin{equation}
 V_{\alpha=1}(\sqrt{<r^2>},r)=(GM^2/R)( d^2/2( \tilde d)^3)={ GM^2r^2\over 2(\sqrt{<r^2>})^3},
 \end{equation}
 when $\sqrt{<r^2>}>2R$. This potential, as well as the associated conserved quantity $V^{eff}(\sqrt{<r^2>})$ which obeys (by construction) equation (\ref{ouf}), interpolate between the parabolic and hyperbolic regimes, around the mesoscopic transition.
It,  of course, also possesses the same properties as the potential considered at the level of equation (\ref{TD}) (simple behaviour of gaussian wave functions, conserved energy and so on), because it remains quadratic in $r$, despite the complicated dependence of its spring constant on $\sqrt{<r^2>}$ (which reflects the intrinsic complexity of the gravitational self-interaction due to the contributions related to the size of the nanosphere).
 
If, in the above, a value of $\alpha$ different from 1 is needed, the above effective potential can be easily generalized 
and corrections due to the discrete, atomic, structure of the nanosphere can also be treated in a similar fashion. In the latter case however, $R$ ought to be replaced by $10^{-15}$ m, and $G$ by $G^{atomic}=G/N$, where $N$ is the number of nuclei of the nanosphere. The contributions to the effective potentials by a homogeneous nanosphere and by the nuclei are illustrated  in figure \ref{figbla}. These results are complementary to those plotted in figure 1, and were all obtained for the same nanosphere. As mentioned before, in the regimes in which we work, the influence on the trajectories due to the discrete, atomic, structure of the nanosphere, is negligible (as can be seen e.g. in figures \ref{simulation2} and \ref{simulation1}).
\begin{figure}
\centering
\includegraphics[width=1.0\textwidth]{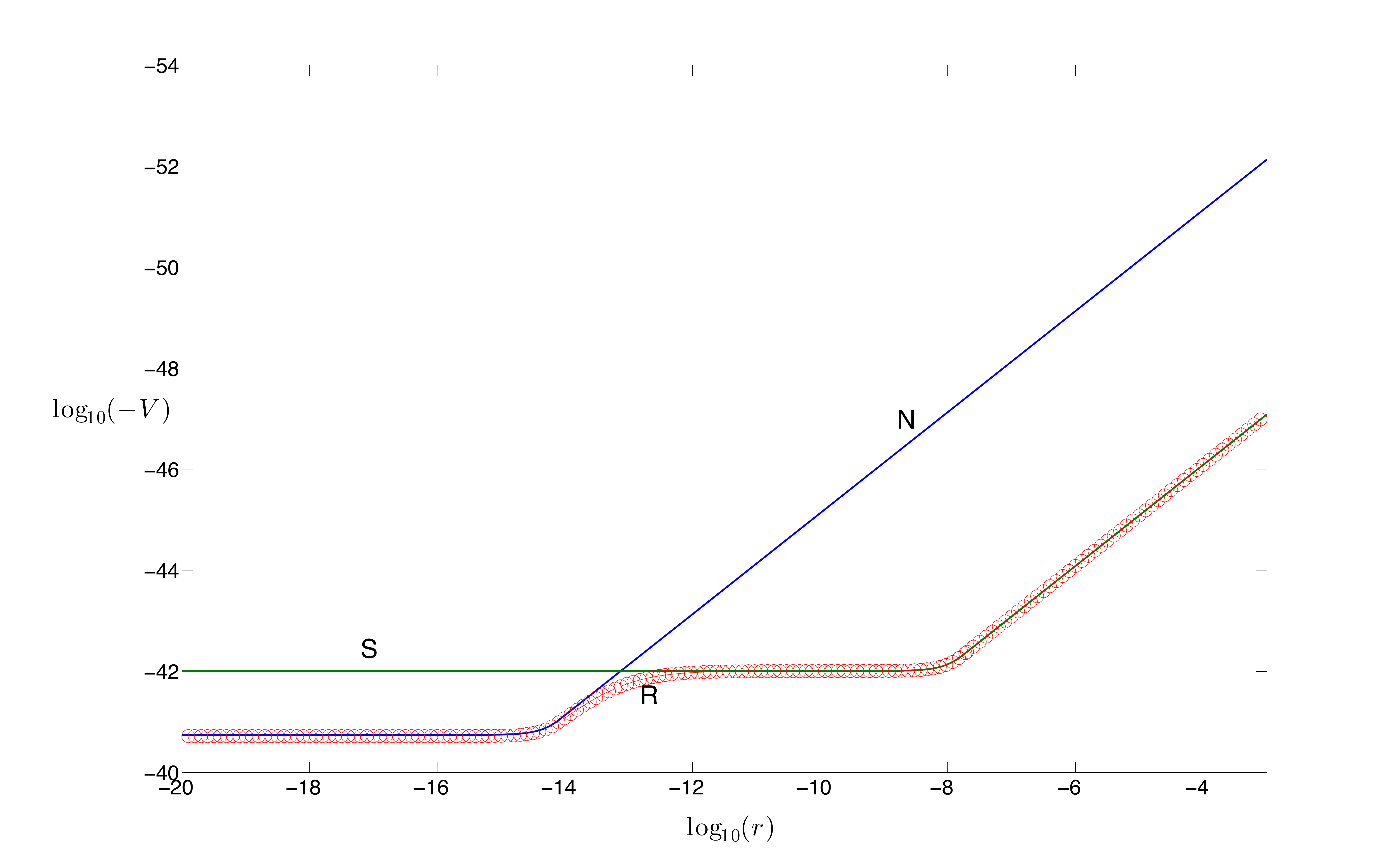}
\caption{\label{figbla} Same as in figure 1, with separated plots for the contributions of a homogeneous sphere (S) and of the nuclei (N).
The curve with the circles (R) is the sum of the two contributions.}
\end{figure}

\subsection*{Numerical simulations}
We assume  that equation (\ref{TD}) (at $\alpha=1$) admits complex spherically-symmetric Gaussian functions of unit norm as solutions.
That is, we write $\Psi(t,{\bf x})$ as $\psi(t,x)\psi(t,y)\psi(t,z)$ where
\begin{equation}\label{gau}
\psi(t,u)=e^{-\frac{a(t) u^2}{2L^2}} e^{-i\frac{b(t) u^2}{2L^2}}e^{-i \mathfrak{Et}(t)/\hbar}\sqrt{\frac{1}{L}\sqrt{\frac{a}{\pi}}}~,
\end{equation}
(for $u$ either $x$, $y$ or $z$) where $a$, $b$ and $\mathfrak{Et}$ are real functions of $t$ and where $L$ is a characteristic length that can be freely adjusted. 
Note that $\sqrt{<u^2>}=L/\sqrt{2a}$ and therefore $\sqrt{<r^2>}=\sqrt{3}L/\sqrt{2a}$. 

If we insert the above expression for $\Psi(t,{\bf x})$ in equation (\ref{TD}), we obtain three decoupled differential equations for $x$, $y$ and $z$
\begin{equation}\label{TDu}
i\hbar\frac{\partial\psi(t,u)}{\partial t}=-\frac{\hbar^2\partial^2_u\psi(t,u)}{2M}+k(t)\frac{u^2}{2}\psi(t,u)~.
\end{equation}
Furthermore, inserting  the ansatz (\ref{gau}) into equation (\ref{TDu}), we obtain the following system of differential equations:
\begin{align}
\frac{\partial a}{\partial t} L^2 M-2 a b \hbar=0\label{redsys1}\\
4\frac{\partial\mathfrak{Et}}{\partial t} L^2 M-2 a\hbar^2=0\label{redsys2}\\
\frac{\partial b}{\partial t}\hbar L^2 M-k M L^4+(a^2-b^2)\hbar^2=0~,\label{redsys3}
\end{align}
which can be solved numerically without great difficulty (using for instance the Runge-Kutta-Felhberg method with time-adaptive step). 
In order to get rid of the appearances of $\hbar$, for which we do not have sufficient numerical precision, we define $k_0=\frac{GM^2}{R^3}$ and 
$L=\sqrt{\frac{\hbar}{\sqrt{k_0 M}}}$. This yields the system of equations:
\begin{align}
\frac{\partial a}{\partial t}=2 a b \sqrt{\frac{k_0}{M}}\\
\frac{\partial(\mathfrak{Et}/\hbar)}{\partial t}=\frac{a}{2}\sqrt{\frac{k_0}{M}}\\
\frac{\partial b}{\partial t}=\frac{k}{k_0}\sqrt{\frac{k_0}{M}}+(b^2-a^2)\sqrt{\frac{k_0}{M}}~.
\end{align}

Note that we obtain a bound state for $\frac{\partial a}{\partial t}=0$, which implies that $b_{BS}=0$ and that $\frac{\partial b}{\partial t}=0$, which in turn implies that 
$a^2_{BS}=\frac{k}{k_0}$. The function $k$ depends on $a$ through the relations
\begin{align}
k=\frac{GM^2}{R^3}(1-\frac{9}{16}\frac{\sqrt{<r^2>}}{R}+\frac{1}{32}\left(\frac{\sqrt{<r^2>}}{R}\right)^3)\quad\textrm{when}~~\sqrt{<r^2>}< 2R\nonumber\\
k=\frac{GM^2}{(\sqrt{<r^2>})^3}\quad\textrm{otherwise}~.
\label{rel}\end{align}
with $\sqrt{<r^2>}=\sqrt{3}\frac{L}{\sqrt{2a}}$. Information on the bound states can be obtained from the constraint $a^2_{BS}=k/k_0$, as a function of $R$, where $k$ depends on $a$ through (\ref{rel}), together with the constraints $\sqrt{<r^2>}=\sqrt{3}\frac{L}{\sqrt{2a}}$ and $L=\sqrt{\frac{\hbar}{\sqrt{k_0 M}}}$.
In particular, if $\sqrt{<r^2>}\geq 2R$, one has a simple expression for the width of the bound state
\begin{equation}
\sqrt{<r^2_{BS}>}=\frac{9}{4}\frac{\hbar^2}{GM^3}~. 
\end{equation}
Otherwise, the width of the bound state has to be obtained numerically, the results of which are shown in Fig. (\ref{figbound}).
\begin{figure}
\centering
\includegraphics[width=1.0\textwidth]{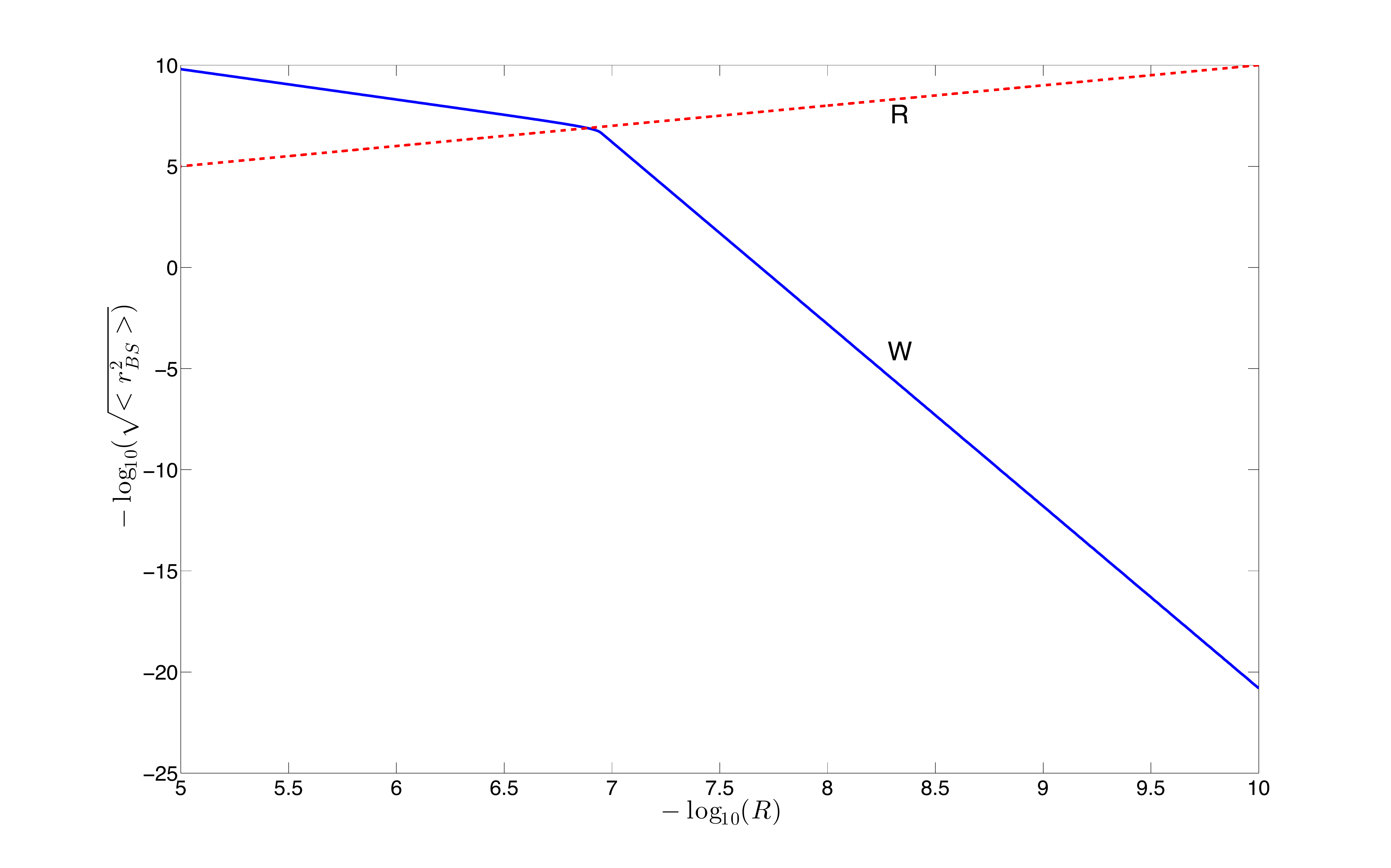}
\caption{\label{figbound}Width of the bound state as a function of the radius of the sphere $R$. The point at which the R line (radius of the sphere) 
intersects the W curve corresponds to the mesoscopic transition. }
\end{figure}

{\section*{Appendix 2: derivation of an effective self-interaction potential in the many particle case.}
Although the single particle Schr\"odinger-Newton equation (\ref{NS}) has attracted a lot of attention in the past, it is its many particle generalisation (\ref{NSgen}) that is the fundamental equation describing self-gravity in the case of a composite system. Now, equation (\ref{NSgen}) is even more complex than its single particle counterpart (\ref{NS}), amongst others because it necessarily entails entanglement between the positions of the different particles. Therefore, approximations must be made in order to be able to infer interesting information about the temporal evolution of the variables under interest. The priviledged observable in our approach is the center of mass of the system. Similar to interferometry, where it is common to center the observations around a dark fringe in order to improve the sensitivity of the interferometer, the center of mass is optimal if we desire to reveal the existence of self-gravitational interaction because it is unaffected by internal (e.g. Coulombian) interactions. This is because, as is well-known, the Schr\"odinger equation is separable into relative and center of mass (C.M.) coordinates whenever interactions between the subsystems respect action-reaction, as is the rule in the standard linear case. The situation changes dramatically when self-gravitation is present, as the C.M. position {\em is} affected. Moreover, what makes the C.M. position a particularly good candidate for probing the existence of self-gravitation is that it is insensitive to all kinds of internal (in particular electro-magnetic) interactions that would otherwise pollute its evolution \cite{penrose}. It is therefore crucial to justify why it is legitimate to assume, as we did throughout the paper, that
\begin{itemize}
\item[(1)] equation (\ref{NSgen}) admits factorisable solutions of the form 
\begin{equation}\Psi(t,{\bf x}_1,{\bf x}_2,\hdots,{\bf x}_i,\hdots{\bf x}_N)= \Psi_{CM}(t,{\bf x}_{CM})\Psi_{rel}(t,{\bf x}_{rel,1},\hdots,{\bf x}_{rel,i},\hdots{\bf x}_{rel,N})~.\label{factfact}\end{equation}
\item[(2)] equation (\ref{NSgen}) splits into the two equation system (\ref{NSgengen}) which consists of the usual Schr\"odinger equation without self-gravity for the relative coordinates and of a reduced Schr\"odinger for the C.M. position in which there appears an effective potential obtained by averaging the self-gravitational potential over the relative coordinates. 
\item[(3)] In the case of a rigid sphere, the effective potential to which the C.M. position is subjected, can be cast in the form (\ref{self-sphere}).
\end{itemize}

{Di\'osi already wrote an equation similar to  (\ref{self-sphere}) in 1984 \cite{Diosi84}, but, to the best of our knowledge, no solid argumentation can be found in the literature explaining why and how the above separation occurs. Actually, there are several puzzles that need to be addressed in the present context. Firstly, is it legitimate to factorize the solutions into relative and C.M. coordinates? If the answer is yes, then what is the form of the two reduced equations that we obtain? We shall adopt two strategies to tackle these questions: a first, qualitative, approach where we consider the C.M. degrees of freedom as a test-particle relative to the relative coordinates in the same sense that in classical gravitation theory, for instance, the earth is a test particle relatively to the sun (to the extent that we may neglect its gravitational back reaction or feedback on the sun's trajectory). We shall also develop a second , more quantitative, approach where factorisability and its corollary, entanglement, are the key concepts. In the first case we shall assume, for simplicity, that the system is a rigid sphere, but generalisations are straightforward. Let us assume that at time $t$ the wave function of the system factorizes according to equation (\ref{factfact}). The main contribution to the self-gravitational energy }

\begin{multline*}
V^{self}(t,{\bf x}_1,{\bf x}_2,\hdots,{\bf x}_i,\hdots{\bf x}_N)=\\
- G\!\sum_{i,j=1,2...N}m_im_j\int d^{3N} x'\frac{|\Psi(t,{\bf x}'_1,{\bf x}'_2,\hdots,{\bf x}'_j,\hdots{\bf x}'_N)|^2}{|{\bf x}_i -{\bf x}'_j|}\Psi(t,{\bf x}_1,{\bf x}_2,\hdots,{\bf x}_i,\hdots{\bf x}_N)
\end{multline*}
{is due to nuclei and we can neglect electronic contributions. Nuclei are well localized and vibrate around their equilibrium positions which form a crystalline lattice. As we show in sections \ref{cristal} and \ref{nuclear2}, when $N$ (the number of atoms) is sufficiently high (this concerns the mesoscopic transition which is the regime of interest for studying effects of self-gravity), the gravitational self-energy of the crystal is very close to the gravitational self-energy of a homogeneous spherical distribution, due to the fact that its main contribution scales like $N^2$. It can thus be considered to be independent of the small variations of the positions of the nuclei inside the crystal, relative to the C.M.. Moreover, it is consistent to neglect the dependence of $V^{self}$ in the relative coordinates because, obviously, self-gravity is overwhelmed by interactions of the Coulomb type. This legitimates our main approximation: }

$$V^{self}(t,{\bf x}_1,{\bf x}_2,\hdots,{\bf x}_i,\hdots{\bf x}_N)\approx V^{self}(t,{\bf x}_{CM}).$$

{In order to evaluate its numerical value, we may consistently average $V^{self}(t,{\bf x}_1,{\bf x}_2,\hdots,{\bf x}_i,\hdots{\bf x}_N)$ over the wave function $\Psi_{rel}(t,{\bf x}_{rel,1},\hdots,{\bf x}_{rel,i},\hdots{\bf x}_{rel,N})$ and we find (introducing the convenient notation ${\bf x}_{rel,1},\hdots,{\bf x}_{rel,i},\hdots{\bf x}_{rel,N}={\bf x}_{rel}$) that }

$V^{self}(t,{\bf x}_{CM})\approx$

$- G
\sum_{\tiny i,j=1}^N\!\!\!m_im_j\int d^{3}x'_{CM} |\Psi_{CM}(t,{\bf x'}_{CM})|^2 \int d^{3N-3} x_{rel}|\Psi_{CM}(t,{\bf x_{rel}})|^2 \int d^{3N-3} x_{rel}' ~\frac{|\Psi_{rel}({\bf x'}_{rel})|^2}{|{\bf x}_i -{\bf x}'_j|}$ 

$\approx -G({M\over {4\pi R^3\over 3}})^2\int_{|\tilde{\bf x}| \leq R, |\tilde{\bf x}'| \leq R} {d}^3 \tilde x {d}^3 \tilde x'\frac{1}{|{\bf x}_{CM} +\tilde{\bf x}-({\bf x}_{CM}'+\tilde{\bf x}')|}$

{in accordance with equation (\ref{self-sphere}). The question of factorisability is now easy to tackle. Indeed, the potential can be split into the self-gravitational potential which acts only on the C.M. coordinates and the internal interaction potential which acts only on the relative coordinates:}
\begin{equation} V({\bf x}_{CM},{\bf x}_{rel})=V^{self}_{C.M.}\otimes Id._{rel}+Id._{C.M.}\otimes V^{int}_{rel}.\end{equation}
{In virtue of the Leibniz rule, it is straightforward to check that factorisability into C.M. and relative coordinates is preserved over time, in accordance with the system of equations (\ref{NSgengen}).}

{The second approach is based on the study of entanglement between C.M. and relative coordinates. As was shown by one of us \cite{DurtZeit}, if at time 0 the state of the system factorizes into the product of individual wave functions assigned to two subsystems (in our case these are the relative and C.M. degrees of freedom), the first derivative relative to time of the rate of increase of the entanglement (measured by $Tr.\rho_{red}-Tr.\rho_{red}^2$, which is equal to $1$ minus the purity of the reduced density matrix of one of the subsystems) is equal to zero, while its second derivative is equal (up to a multiplicative factor $\hbar^{-2}$) to the modulus squared of the $L^2$ norm of the projection of $H\Psi_{CM}(t=0,{\bf x}_{CM})\Psi_{rel}(t=0,{\bf x}_{rel})$ onto the sub-Hilbert space bi-orthogonal to $\Psi_{CM}(t=0,{\bf x}_{CM})\Psi_{rel}(t=0,{\bf x}_{rel})$. Let us denote this projection $(H\Psi_{CM,rel.})_{bi-ortho}$. It is easy to check that $(H\Psi_{CM,rel.})_{bi-ortho}=(V^{self}\Psi_{CM,rel.})_{bi-ortho}$, because neither the kinetic energy operators nor the internal interaction potential couple the initial state to its bi-orthogonal space.}

{In the previous paragraph we showed that, to a good approximation, $V^{self}({\bf x}_{CM},{\bf x}_{rel})=V^{self}_{C.M.}\otimes Id._{rel}$, in which case the coupling to the bi-orthogonal space remains equal to 0. Now, this was an approximation and one might wish to have a more quantitative argument at ones disposal. This suggests to estimate the rate of generation of entanglement between relative and C.M. coordinates, using its Taylor development, in which case we get}

$$Tr.\rho_{red}-Tr.\rho_{red}^2\approx {\big\|(V^{self}\Psi_{CM,rel.})_{bi-ortho}\big\|_{L^2}^2~\!t^2\over \hbar^2}~.$$}

{It is not an easy task to evaluate $\big\|(V^{self}\Psi_{CM,rel.})_{bi-ortho}\big\|_{L^2}^2$ but clearly, 
$$\big\|(V^{self}\Psi_{CM,rel.})_{bi-ortho}\big\|_{L^2}^2\ll\big\|(V^{self}\Psi_{CM,rel.})\big\|_{L^2}^2$$
so that it is easy to derive a (very crude) upper bound for the entanglement rate at time $t$. Indeed, as explained in sections \ref{cristal} and \ref{nuclear2}, the gravitational self-energy of a rigid sphere of radius $R$ and mass $M$ is bounded by below in the meso and macro regimes\footnote{The gravitational self-energy of a rigid sphere obviously goes to zero in the quantum regime which, of course, justifies why it is fully legitimate to resort to the single particle N-S equation in this limit case.}, by the quantity $-(6/5)GM^2/R$. For instance, for a nanosphere of radius equal to 100 nanometer, at the mesoscopic transition, we find }{$Tr.\rho_{red}-Tr.\rho_{red}^2\ll ({6GM^2 t\over 5\hbar R})^2$ which guarantees that no entanglement will be generated between the C.M. and relative degrees of a freedom during at least 1200 seconds. In realistic (extreme) vacuum conditions, a collision between an atom/molecule of residual gas occurs every 200-300 seconds \cite{GRWSELF}, which imposes that the duration of the experiment is bounded by, say, 300 seconds. The entanglement rate after this period is thus at most of  6 \% and it is legitimate to suppose that the factorisation between C.M. and relative coordinates is preserved during that period.}






{\section*{Appendix 3: Newtonian self-energy of a rigid sphere.}
\subsection*{Rigid homogeneous spheres.}
In order to evaluate the potential of self-interaction between two spheres in function of the distance $d$ between their centres, one must integrate the Newtonian potential resulting from the gravitational attraction caused by one sphere (the left hand side one in figure \ref{picasso1}) over the second sphere (on the right). We integrate along slices of constant $r$, in red in figure \ref{picasso1}, distinguishing two regions, $r$ $<$ $R$ (above in figure \ref{picasso1}) and $r$ $>$ $R$ (below). When $r$ $<$ $R$, the potential is proportional to $-(3/2)+(1/2)(r^2/R^2)$, otherwise it varies like $1/r$ in accordance with Gauss's theorem. The solid angular opening of a slice of constant $r$ is equal to $(2\pi(R^2-(r-d)^2)/2dr)$. The resulting energy of interaction obeys therefore}
$$V^\mathrm{eff}(d) = \frac{-3GM^2}{2R^4}~\!\left(\int_{r=d-R}^{r=R}dr r {R^2-(r-d)^2\over 2d} ({3\over 2}-{r^2\over 2R^2})+\int^{r=d+R}_{r=R}dr r {R^2-(r-d)^2\over 2d} {R\over r}\right).$$
{A lengthy but straightforward integration leads to the compact expressions (\ref{fullpot},\ref{fullpot2}).}
\begin{figure}
\centering
\includegraphics[width=0.4\textwidth]{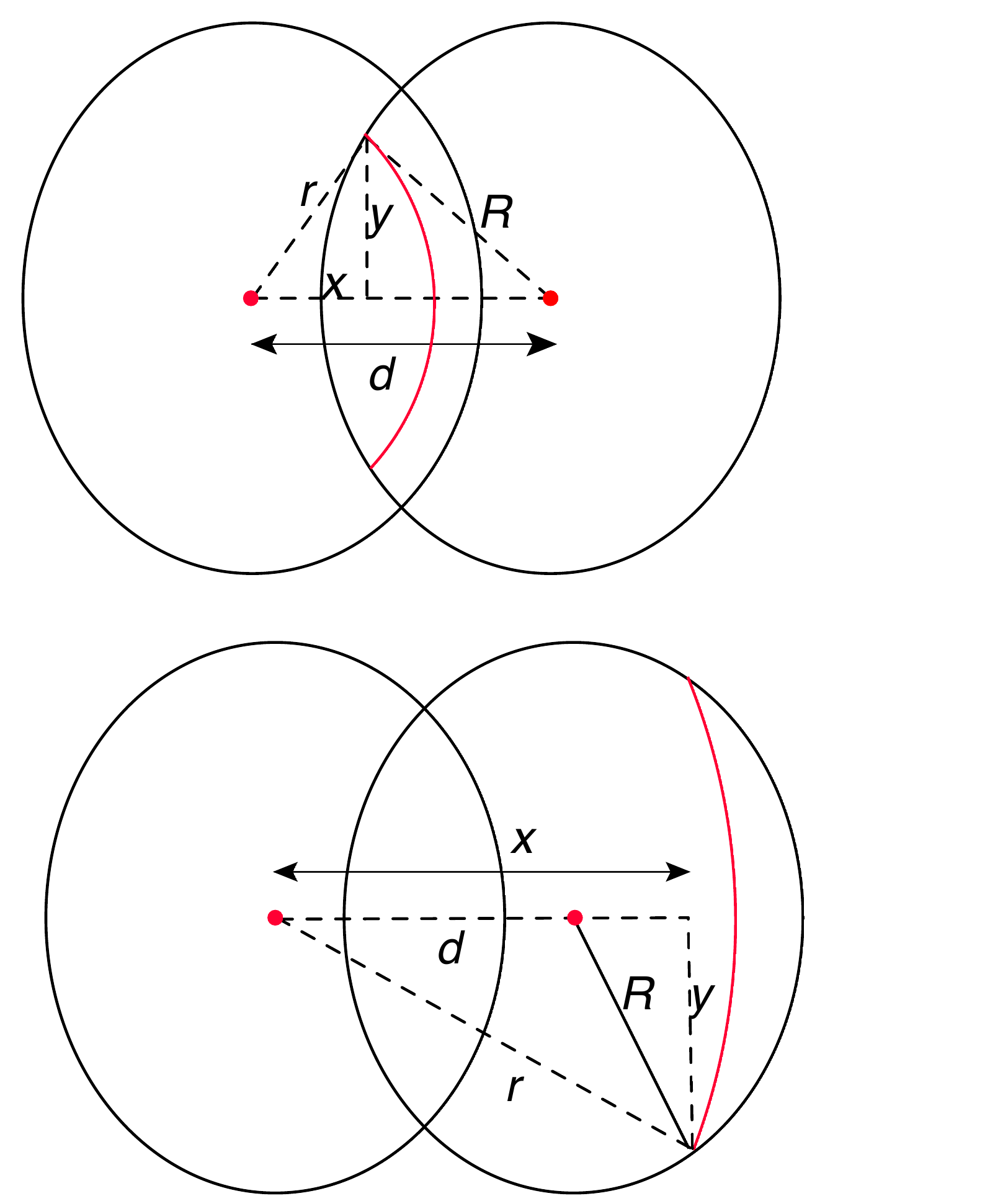}
\caption{\label{picasso1}Parametrisation of the overlapping (above) and non overlapping (below) regions between the two spheres}\end{figure}
{\subsection*{Structure corrections.}
In order to take into account possible inhomogeneities of the mass distribution inside the spheres, we modelize it as a collection of spherical homogeneous nuclei (see e.g. the labels $i,j$ ($i',j'$) in figure \ref{picasso2}). The exact self-interaction is a sum of self-interactions between homogeneous spheres of radius $\Delta x_{zp}$ where, as discussed in section \ref{compa}, for a Si crystal,  $\Delta x_{zp}$ is of the order of $5\cdot 10^{-12}$ meter. Now, the interaction potential between two non-overlapping nuclei (like $i$, $j'$ on figure  \ref{picasso2}) is the same  (in virtue of Gauss's theorem) as the interaction potential between two non overlapping atoms ($k$, $l'$ on figure  \ref{picasso2}) of homogeneous density  and of same mass (we consistently neglect electronic contributions), so that in good approximations we can write }

$$V^\mathrm{eff}_{exact}(d) = \sum_{i} V_{nucleic}^\mathrm{eff}(d_{ii'}) +  \sum_{j\not= i} V_{nucleic}^\mathrm{eff}(d_{ij'})=\sum_i V_{nucleic}^\mathrm{eff}(d_{ii'}) +  \sum_{k \not= l} V_{atom}^\mathrm{eff}(d_{kl'}).$$
{Besides, the interaction between overlapping atoms is quite smaller than the interaction between overlapping nuclei (in a ratio $\Delta x_{zp}$ /$a_0$ where $a_0$ is the Bohr radius) so that}

$$V^\mathrm{eff}_{exact}(d) \approx \sum_{i} V_{nucleic}^\mathrm{eff}(d_{ii'}) +  (\sum_{k = l} V_{atom}^\mathrm{eff}(d_{kl'})  +\sum_{k \not= l} V_{atom}^\mathrm{eff}(d_{kl'})).$$
{Now, $(\sum_{k = l} V_{atom}^\mathrm{eff}(d) + \sum_{k \not= l} V_{atom}^\mathrm{eff}(d))\approx V_{hom}^\mathrm{eff}(d) $, where $V_{hom}^\mathrm{eff}(d) $ is the potential of self-interaction obtained for spheres of homogeneous density (more or less $(\Delta x_{zp}/a_0)^3\approx 1/8000$ times the nucleus density), so that $V^\mathrm{eff}_{exact}(d) \approx V^\mathrm{eff}_{hom}(d)+\sum_{i=j} V^\mathrm{eff}_{nucleic}(d)$, which leads to equation (\ref{chen}).}
\begin{figure}
\centering
\includegraphics[width=0.4\textwidth]{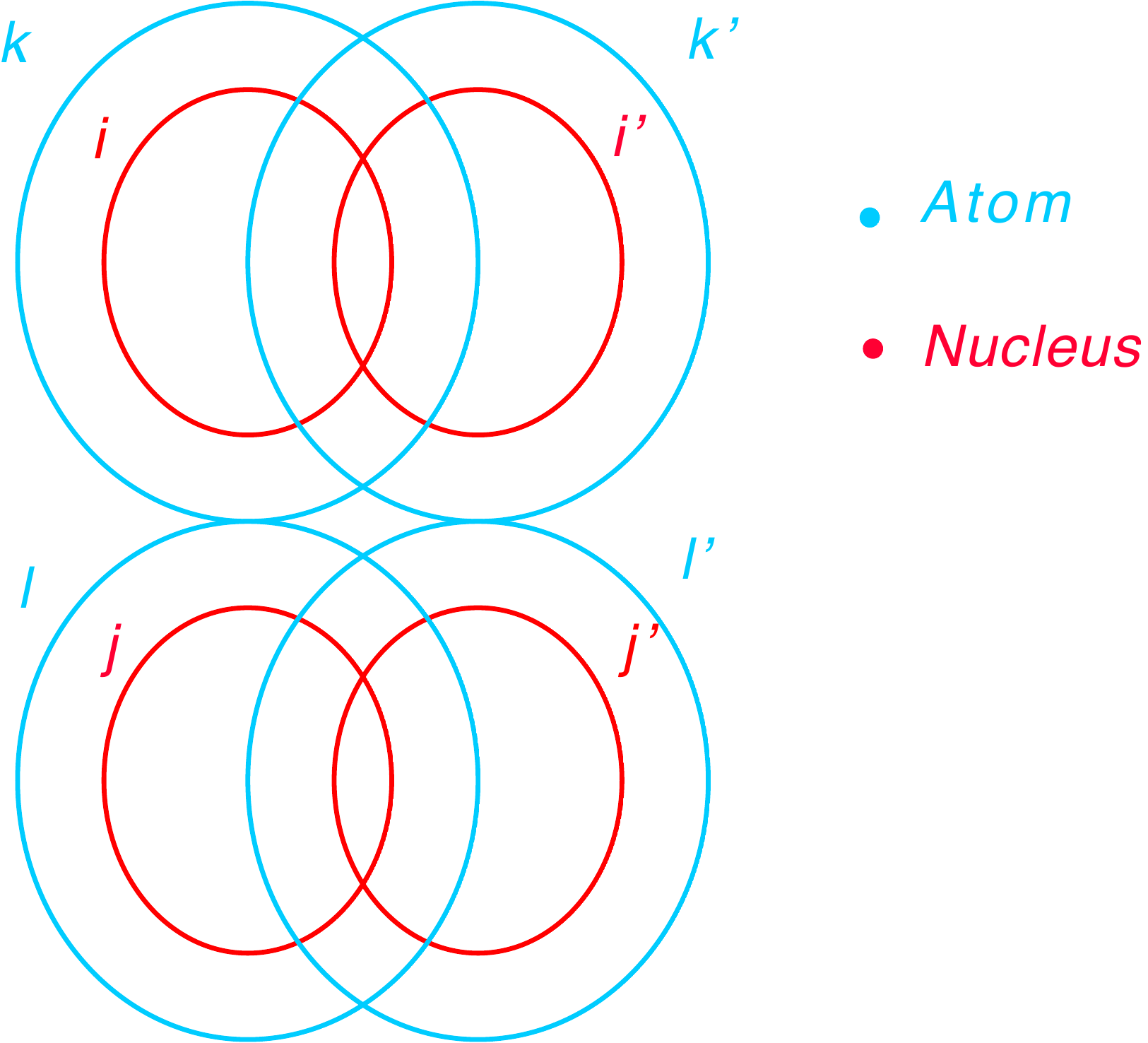}
\caption{\label{picasso2}Schematic representation of the atoms (k,l,k',l') and nuclei (i,j,i',j') inside two localisations of a same rigid nanosphere.}\end{figure}

\end{document}